\documentclass[10pt,fleqn]{article}
\usepackage{arxiv-style}
\hypersetup{
 pdftitle={GroArnold: A Framework for Constructing Action-based Models for Triaxial Galaxies and Dark Matter Halos},
 pdfauthor={Jian-Yu Gu, Xiao-Bo Dong, Shi-Yan Zhong, Run-Yu Meng},
 pdfsubject={Research article accepted by Research in Astronomy and Astrophysics}
}

\title{GroArnold: A Framework for Constructing Action-based Models for Triaxial Galaxies and Dark Matter Halos}
\author[1,2]{Jian-Yu~Gu}
\author[1,2]{Xiao-Bo~Dong}
\author[3]{Shi-Yan~Zhong}
\author[1,2]{Run-Yu~Meng}

\affil[1]{Yunnan Observatories, Chinese Academy of Sciences, Kunming, Yunnan 650216, China; \mbox{\it xbdong@ynao.ac.cn}, \mbox{\it gujianyu@ynao.ac.cn}}
\affil[2]{School of Astronomy and Space Sciences, University of Chinese Academy of Sciences, 19A Yuquan Road, Beijing 100049, China}
\affil[3]{South-Western Institute for Astronomy Research, Yunnan University, Kunming, Yunnan 650500, China}
\date{}

\begin{document}

\twocolumn[
\begin{@twocolumnfalse}
\maketitle

\begin{abstract}
Action-based distribution functions $f(\mathbf{J})$ provide a powerful description of stellar systems 
and are increasingly used to model the Milky Way and external galaxies in the era of \textit{Gaia} and large spectroscopic surveys.
%% The wealth of observational data for the Milky Way and external galaxies has enabled advanced modeling techniques, 
%% particularly the fitting of data using distribution functions expressed in the space of action variables, $f(\mathbf{J})$.
%%
Consequently, the construction of physically accurate and dynamically stable $f(\mathbf{J})$ for galactic components has emerged as a key challenge in the field. 
In this study, we present \texttt{GroArnold}, a computational framework designed to construct 
analytical $f(\mathbf{J})$ expressions 
for axisymmetric and triaxial galaxies and dark matter halos. 
The framework integrates modules for $N$-body initial condition generation, $N$-body simulation, angle--action variable estimation, 
and distribution-function fitting into a unified workflow. %% , initiated by a single user configuration file. 
It supports a wide range of density profiles and kinematic properties, 
and is uniquely designed to generate stable $N$-body models for user-specified galactic systems, 
with the aim of obtaining an $f(\mathbf{J})$ that approximates a dynamically stable and physically consistent state. 
%%
%To this end, we extend the Triaxial St\"ackel Fudge (TSF) method of the \texttt{TACT} code of Sanders \& Binney (2015), 
% implemented in our \texttt{Nbody-TACT} module, to enable action estimation directly from $N$-body gravitational potentials.
To this end, we extend the triaxial St\"ackel Fudge method of the \texttt{TACT} code,
implemented in our \mbox{Nbody-TACT} module, to enable action estimation directly from $N$-body gravitational potentials. 
%%%%
\texttt{GroArnold} provides a robust and flexible tool for deriving
numerically empirical, explicitly analytical distribution functions
for user-specified galaxies or halos, thereby facilitating a wide range of practical $f(\mathbf{J})$-based modeling applications.
In addition, it serves as a powerful tool for investigating the dynamics of realistic triaxial galaxies and halos 
beyond the scope of traditional spherical or axisymmetric models.
We assessed the accuracy of the output $f(\mathbf{J})$ 
	by comparing its predicted spatial and velocity distributions with those of the input $N$-body snapshot,
	and confirmed that it provides a reasonably accurate representation of the input.
The code is modular, publicly available, and designed to support future extensions and adaptations.
\end{abstract}

\vspace{1em}
\end{@twocolumnfalse}
]

\section{Introduction}\label{Sec1}

The rapid growth of high-precision, large-volume observational data 
has revolutionized our ability to model galactic structure and dynamics. 
In particular, the exquisite astrometric measurements from \mbox{\textit{Gaia} \citep{Gaia2016_mission}}, %(Gaia Collaboration et al. 2023),
combined with large-scale spectroscopic surveys, 
have provided detailed six-dimensional phase-space information for millions of stars. 
%\Authorfootnote %scpma
This data deluge has created an urgent need for advanced modeling techniques 
that connect spatial structure and kinematics through dynamics, 
rather than prescribing kinematics through \textit{ad hoc}\/ assumptions (cf. \citealt{B20,Binney2026_spheroids}).
\par 

Among the most promising approaches is the construction of dynamical models using distribution functions (DFs) 
with action variables as their arguments, $\DFJ$ (see, e.g., \citealt{B20,Binney2026_spheroids}). 
In this respect, actions are powerful, particularly in modeling a system with multiple components. 
%In recent years, several methods for estimating angle--action variables coordinate $(\mathbf{\theta}, \mathbf{J})$ from position--velocity phase space coordinate $(\mathbf{x}, \mathbf{v})$ has been proposed (see \citet{SB16} for a review). 
In recent years, several methods have been proposed for estimating angle--action coordinates $(\boldsymbol{\theta},\mathbf{J})$ 
from phase-space coordinates $(\mathbf{x},\mathbf{v})$ (see \citet{SB16} for a review). 
\citet{SB15} introduced a method for action estimation in spherical, axisymmetric or triaxial confocal ellipsoidal coordinates 
under a spherical, axisymmetric or triaxial potential in analytical forms -- especially the Triaxial St\"ackel Fudge (TSF) method -- in 
their \texttt{TACT} code (\url{https://github.com/jls713/tact}; \citealt{SB16_TACT_code}), 
which trades some accuracy for computational efficiency. 
This made genuinely triaxial action-based modeling practical. 
%% because they are adiabatic invariants and have clear physical interpretations: 
%% in an axisymmetric system, $(J_r, J_z, J_\phi)$ quantify the amplitudes of
%% radial oscillation, vertical oscillation, and the azimuthal angular momentum, respectively.
%% Moreover, the phase-space volume element is simple in action--angle coordinates,
%% $\mathrm{d}^3\mathbf{x}\,\mathrm{d}^3\mathbf{v} = \mathrm{d}^3\mathbf{\theta}\,\mathrm{d}^3\mathbf{J}$,
%% making action-based DFs convenient for forward modeling, likelihood evaluation, and
%% self-consistent model construction.
\par 

%Substantial progress has also been made in developing analytical $\DFJ$ forms and the associated software tools. 
Substantial progress has also been made in developing analytical forms for $\DFJ$ and the associated software tools. 
For spheroidal components, flexible double-power-law DFs can reproduce a wide range of density profiles, including NFW-like halos and Hernquist models \citep{Posti15}. 
In his latest study, 
\mbox{J.\,Binney} emphasized that analytical DFs for spheroidal or other hot components 
must satisfy regularity conditions near $J_\phi \to 0$ in order to avoid unphysical velocity distributions, 
especially when velocity anisotropy or flattening is introduced \citep{Binney2026_spheroids}. 
For discs, the quasi-isothermal DF \citep{BM11} has been widely used, while
more recent developments have introduced disc DFs that depend on the actions alone and are
therefore better suited to self-consistent modeling \citep{BV23}.
%%%
On the computational side, public libraries such as \texttt{AGAMA} provide efficient tools for action-based modeling and iterative construction of self-consistent multicomponent galaxies \citep{V19}, albeit currently focusing on (oblate) axisymmetric systems.% 
\footnote{
The latest extension of \texttt{AGAMA}, namely \texttt{AGAMAb} \citep{Binney2026_AGAMAb}, 
improves torus generation and implements a torus-based action finder for axisymmetric potentials. 
This removes the specific oblate-only limitation of \texttt{AGAMA} 
and should in principle allow applications to prolate axisymmetric potentials. 
However, \texttt{AGAMAb} does not provide a general action-estimation method for triaxial potentials.
% }.
}
\par 

Despite those advances,% 
\footnote{
	Please see Appendix A for the detailed account of progress in action estimation methods and \DFJ\ forms,
	as well as an introduction to angle--action variables, the advantages of \DFJ\ modeling, and knowledge concerning triaxial St\"ackel Fudge.
} 
several practical challenges remain for $\DFJ$ modeling.
First, most existing analytical $\DFJ$ forms have been derived 
by ``deforming'' or generalizing from spherically symmetric gravitational potentials with known analytical DFs. 
This essentially amounts to %%``guessing'' an approximate functional form for the DF of real galaxies, -Apr.1
adopting an ansatz (treating the functional forms thus approximated 
as the DFs for realistic galaxies), 
%%%  
guided by theoretical expectations rather than by empirical constraints.
Second, such approximation formulae do not necessarily guarantee the dynamical stability of the resulting galaxy
or halo models \citep{BV23}. 
Third, realistic elliptical galaxies, galactic bulges and dark matter halos are often triaxial \citep{Jing2002,LawMajewski2009}, 
%%Vera-Ciro et al. 2011), ; replaced, by the Law+2009, observational triaxiality of MW halo. 
yet both the construction of DFs and the estimation of actions become significantly more difficult 
beyond spherical or axisymmetric potentials.
%% Extending action-based modeling to triaxial systems is essential for 
%% capturing galactic dynamics more realistically and to the full complexity.
\par 

A complementary strategy, which we pursue here and refer to as a \emph{numerically empirical} approach, 
is to infer $\DFJ$ directly from the output of $N$-body simulations.
In this approach, one evolves a galaxy model to a relaxed, quasi-stationary state,
then estimates actions for all particles moving in the gravitational potential (inferred from $N$-body snapshots), 
and finally selects a best \DFJ\ by fitting explicitly analytical forms 
to the resulting distribution of actions of the $N$-body particles.
If the simulated system is indeed stable, the best-fitting $\DFJ$ 
% has a high probability to represent
has a high probability of representing
a dynamically stable model. 
%%  and naturally incorporates triaxial geometries without requiring \textit{ad hoc} deformations of spherical models.
\par 

% -- In the spirit of the \emph{numerically empirical} approach,
In this work, we present \texttt{GroArnold}, a modular framework for constructing stable action-based
dynamical models of axisymmetric and triaxial galaxies and their components, with the main
goal of facilitating the user's practical $\DFJ$-based modeling applications.
In particular, we extend the \texttt{TACT} code \citep{SB16} 
to estimate actions for the output data of $N$-body simulations (e.g., generated by \texttt{Gadget}), 
using the Triaxial St\"ackel Fudge (TSF) for particles moving in a triaxial potential. 
We call the modified code \texttt{Nbody-TACT}, which is incorporated into
\texttt{GroArnold} as Module~3 (see below).%
\footnote{The open-source code of \texttt{Nbody-TACT} is available at \url{https://github.com/starlifting1/Nbody-TACT}\/.}
\par 

Section~\ref{Sec2} presents a user-oriented overview of \texttt{GroArnold}.
Section~\ref{Sec3} describes the programs for generating $N$-body initial conditions and simulation. 
Section~\ref{Sec4} describes the preprocessing of simulation snapshots and the estimation of angle--action variables 
for $N$-body particles moving in a triaxial potential. 
Section~\ref{Sec5} presents DF fits for a multi-component galaxy.
% Section~\ref{Sec6} discusses the applicability, limitations, and future developments of this software framework.
Section~\ref{Sec6} compares the spatial and velocity distributions predicted by the fitted halo DF with those of the input $N$-body snapshot and discusses the applicability, limitations, and future developments of this software framework.
Finally, Section~\ref{Sec7} summarizes the paper. 
\par 

%
%%%
%% %However, deriving the DF directly from the CBE for a self-gravitating system is substantially more difficult than evolving $f$ in a prescribed potential, as  a self-gravitating system has to be made self consistent. 

%% Sec2 begin
\section{Overview of \texttt{GroArnold} } \label{Sec2}

\texttt{GroArnold}%
\footnote{The open-source code of \texttt{GroArnold} is available at \url{https://github.com/starlifting1/GroArnold}\/.}
is organized as a modular pipeline that takes the user
from an observation-level description of a user-specified gravitating system (here termed observational initial conditions) --- for instance,
a multi-component galaxy specified by its density profiles and kinematics ---
all the way to a best-fit $\DFnJ$ in action space. 
%The input specification may, for example, consist of a multicomponent galaxy model defined by density profiles, velocity structure, and component parameters. 
This section provides a ``user's guide'' to each module; technical details are deferred to later sections.
\par

The workflow proceeds as follows. Module~1 generates an $N$-body realization of initial conditions from the user's specification 
and evolves it to a dynamically relaxed and stable state.
%Module~2 pre-processes each snapshot by re-centering the system and aligning it with the principal axes of its mass distribution.
Module~2 preprocesses each snapshot by recentering the system and aligning it with the principal axes of the mass distribution.
Module~3 estimates angle--action variables for the particles using the triaxial St\"ackel Fudge, 
adapted to work directly with the gravitational potential from the $N$-body snapshot.
Finally, Module~4 constructs the numerical $\DFJ$ from the particle data via kernel density estimation (KDE) 
and fits it with a library of analytical functional forms. 
The primary output is the best-fitting analytical $\DFnJ$; optional diagnostic outputs can also be produced.
\par

%% It is organized into several independent modules, each responsible for a distinct stage of the workflow: 
%% initial condition generation, $N$-body simulation, snapshot preprocessing, 
%% action estimation (via the modified \texttt{Nbody-TACT} module), 
%% construction of the numerical $\DFJ$ via KDE,  and finally $\DFnJ$ fitting. 
%%% \texttt{GroArnold} integrates several independent modules: an initial conditions (IC) module, an $N$-body simulation module, 
%%% a data preprocess module, an action estimation module (the aforementioned \texttt{Nbody-TACT}), and a data analysis module.  }

\texttt{GroArnold} builds upon and extends several well-tested open-source codes --- including \texttt{DICE} \citep{Perret14_DICE, DICE_code} 
for $N$-body realization of initial conditions, 
\texttt{Gadget2} \citep{Gadget2} for time evolution of the $N$-body system, 
and \texttt{TACT} \citep{SB16_TACT_code} for action estimation --- but 
\texttt{GroArnold}'s contribution goes beyond simple aggregation. 
The framework introduces a systematic methodology for deriving $\DFnJ$ from stable $N$-body realizations 
of user-specified galaxies or dark matter halos. 
%%%
Following the original open-source codes, modules are implemented in a mix of C/C++, Fortran, and Python. We use Python to control the overall workflow and user interface.
Each module is designed to operate independently, allowing the user to replace or enhance any component as needed. 
By building on familiar, well-established tools (\texttt{DICE}, \texttt{Gadget2} and \texttt{TACT}), 
\texttt{GroArnold} lowers the barrier for the user who wishes to modify or extend specific components to address their own scientific questions.
%%%
The key innovations of \texttt{GroArnold} are threefold: 
(i) it integrates existing methods into a comprehensive, end-to-end framework; 
(ii) it extends \texttt{TACT} to \texttt{Nbody-TACT}, enabling action estimation directly from $N$-body snapshots in triaxial potentials; 
and (iii) it provides a flexible library of analytical $\DFJ$ forms to fit the numerical distribution.
\par

%
%% \texttt{GroArnold} %
%% \footnote{The open-source code of \texttt{GroArnold} is available at \url{https://github.com/starlifting1/GroArnold}\,.}
%% is organized as a modular pipeline that carries the user from an initial multi-component galaxy all the way through 
%% to the fit of $\DFnJ$ in action space. In this section we present a ``user guide'' to each module; 
%% mathematical details appear in the subsequent sections. \par 

\subsection{Data-Flow Architecture of \texttt{GroArnold}}
\begin{figure*}[tbp]
	\centering
	\includegraphics[width=\textwidth,height=0.78\textheight,keepaspectratio]{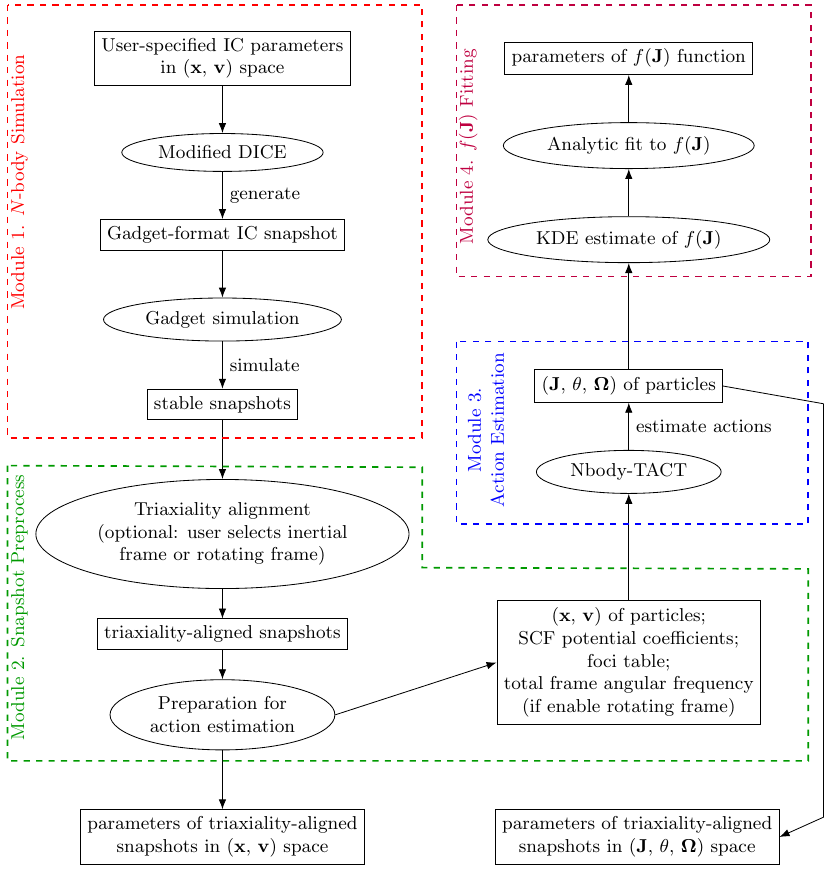}
	\caption{
		Overview of the data flow in the \texttt{GroArnold} framework. 
		``IC'' denotes the initial condition, ``triaxiality-aligning'' denotes the procedure that aligns the principal axes of the snapshot with the coordinate axes, ``SCF'' denotes the self-consistent field method used to estimate the potential from the particle positions, and ``KDE'' denotes the kernel density estimation used to estimate $\DFnJ$.
		After these steps, the user can compare the properties of snapshots with triaxiality-alignment in both $(\mathbf{x},\mathbf{v})$ space ( positions and velocities) and $(\mathbf{J},\boldsymbol{\theta},\boldsymbol{\Omega})$ space ( actions, angles, and frequencies).
	}
	\label{fig_data_flow} 
\end{figure*}

Fig.~\ref{fig_data_flow} shows the end-to-end data flow diagram of the \texttt{GroArnold} framework.
See Sections~\ref{Sec3}, \ref{Sec4} and \ref{Sec5} for more details. 
Note that there are various data such as $\{\mathbf{x},\, \mathbf{v}, \Phi,\, ...\, t\}$ in the snapshot data (here $\Phi$ is 
%the potential where the particle locates, 
the potential at the particle position, 
$t$ is the galaxy simulation time). 
The dashed boxes are modules in \texttt{GroArnold}. These steps in the data flow diagram can be repeated for each snapshot.
%and the $\DFnJ$ function of each snapshot are compared. 
\par 

%plot data_flow_diagram, Figure 1:
%[p]%[h]             % 'h' means the image is inserted at the current location %do not use ``\begin{figure*}[h]''
% ``IC'' denotes the initial condition, ``triaxiality-aligning'' denotes the procedure that aligns the principal axis of the snapshot with the coordinate axis, ``SCF'' denotes the self-consistent field method used to estimate the potential from the particle positions, and ``KDE'' denotes the kernel density estimation used to estimate $\DFnJ$.
% After these steps, the user can compare the properties of snapshots with triaxiality-alignment in both $(\mathbf{x},\mathbf{v})$ space (position--velocities) and $(\mathbf{J},\boldsymbol{\theta},\boldsymbol{\Omega})$ space (angle--actions).
\par

%\clearpage

\begin{enumerate}
\sloppy
	\item \textbf{Module 1: User inputs (initial conditions) \& $N$-body Simulation}\\
	
	\emph{Input:} User-specified parameters of triaxial components (density profile, axis ratios, spin, etc.).\\
		For example, the user can specify the density profile (e.g., double power-law or Einasto),
	 axis ratios $(q_x,q_y,q_z)$, particle count $N$, and the spin parameter $\lambda_\mathrm{spin}$ for a halo component
	 in the driver script file, \path{unified_settings.yaml}\/.\\
	\emph{Process:}
	An $N$-body initial condition is generated and simulated until it reaches a stable state (the so-called 
	dynamic quasi-stationary state, QSS).
	The phase space information of particles is then obtained. 
 %%%%
	\begin{itemize}
		\item In the workflow controller \path{GroArnold_framework/install_and_run/workflow_wrapper.py}, \path{module_1_ic_and_sim}\allowbreak\path{(paths, model_name, ...)}:\\
		create or refresh directory \path{galaxy_general/}, establish subdirectories (e.g. \path{init/}, \path{snapshot/}, \path{aa/}, \path{fit/}), construct initial conditions, and run the $N$-body evolution (e.g. \texttt{Gadget2}) to a stable snapshot.
	\end{itemize}
	\emph{Output:} One or more stable snapshots.
	
	\item \textbf{Module 2: Triaxiality Alignment \& Preprocess}\\
	\emph{Input:} Snapshot from Module 1.\\
	\emph{Process:}
		In this module, the galaxy snapshot is preprocessed to obtain a triaxiality-aligned snapshot, by recentering and aligning the coordinates with the principal axes of the mass distribution. 
		This includes subtracting the center-of-mass position and velocity and aligning the principal axes of the moment-of-inertia tensor with the coordinate axes.
	\begin{itemize}
		\item \path{module_2_triax_alignment}\allowbreak\path{(paths, snapshot_actions, ...)}:\\
		recenter ( position and velocity), align principal axes to the coordinate axes and build foci tables
		(see \S~\ref{sec_M_foci} for the meaning of foci).
	\end{itemize}
	\emph{Output:} Triaxiality-aligned snapshot, SCF potential coefficients, foci tables, etc.
	
	\item \textbf{Module 3: Angle--Action Estimation}\\
	\emph{Input:} Triaxiality-aligned snapshot.\\
	\emph{Process:}
	 The \texttt{Nbody-TACT} code is modified to estimate actions under a triaxial potential from the $N$-body distribution. The foci table for actions is established, and then the actions of each particle are estimated.
	\begin{itemize}
		\item \path{module_3_actions_calc}\allowbreak\path{(paths, snapshot_actions, ...)}:\\
		compute actions $(J_\lambda,J_\mu,J_\nu)$, frequencies $(\Omega_\lambda,\Omega_\mu,\Omega_\nu)$, angles $(\theta_\lambda,\theta_\mu,\theta_\nu)$ via the triaxial St\"ackel-Fudge on the SCF potential.
	\end{itemize}
	\emph{Output:} Action and frequency data for each particle.
	
	\item \textbf{Module 4: $\DFnJ$ Construction \& Fitting}\\
	\emph{Input:} Action and frequency data from \mbox{Module~3}.\\
	\emph{Process:}
	The numerical $\DFnJ$ is computed using Kernel Density Estimation (KDE). 
	% Various candidate function forms for the $\DFnJ$ are provided.
	Various candidate functional forms for the $\DFnJ$ are provided.
	\begin{itemize}
		\item \path{module_4_df_fit_and_plots}\allowbreak\path{(paths, snapshot_to_fit_DF, ...)}:\\
		compute and fit $\DFnJ$, and then plot figures.
	\end{itemize}
	\emph{Output:} Best-fit $\DFJ$ parameters, goodness-of-fit plots.
\end{enumerate}

%The 4 modules are in a loop for a galaxy model. 
With these modules in place, \texttt{GroArnold} gives the user a complete route: 
generating $N$-body initial conditions of any user-specified (triaxial) system, 
evolving this system to a dynamically stable QSS state, estimating actions of all the particles, 
calculating the numerical DF of the particles' actions, 
and finally outputting a best-fitting analytical \DFJ\ form for this system.
%%%%
Subsequent sections explain the algorithms and mathematical underpinnings in detail. 
\par

\subsection{A simple guide}
The four modules are wrapped into a workflow controlled by the shell-like driver script \path{GroArnold_framework/install_and_run/}\allowbreak\path{workflow_wrapper.py}. 
The user can find more detailed instructions for running the program in \path{GroArnold_framework/README.md}. 
The user can write some settings (like specified parameters of triaxial components: density profile, axis ratios, spin) in the settings YAML file \path{unified_settings.yaml}. 
An example of running the workflow is shown in Appendix~\ref{appendix_guide}. 
\par 

For illustration in the remainder of this paper, we adopt one representative simulation as an example dataset. 
We generate an initial realization with $2.1\times10^6$ particles (all particle types), a virial mass of $137.0\times10^{10}\,M_\odot$, and a scale length of $19.6\,\mathrm{kpc}$ in \texttt{DICE}. 
We evolve the system with \texttt{Gadget2} and analyze the snapshot at $t=3.0\,\mathrm{Gyr}$. 
After applying the preprocessing steps described below, we refer to this example galaxy model as GAL0 (corresponding to model \texttt{Ein\_spinL\_axisLH0} in Table~\ref{table_galaxy_params}). 
\par

%% Sec3 begin
\section{\texorpdfstring{Initial conditions and $N$-body simulation}{Initial conditions and N-body simulation}}\label{Sec3}

In the initial conditions module, we use \texttt{DICE} to generate an $N$-body initial condition, with the observational properties (e.g., the density profile and velocity-field properties of the dark halo component) as specified by the user. 
%For the user's manual, please see the above section and Appendix~\ref{appendix_guide}. 
For the user's manual, see Section~\ref{Sec2} and Appendix~\ref{appendix_guide}. 
Below some technical details are presented.
\par

% \texttt{DICE} is a program to generate initial conditions of galaxies or streams for gadget simulation
\texttt{DICE} is a program to generate initial conditions of galaxies or streams for \texttt{Gadget} simulations
(\url{https://bitbucket.org/vperret/dice/wiki/browse/}, by V. Perret in C code; \citealt{Perret14_DICE, DICE_code}). 

We modified some of the \texttt{DICE} code to provide more types of radial density profiles. 
These include a double-power-law profile 
\begin{equation} \label{eq_DPL}
\rho_\mathrm{DPL} = \frac{\rho_\mathrm{s}}{r_\mathrm{sq}(1+r_\mathrm{sq})^s} ~, 
\end{equation}
and an Einasto profile (simplified formula)
\begin{equation} \label{eq_Ein}
\rho_\mathrm{Ein} = \rho_\mathrm{s} \exp \left[-b_n \left(r_\mathrm{sq}^{1/n} - 1\right)\right] ~.
\end{equation}
Here, $\rho_\mathrm{s}$ is the scale density, $r_\mathrm{s}$ is the scale radius, 
$q_i\,(i=x,y,z)$ are the axis ratios, and 
\begin{equation} \label{eq_rsq}
r_\mathrm{sq} = \sqrt{ (\frac{x}{q_x r_\mathrm{s}})^2 + (\frac{y}{q_y r_\mathrm{s}})^2 + (\frac{z}{q_z r_\mathrm{s}})^2 } ~
\end{equation}
%is the flattened scale radius. 
is the dimensionless flattened ellipsoidal radius. 
Here, $s$ controls the outer slope of the double-power-law profile. 
%We adopt $b_n \approx 3n-1/3+0.0079/n,\ n\,\in\,(1,10)$ \citep{Baes22_Einasto}. 
Here, $n$ is the Einasto index, and $\rho_\mathrm{s}$ in Equation~(\ref{eq_Ein}) is the density at $r_\mathrm{sq}=1$.
We adopt $b_n \approx 3n-1/3+0.0079/n$ for $n\,\in\,(1,10)$ \citep{Baes22_Einasto}.
%For any finite $n$, the Einasto profile is non-singular at the origin (as $r\to 0$, its central density is finite and the logarithmic slope tends to zero, namely ``cored''), 
%but the Einasto index $n$ controls how rapidly the slope steepens with radius: smaller $n$ yields a more extended core-like region (broadly with the logarithmic slope being $< 1$ and even asymptotic to 0), 
%whereas larger $n$ produces a more centrally concentrated, cuspy profile over the resolved radial range \citep{Baes22_Einasto}. 
For any finite $n$, the Einasto profile is non-singular at the origin: as $r\to 0$, the central density remains finite and the logarithmic slope tends to zero. 
The Einasto index $n$ controls how rapidly the slope steepens with radius. 
Smaller $n$ produces a more extended core-like region, whereas larger $n$ gives a more centrally concentrated profile that can appear cuspy over the resolved radial range \citep{Baes22_Einasto}. 
\par

% Although the simulation is unit independent, we set most of the IC parameter values of the example galaxies provided in the \texttt{GroArnold} package to be near the Milky Way's values and in units of ${\mathrm{kpc}, \mathrm{km\,s^{-1}}, \mathrm{Gyr}, 10^{10}\mathrm{M_\odot}}$, for ease of reference. 
% The total mass and scale length of in the initial condition refers to the Milky Way, the particle count $N_\mathrm{particles}$ is usually around $1\mathrm{e}6$ and the mass of each particle is the same for each component. 
The example galaxies use Milky-Way-like mass and length scales for reference, while their halo shapes and spins are specified independently through the initial-condition parameters. 
We use units of $\mathrm{kpc}$, $\mathrm{km\,s^{-1}}$, $\mathrm{Gyr}$, and $10^{10}\,\mathrm{M}_\odot$.
The particle number is typically of order $10^6$, 
with each component consisting of equal-mass particles, although the particle mass may differ from component to component. 
Note that while the user-specified axis ratio fed to \texttt{DICE} is fixed (i.e., the model consists of concentric ellipsoidal surfaces), the axis ratios in the subsequent Gadget-output snapshots may vary with radius. 
Fortunately, the TSF algorithm in Module~3 is capable of handling such radially varying axis ratios. 
\par 

%We first translate the system to the center-of-mass frame by removing the position and velocity offsets (see also Operation~(i) in Section~\ref{sec_M_aligenment}). 
We first translate the system to the center-of-mass frame by removing the bulk position and velocity offsets (see also Operation~(i) in Section~\ref{sec_M_aligenment}). 
\par 

The $N$-body IC generated by the modified-\texttt{DICE} module is then evolved with the \texttt{Gadget2} code (\url{https://wwwmpa.mpa-garching.mpg.de/galform/gadget/}, in C code; \citealt{Gadget2}), usually for several gigayears, to guarantee that the $N$-body system is virialized and stable. 
Note that the DICE-generated N-body ICs may not be stable enough to preserve the user-specified properties exactly; 
in turn, the Gadget-evolved snapshot may not retain the exact parameter values of the ICs.
During this dynamical readjustment, the halos 
in our examples 
may become rounder or closer to axisymmetry than their user-specified initial configurations.
The structural and kinematic parameters relevant to the subsequent analysis are therefore measured from the evolved snapshots rather than taken from the initial-condition settings. 
\par

%Example code X. (pl and einasto, axis ratio, spin) \par 

%Example code X. (python control C exe to run gadget2 simulation) \par 

%Especially, if an IC with a high spin desired, we can re-orient each particle's velocity towards the local azimuthal direction while preserving its speed. 
Especially, for models requiring higher spin, we optionally reorient each particle velocity toward the local azimuthal direction while preserving its speed. 
For a particle at position $\mathbf{x}$ with initial velocity $\mathbf{v}_0$ and speed $v_0\equiv\|\mathbf{v}_0\|$, let $\hat{\mathbf e}_\phi(\mathbf{x})$ denote the unit vector in the positive azimuthal direction (about the chosen spin axis), and define the purely azimuthal velocity $\mathbf{V}_\phi \equiv v_0\,\hat{\mathbf e}_\phi$. 
We define the updated velocity $\mathbf{v}_1$ to have the direction of $\mathbf{v}_0+\mathbf{V}_\phi$ and the original speed $v_0$: 
\begin{equation}\label{eq_change_v_azimuthal}
\mathbf{v}_1 = v_0\,\frac{\mathbf{v}_0+\mathbf{V}_\phi}{\|\mathbf{v}_0+\mathbf{V}_\phi\|} ~ .
\end{equation}
Repeating this operation progressively aligns the velocities with $\hat{\mathbf e}_\phi$, thereby increasing the total angular momentum and the spin parameter. 
In our tests, the spin saturates after $\sim 6$ iterations; in the examples presented here, however, we apply only one iteration. 
Since the particle positions are unchanged and each update preserves $|\mathbf v|$ before any residual bulk motion is removed, this operation leaves the spatial density profile essentially unchanged while increasing the degree of ordered rotation. 
In the simulated galaxy examples presented in this work (see Table~\ref{table_galaxy_params} and \S~\ref{sec_M_fit-DFJ}), 
the fraction of unbound particles remains below 2\%, which provides a basic check that the galaxy is not strongly disrupted by this velocity reorientation. 
\par

%% Sec4 begin
\section{Action estimation method}\label{Sec4}

In this section, we describe the technical details of action estimation from $N$-body snapshots. 
We first present Module~2, which preprocesses the snapshot data for action estimation (\S~\ref{sec_M_aligenment} and \S~\ref{sec_M_pot}) and determines the foci used in the triaxial St\"ackel approximation (\S~\ref{sec_M_foci}). 
We then describe the core algorithm of Module~3, \texttt{Nbody-TACT} (\S~\ref{sec_M_AA}). 
% Background material on the triaxial St\"ackel Fudge is provided in \mbox{\ref{sec_M_TSF}}.
Background material on the triaxial St\"ackel Fudge is provided in Appendix~\ref{sec_M_TSF}.
\par

\subsection{Alignment of the simulated snapshot}\label{sec_M_aligenment}

The St\"ackel Fudge works in a locally St\"ackel-like potential which is triaxially symmetric. 
%The potential of a simulated galaxy is approximately triaxial, so in the module 2 (triaxial alignment) of \texttt{GroArnold}, we perform some operations to align the simulated data before estimation of the actions -- the preprocess of triaxiality-alignment to the galaxy: 
Because the potential of the simulated galaxy is only approximately triaxial and may not be aligned with the coordinate system of the snapshot, Module~2 of \texttt{GroArnold} performs a preprocessing alignment procedure before action estimation: 
\par 

%(i) Move the position center and momentum center. 
(i) Recenter the system in position and velocity space. 
We compute the original center of mass 
$\mathbf x_c = \frac{1}{M_\mathrm{total}} \sum_{j}^{N_\mathrm{particles}} m_j \mathbf x_j$, where $M_\mathrm{total}$ is the total mass, and translate the coordinates to obtain the new position, 
\begin{equation}\label{eq_x_c}
\mathbf x'_j = \mathbf x_j-\mathbf x_c, 
\end{equation}
where $j$ represents the index of a particle. 
We compute the original 
center-of-mass velocity 
$\mathbf v_c = \frac{1}{M_\mathrm{total}} \sum_{j}^{N_\mathrm{particles}} m_j \mathbf v_j$ and translate to 
\begin{equation}\label{eq_v_c}
\mathbf v'_j = \mathbf v_j-\mathbf v_c. 
\end{equation}
%This operation makes the self-consistent potential more standard. 
This operation places the system in a centered frame for potential reconstruction. 
Note that the frame is still inertial after this operation. 
\par 

%(ii) Rotate galaxy and align the principal axis of moment of inertia at each snapshot. 
(ii) Rotate the galaxy to align the principal axes of the moment-of-inertia tensor. 
We compute the eigenvectors of the total moment-of-inertia tensor $I$, 
order the eigenvectors by decreasing eigenvalue 
and invert the matrix formed by these ordered eigenvectors to obtain the rotation matrix $T$
%which means align the principal axis of moment-of-inertia tensor. 
that aligns the coordinate axes with the principal axes of the mass distribution. 
Then the new phase space coordinates of each particle are 
\begin{equation}\label{eq_xv_rot_I}
\begin{split}
\mathbf x' &= \mathbf x T, \\
\mathbf v' &= \mathbf v T.
\end{split}
\end{equation}
Note that the frame is still inertial after this operation. 
\par

%
%we should eliminate the total rotating in rotating frame. 
%\footnote{
%Note that in Fig. \ref{fig_potential_compare_x}, the potential in rotating frame beyond the radius range of about 40 $\mathrm{kpc}$ has large difference from the potential in inertial frame (the median of radius is about 15 \~ 20 $\mathrm{kpc}$). 
%This is because in (\ref{eq_Phi_rot_L}) the higher total rotation angle frequency, the larger radius, the higher difference of potential to modified. 
%For location with very large radius this would be larger than the original potential in the inertial frame and lead to the distortion of the effective potential. However, it might be acceptable for most of the particles because about 90\% of particles is within the range of 41.9 $\mathrm{kpc}$.
%}.

%

%Example code X. (preprocess) \par 

%Example code X. (mass contour) \par 

%
% Fig.~\ref{fig_mass_contour} of \ref{contours_of_mass_and_potential} demonstrates 
% the mass density distributions in the $x$--$y$ plane
% at different $z$ positions for the triaxiality-aligned galaxy GAL0,
% with the $(x, y)$ range of each panel enclosing 98\% of the particles.
%% bad english: the bound of $x,y$ we plot in each panel is equal to the radius with 98\% particles. 
Fig.~\ref{fig_mass_contour} in Appendix~\ref{contours_of_mass_and_potential} shows the mass-density distribution in the $x$--$y$ plane at different $z$ slices for the triaxiality-aligned galaxy GAL0. The plotted $(x,y)$ range in each panel encloses 98\% of the particles.
\par

\subsection{Numerical potential for action estimation}\label{sec_M_pot}

% In the action estimation module, we modified some code of \texttt{TACT} into \texttt{Nbody-TACT}, to use the potential from the self-consistent field (SCF) method code (by Chris Mihos et al. in Fortran and compiled in C++, using the Hernquist-Ostriker basis function) and make an interface to snapshot data. 
% The SCF method is an approach to solve Poisson's equation by expanding both the density and potential of a system in terms of a set of basis functions. 
In the action-estimation module, we modify parts of \texttt{TACT} to form \texttt{Nbody-TACT}, which interfaces the action-estimation routines with snapshot data and with a self-consistent-field (SCF) potential solver. 
The SCF solver used here is based on the Fortran implementation by Chris Mihos et al., called from the C++ interface, and uses the Hernquist--Ostriker basis functions. 
The SCF method solves Poisson's equation by expanding the density and potential of the system in a chosen set of basis functions. 
%In practice, the SCF expansion provides a smooth approximation to the potential implied by a discrete particle snapshot, suppressing small-scale noise and enabling efficient orbit integration and action estimation \citep{Herquest92_SCF}. 
%%  future work may use Tree method for less particles}
In practice, the SCF expansion is intended to provide a smooth approximation to the potential implied by a discrete particle snapshot, 
suppressing small-scale particle noise and enabling efficient orbit integration and action estimation \citep{Herquest92_SCF}.
%%  when the expansion is well behaved
\par

%One can select a virialized and stable snapshot of the user-specified galaxy and perform the pre-processing operations described in Section~\ref{sec_M_aligenment} to obtain a triaxiality-aligned snapshot, then obtain the SCF potential coefficients from the snapshot data. 
We select a virialized and stable snapshot of the user-specified galaxy, perform the preprocessing operations described in Section~\ref{sec_M_aligenment} to obtain a triaxiality-aligned snapshot, and then compute the SCF potential coefficients from the snapshot data. 
\par

\begin{figure*}[tbp]
	\centering
	\includegraphics[width=0.8\textwidth]{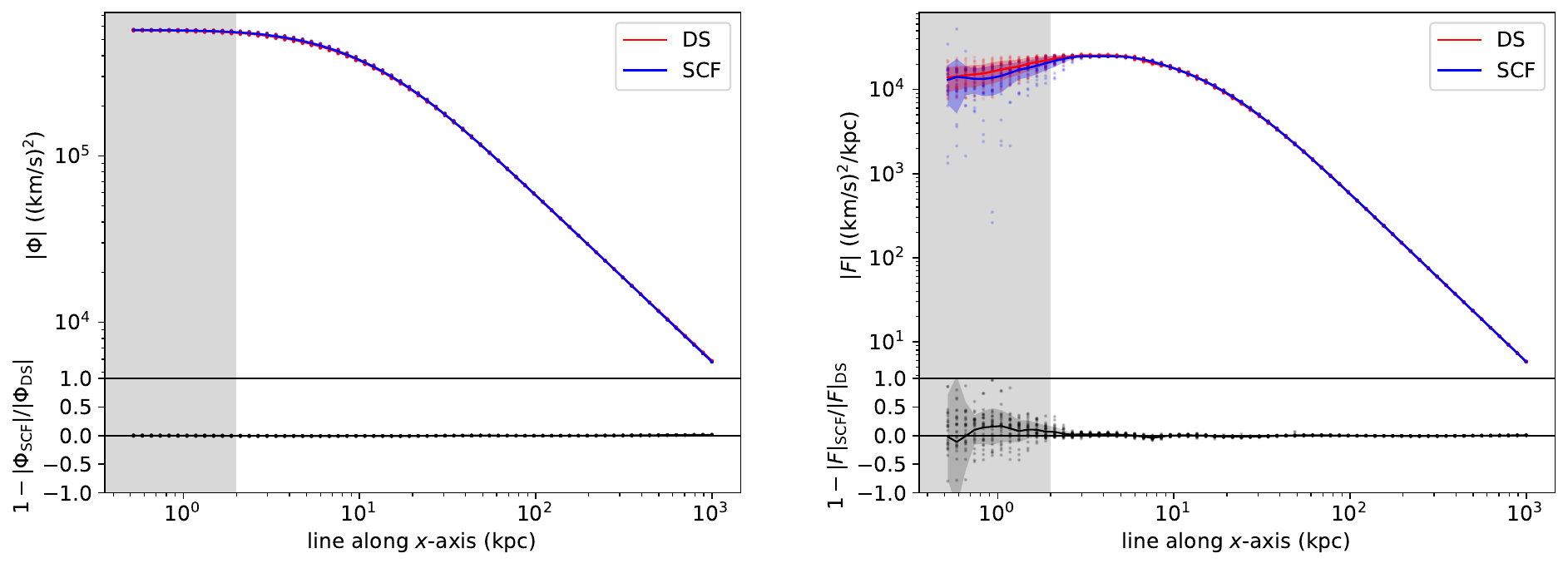}
	\caption{
		Comparison of the gravitational potential and force magnitude obtained with the DS (direct summation) and SCF (self-consistent field) methods along the positive $x$-axis. 
		The red curves show the DS results, and the blue curves show the SCF results. 
		The upper-left panel shows the absolute value of the potential, $|\Phi|$, evaluated after the preprocessing procedure described in Section~\ref{sec_M_aligenment}, while the upper-right panel shows the magnitude of the total force, $|F|$. 
		The lower panels present the corresponding relative differences, defined as $1-|\Phi_{\rm SCF}|/|\Phi_{\rm DS}|$ and $1-|F_{\rm SCF}|/|F_{\rm DS}|$, respectively. 
		The scattered points represent values evaluated at 
		the 27 sampling positions (the center point and its surrounding offset positions) $(x+i h_{\rm offset},\,y+j h_{\rm offset},\,z+k h_{\rm offset})$, with $i,j,k\in\{-1,0,1\}$, 
		where $h_{\rm offset}=0.5~\mathrm{kpc}$, equal to one-half of the halo-particle softening length. 
		The shaded bands indicate the scatter among these offset samples, and the gray-shaded region marks the radial range considered unreliable. 
		Note that \texttt{GroArnold} uses only the potential, rather than the force, for TSF action estimation; 
		the force is shown here for illustrative purposes only, 
		as force estimates are more sensitive to numerical differentiation and particle noise. 
	}
	\label{fig_potential_compare_x}
\end{figure*}

We compare the direct summation (DS) potential computed from the $N$-body particles with the approximate SCF potential. 
Fig.~\ref{fig_potential_compare_x} shows the SCF and DS potentials along the positive $x$-axis (left panel), together with their corresponding force magnitudes (right panel), for the galaxy GAL0. 
%% -- silly: In GAL0, about 98\% of particles is within the range of 110.1 $\mathrm{kpc}$. 
The relative errors in the potential (as well as the total force) 
are sufficiently small over most of the radial range, making the SCF approximation acceptable for action estimation. 
Note that \texttt{GroArnold} uses only the potential, rather than the force, for TSF action estimation; the force is shown here for illustrative purposes only.
%%%
%
%Fig.~\ref{fig_potential_contour} in \ref{contours_of_mass_and_potential} displays the SCF potential contours of the triaxiality-aligned galaxy GAL0, with the plotted $(x,y)$ region extending to twice the median radius of all particles. 
%Appendix~
Fig.~\ref{fig_potential_contour} shows the SCF potential contours of the triaxiality-aligned galaxy GAL0; the plotted $(x,y)$ region extends to twice the median radius of all particles.
\par 

%[h]             % 'h' means the image is inserted at the current location
% Comparison of the gravitational potential and force obtained with the DS (direct summation) and SCF (self-consistent field) methods along the positive $x$-axis. 
%The red curves denote the DS results, and the blue curves denote the SCF results. 
% as its evaluation suffers from notoriously large errors due to numerical differentiation. 
\par

\subsection{Numerical method for foci choice}
\label{sec_M_foci}

%The value of actions depends on semi-axis length $-\alpha, -\beta, -\gamma$ of TCE coordinate, i.e. the foci $\Delta_1,\, \Delta_2$. 
The estimated actions depend on the confocal coordinate parameters $-\alpha$, $-\beta$, and $-\gamma$ of the triaxial ellipsoidal coordinate (TCE) system, 
or equivalently on the two focal distances $\Delta_1$ and $\Delta_2$. 
%\texttt{TACT} program sets $\gamma=-1$ fixed and searches the best foci by minimizing the standard deviation of $J_\nu$ or $J_\mu$ in orbits of different energy $E$ at a certain $y$-cut (the $y_0$ in their paper). 
The original \texttt{TACT} implementation fixes $\gamma=-1$ and searches for the best foci by minimizing the standard deviation of $J_\nu$ or $J_\mu$ along selected orbits at different energies, parameterized by a $y$-cut, denoted as $y_0$ in \citet{SB15}. 
We use an alternative procedure to search for the best foci because the original \texttt{TACT} orbit-search procedure may fail when applied to the numerical potential, especially in regions where the reconstructed potential is not sufficiently smooth (where unexpected NaN, or not-a-number, values can sometimes occur during the orbit-searching steps). 
\par 

In \texttt{TACT}, the confocal coordinate system is selected by first identifying the two planar closed-loop orbits (short-axis and long-axis loops) at a given energy, and then varying the focal parameters to minimize the time variation of the corresponding single non-zero action along each orbit 
(see Section~4.1 of \citealt{SB15}). 
%In \texttt{GroArnold}, the potential is obtained from numerical representations (via SCF), 
In \texttt{GroArnold}, the potential is represented numerically through the SCF expansion, 
and small-scale noise may occasionally cause instabilities in this optimization. 
We therefore adopt a more robust geometry-based procedure that avoids repeated action evaluations during the foci search, while still producing a smooth mapping from orbit family (or equivalently energy) to the confocal parameters. 
\par 

The procedure is based on the fact that, in a true triaxial St\"ackel potential, the planar closed-loop orbits are ellipses whose foci coincide with those of the underlying confocal ellipsoidal coordinate system \citep{SB15}. 
For each $y_{\rm cut}$, we launch a near-planar loop-like trajectory from $(0,y_{\rm cut},0)$, 
tune the single launch-speed component with a 1D golden-section search 
%to minimize the spread in successive $y$-axis intercepts, 
operationally by reducing the spread in successive $y$-axis intercepts, 
and use the resulting orbit projection in the corresponding principal plane. 
We then rotate the projected orbit to align with its principal axes 
%and estimate the focal distance from the envelope extents via the ellipse relation $\Delta^2 \simeq a_{\rm env}^2-b_{\rm env}^2$, 
and estimate the focal distance from the span of the projected orbit points using the ellipse-motivated relation 
$\Delta^2 \simeq a_{\rm env}^2-b_{\rm env}^2$,
%where $a_{\rm env}$ and $b_{\rm env}$ are the semi-axes of the envelope in that plane. 
where $a_{\rm env}$ and $b_{\rm env}$ are span-based estimates of the two semi-axis extents in that plane. 
Repeating this procedure for the two loop families provides estimates of the two focal distances $(\Delta_1,\Delta_2)$---and hence of $(\alpha,\beta,\gamma)$---as functions of $y_{\rm cut}$, or equivalently of energy. We then build a lookup table for subsequent interpolation. 
\par 

%I can see they are equivalent for a true Staeckel potential.}
In the ideal limit where the potential is of St\"ackel form and the selected reference orbits coincide with the true planar closed loops, 
%this construction recovers the correct confocal foci; but in more general potentials it should be regarded as a stable empirical choice. 
this construction reduces to the usual geometric identification of the confocal foci; but in more general potentials it should be regarded as a stable empirical prescription for assigning the TCE foci used by the St\"ackel Fudge. 
%a stable empirical choice whose adequacy can be checked a posteriori by the level of action conservation along test orbits. 
\par

%Finally, we choose the best initial $y$-cut by interpolation from the table. 
For a particle of given energy, we interpolate the tabulated $E\mapsto(\Delta_1,\Delta_2)$ relation 
to assign the confocal parameters used in the St\"ackel Fudge action estimation. %}
\par 

\begin{figure*}[tbp]
	\centering
	\begin{subfigure}[t]{0.3\textwidth}
		\centering
		\includegraphics[width=\linewidth]{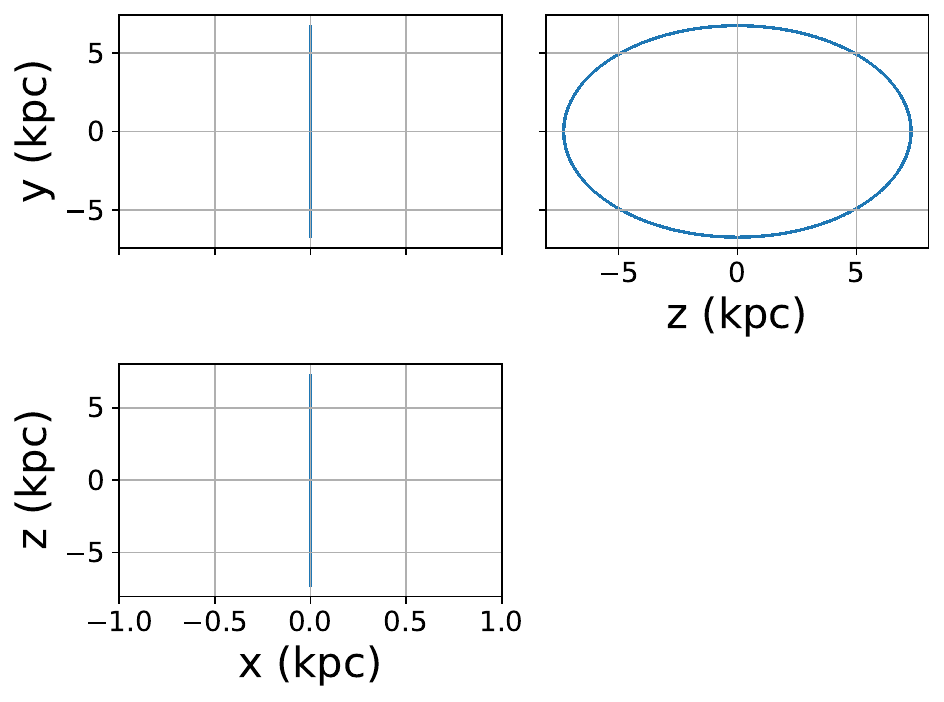}
		\caption{Orbit 2 for $b^2$.}
		\label{fig_orbit_1_b}
	\end{subfigure}
	\hfill
	\begin{subfigure}[t]{0.3\textwidth}
		\centering
		\includegraphics[width=\linewidth]{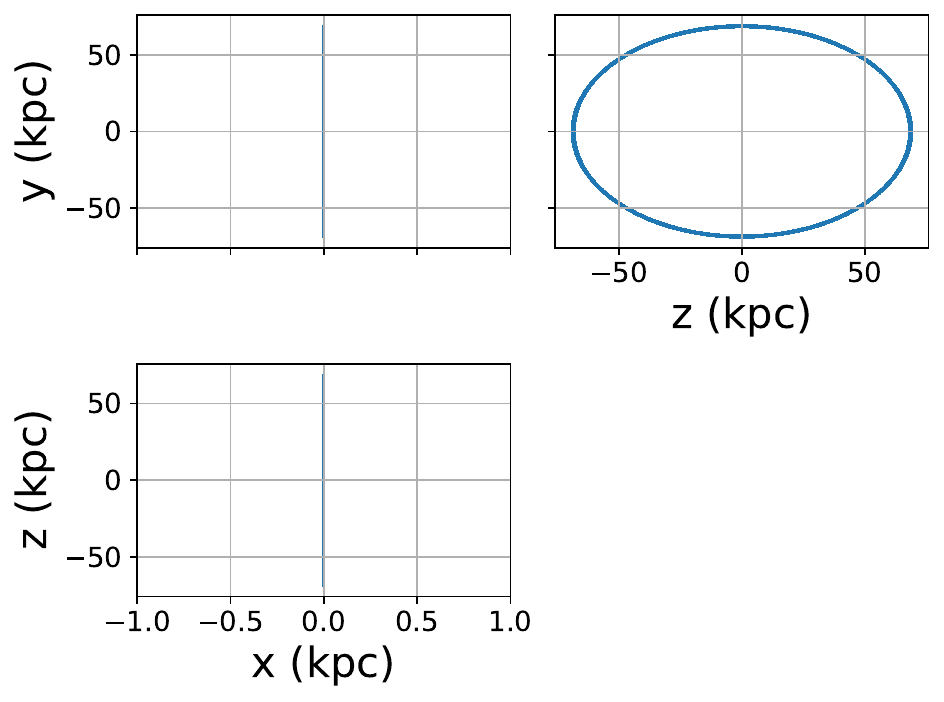}
		\caption{Orbit 12 for $b^2$.}
		\label{fig_orbit_2_b}
	\end{subfigure}
	\hfill
	\begin{subfigure}[t]{0.3\textwidth}
		\centering
		\includegraphics[width=\linewidth]{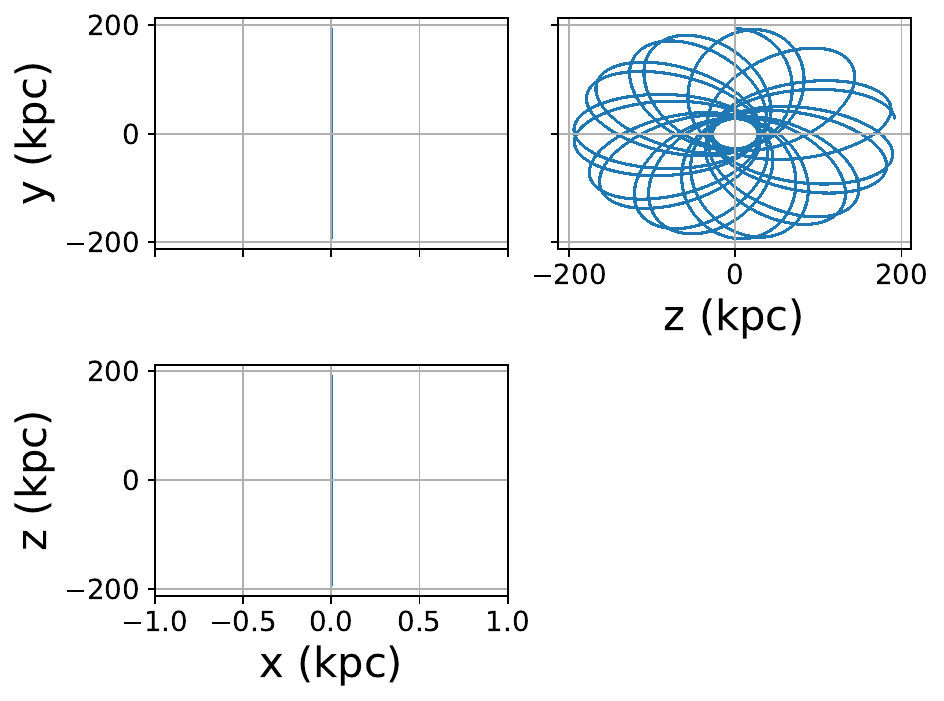}
		\caption{Orbit 32 for $b^2$.}
		\label{fig_orbit_3_b}
	\end{subfigure}
	\hfill\par\medskip
	\begin{subfigure}[t]{0.3\textwidth}
		\centering
		\includegraphics[width=\linewidth]{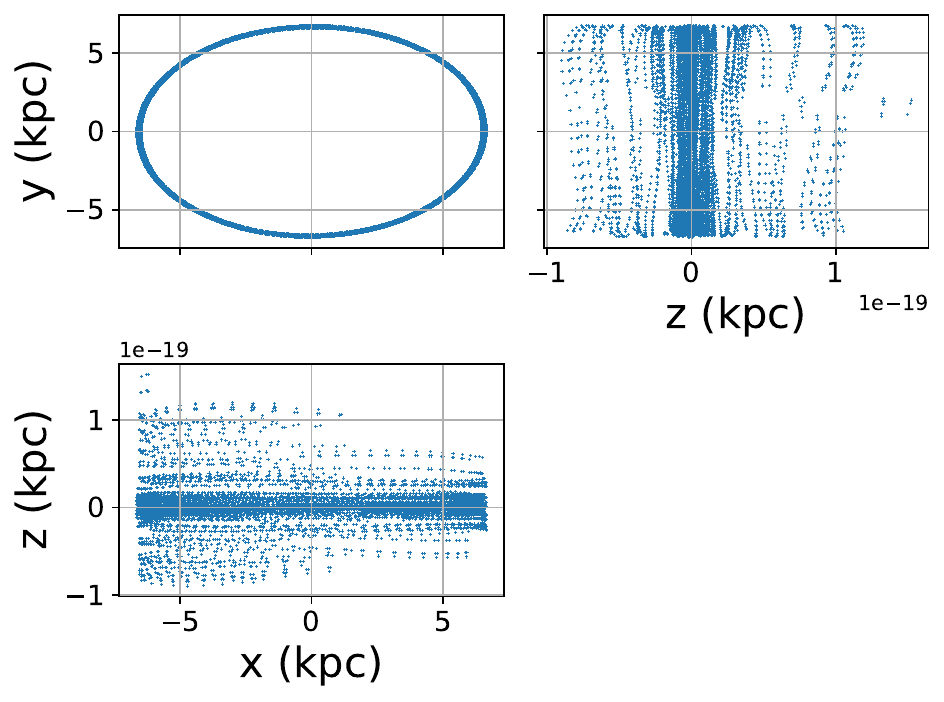}
		\caption{Orbit 2 for $a^2$.}
		\label{fig_orbit_1_a}
	\end{subfigure}
	\hfill
	\begin{subfigure}[t]{0.3\textwidth}
		\centering
		\includegraphics[width=\linewidth]{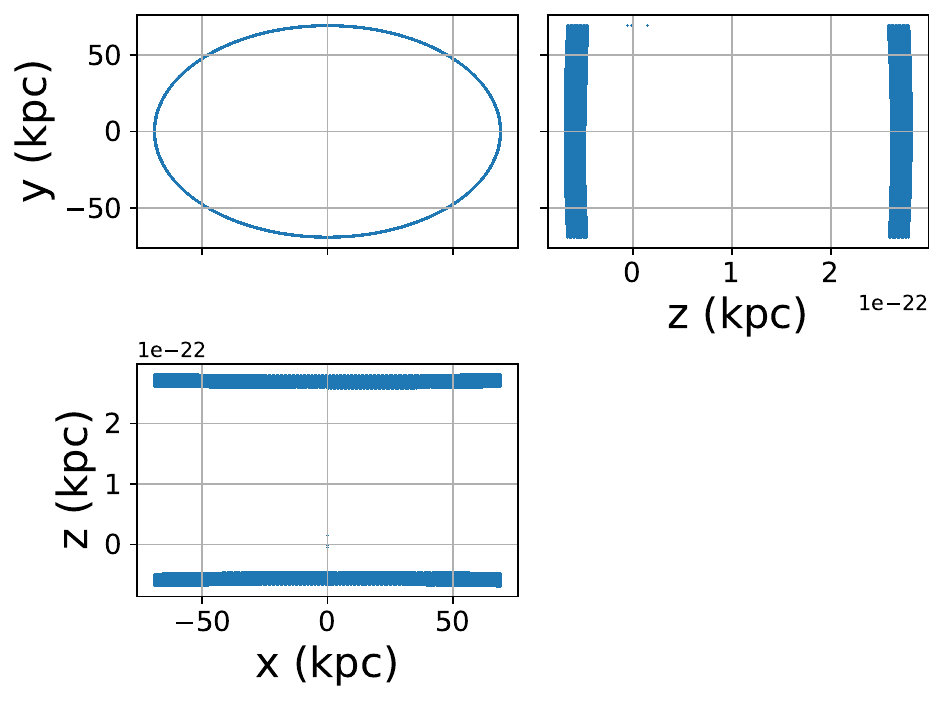}
		\caption{Orbit 12 for $a^2$.}
		\label{fig_orbit_2_a}
	\end{subfigure}
	\hfill
	\begin{subfigure}[t]{0.3\textwidth}
		\centering
		\includegraphics[width=\linewidth]{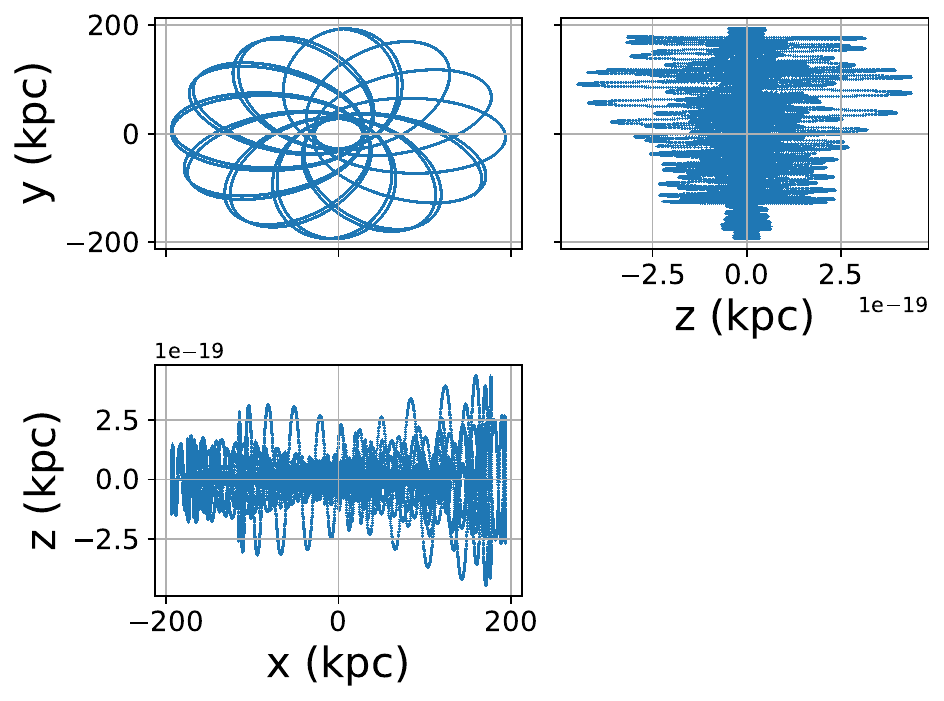}
		\caption{Orbit 32 for $a^2$.}
		\label{fig_orbit_3_a}
	\end{subfigure}
	\hfill
	\caption{
		Representative near-planar reference orbits used in the foci-choice procedure at three values of $y_{\rm cut}$. 
		For each $y_{\rm cut}$ on a 32-point grid, we integrate two candidate loop families launched from $(0,y_{\rm cut},0)$, tuning the launch-speed component to keep the trajectory close to the corresponding loop family. 
		The span of the projected orbit points in the relevant principal plane is then used to estimate the confocal focal distance through $\Delta^2 \simeq a_{\rm env}^2-b_{\rm env}^2$. 
		The top and bottom rows show the reference orbits used to estimate the two focal-distance parameters, labelled here as $b^2$ and $a^2$, respectively. 
		The three columns show examples from the 2nd, 12th, and 32nd $y_{\rm cut}$ grid points.
	}
	\label{fig_foci_orbits} 
\end{figure*}

\begin{figure*}[tbp]
	\centering
	\includegraphics[width=0.46\textwidth]{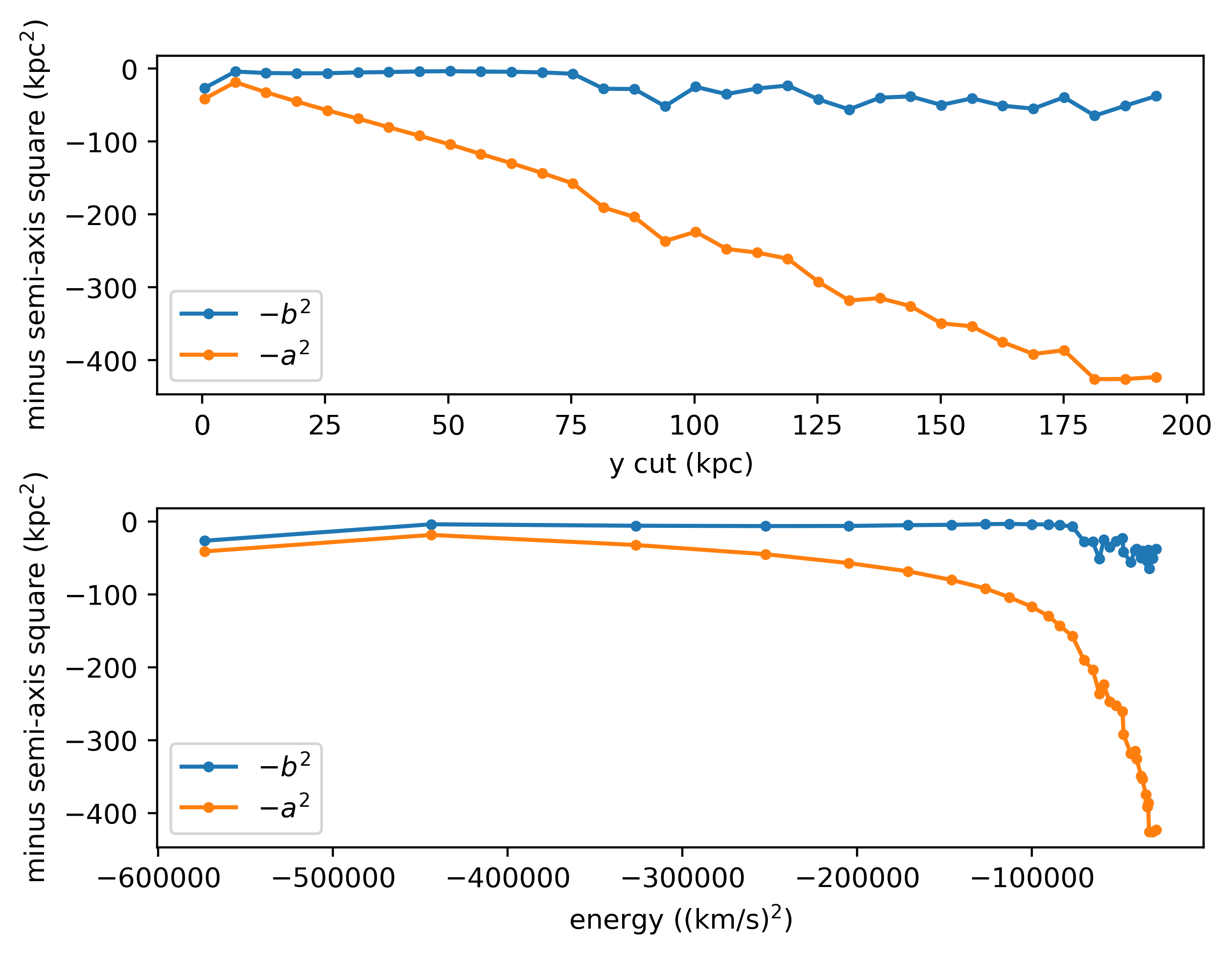} 
	\caption{
		The $y_{\rm cut}$--energy and foci--energy tables, showing the relations between $y_{\rm cut}$, orbital energy, and the fitted confocal parameters. 
		The top panel shows the selected values of $-a^2$ and $-b^2$ versus $y_{\rm cut}$. 
		The bottom panel shows the selected values of $-a^2$ and $-b^2$ versus the orbital energy corresponding to each $y_{\rm cut}$ in the current potential. 
	}
	\label{fig_foci_tables} 
\end{figure*}

Fig.~\ref{fig_foci_orbits} shows representative projections of the chosen reference orbits at different values of $y_{\rm cut}$. 
Fig.~\ref{fig_foci_tables} shows the $y_{\rm cut}$--energy table and foci--energy table. 
Although in an exact triaxial St\"ackel potential the planar closed-loop orbits are ellipses with foci equal to those of the confocal coordinates, 
the orbit projections in Fig.~\ref{fig_foci_orbits} are not expected to be perfectly closed ellipses. 
The potential used here is a numerical representation (SCF) and is not exactly of St\"ackel form, 
so the reference trajectories launched from $(0,y_{\rm cut},0)$ generally remain only near the corresponding parent loop family and may precess, 
producing rosette-like trajectories rather than single closed ellipses. 
%On the other hand, we integrate a robust near-planar loop and use the envelope of its projection to estimate the focal distances. 
We therefore integrate a robust near-planar loop and use the envelope of each near-planar orbit projection to estimate the focal distances. 
%This envelope-based estimation for foci choice does not require the orbits to be simple closed ellipses. 
This foci choice does not require the reference orbits to be simple closed ellipses. 
%\\
\par 
%\bigskip

%[h]             % 'h' means the image is inserted at the current location
% No blank line means no line breaks
%Add horizontal blank space appropriately
% No blank line means no line breaks
%Add horizontal blank space appropriately
% No blank line means no line breaks
%Add horizontal blank space appropriately
% No blank line means no line breaks
%Add horizontal blank space appropriately
% No blank line means no line breaks
%Add horizontal blank space appropriately
% No blank line means no line breaks
%Add horizontal blank space appropriately
%
%For each $y_{\rm cut}$ on a 32-point grid we integrate two candidate loop families launched from $(0,y_{\rm cut},0)$, and we retain the orbit that yields the smallest time variation of the corresponding approximate action. 
%The orbit envelope in the relevant principal plane is then used to estimate the confocal focal distance through $\Delta^2 \simeq a_{\rm env}^2-b_{\rm env}^2$, 
%which provides $b^2$ (top row, selected by minimizing the standard deviation of $J_\nu$) and $a^2$ (bottom row, selected by minimizing the standard deviation of $J_\mu$). 

%\begin{figure}[H]
% figure name
\par

\subsection{Action estimation} \label{sec_M_AA}

\begin{figure*}[tbp]
	\centering
	\includegraphics[width=0.9\textwidth]{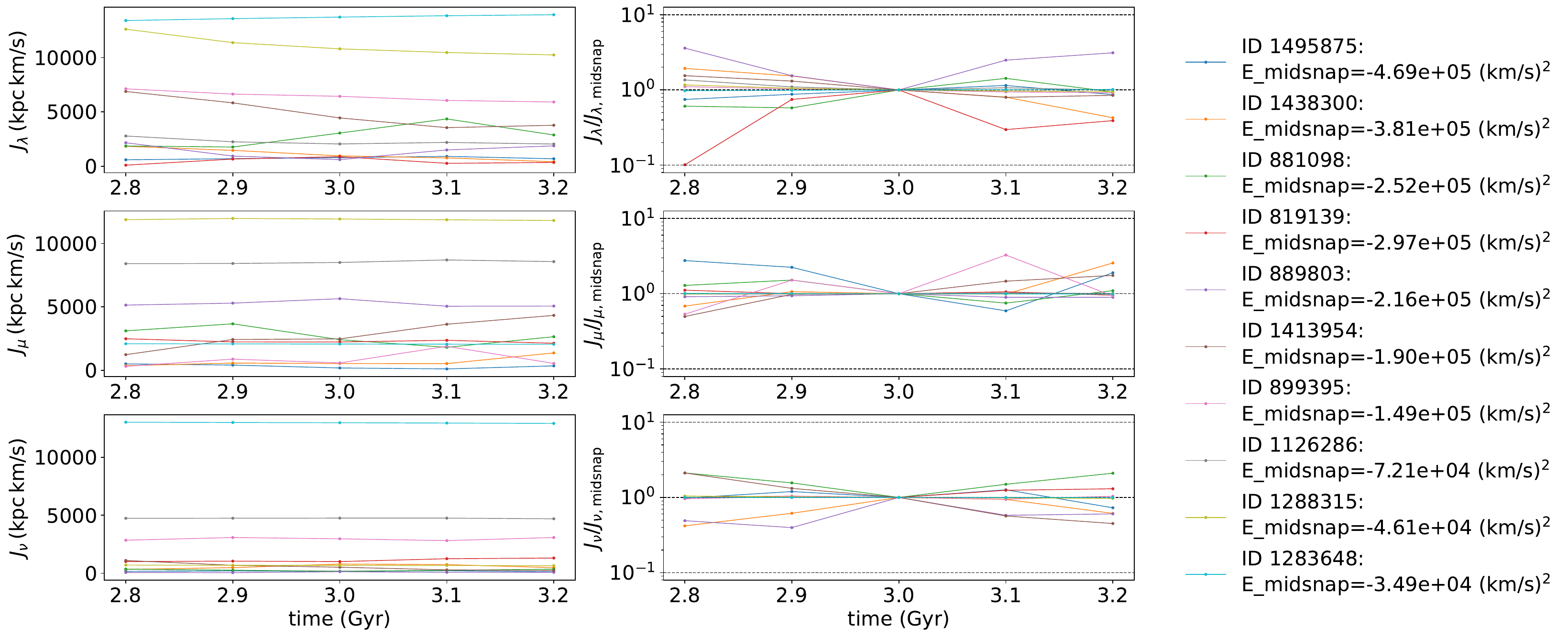} 
	\caption{
		Time variation of the actions for several $N$-body particles. 
		The left panels show the values of $J_\lambda,\, J_\mu,\, J_\nu$ estimated with TSFD for representative particles in snapshots from 2.8 to 3.2 $\mathrm{Gyr}$.
		The plotted particles are selected from different radii in the middle snapshot (namely the 3~Gyr one), with approximately logarithmic spacing. 
		The right panels show the relative variations of those quantities with respect to the middle snapshot. 
		For the full particle sample, roughly $90\%$ of the particles have all three actions within the plotted relative-variation range, $0.1 \leq J/J_{\mathrm{midsnap}} \leq 10$. 
	}
	\label{fig_actions_with_time_one_particle} 
\end{figure*}

After assigning the foci, \texttt{GroArnold} uses the TSF implementation in \texttt{Nbody-TACT} to estimate actions in the SCF potential inferred from $N$-body snapshots; we refer to this application as the Triaxial St\"ackel Fudge in Data potential (TSFD). 
Because TSF and TSFD are approximate, non-convergent action estimators, 
the actions for a given orbit can show substantial periodic time variation.
%The TSFD actions are approximate quantities because the method is fast and suitable for large particle samples, but it is not a convergent orbit-by-orbit action solver.
%We therefore use the recovered actions primarily as statistical coordinates for describing the ensemble distribution of particles in actions space. 
%In Fig.~\ref{fig_energy_with_time_one_particle}, we show the variation of the parameters of $N$-body particle's motion 
%such as energy (in the SCF potential), radius, square of speed and the SCF potential energy of several particles in GAL0. 
In Fig.~\ref{fig_actions_with_time_one_particle}, we show the variation of the TSFD-derived actions ${J_\lambda,\, J_\mu,\, J_\nu}$ 
of several representative particles in GAL0. 
We find that, for the halo particles, 
the median relative time variations of the three action components ($J_\lambda,\, J_\mu,\, J_\nu$), defined as the temporal standard deviation normalized by $\max\!\left(|\bar{J}_j|,\epsilon J_{\rm tot}\right)$ for each particle, 
are 24.2\%, 22.8\% and 10.3\%, respectively. 
\par 

Fig.~2 of \citet{SB15} shows the standard deviation of $J_\mu$ as a function of $\Delta_2$ for the closed long-axis loop orbit %%shown in the inset 
and their Fig.~3 shows the standard deviation of $J_\nu$ as a function of $\Delta_2$ for a closed short-axis loop orbit. %% shown in the inset. 
In Fig.~2 of \citet{SB16}, the curve labelled ``Fudge v2'' shows the variation of the SF-derived actions with time; 
the typical relative errors in the actions are roughly 10--20\%. 
\par

%%the TSF can have large errors. And here that is exacerbated by a non-smooth, $N$-body potential?}
The triaxial St\"ackel Fudge is a non-convergent action estimation method: it returns an approximate action vector $\mathbf{J}$ from a local St\"ackel approximation at a given phase-space point. 
Consequently, when the same orbit is sampled at successive times and the actions are re-estimated, 
Ref.~\citet{SB15} found that the recovered $(J_\lambda,J_\mu,J_\nu)$ can show substantial scatter, especially for box orbits and for regions of phase space where the potential is not well approximated by a St\"ackel form. 
In our implementation, this scatter for an orbit may be exacerbated by additional numerical effects. 
%Although energy is an exact integral in a time-independent potential, along numerically integrated trajectories the computed $E(t)$ can fluctuate because of finite time steps and because the adopted numerical potential representation (here we use the SCF potential which is very large particle count). 
Although energy is an exact integral in a time-independent potential, 
the computed $E(t)$ along a numerically sampled trajectory can fluctuate 
because of finite time resolution, finite integration accuracy, and the finite-order SCF representation of the snapshot potential. 
%%%% 
The intermediate quantities used by the St\"ackel approximation, including the pseudo-integrals $A_\tau$ and $B_\tau$, depend on the instantaneous phase-space point $(\mathbf{x},\mathbf{v})$ and can vary appreciably along an orbit in a non-St\"ackel potential. 
In addition, for some particles and some choices of the confocal parameters, the numerical evaluation of $p_\tau$ can become ill-conditioned within the $\tau$ integration interval, producing occasional large outliers in the estimated actions. 
These numerical and modeling effects can increase the apparent time variation of the recovered actions compared with the behavior seen in smooth analytic test potentials. 
\par 

Despite the orbit-by-orbit variability discussed above, 
the subsequent DF construction uses the distribution of a large particle sample. 
The scatter mainly broadens the numerical distribution in action space and 
does not make significant influence to $\DFnJ$ in statistics.
% does not significantly affect the statistical estimate of $\DFnJ$. 
\par

%[h]             % 'h' means the image is inserted at the current location
% figure name
%% by the geomspace between 0.01 to 100.0 times of median radius of the middle snapshot. 
% figure title

%[h]             % 'h' means the image is inserted at the current location
% figure name
%The particles we plot are chosen to be at different locations in the middle snapshot (namely the 3~Gyr one), linearly in the log-scale. 
%% by the geomspace between 0.01 to 100.0 times of median radius of the middle snapshot. 
\par

%\newpage
%% Sec5 begin
\section{\texorpdfstring{The $\DFnJ$ fitting module}{The f(J) fitting module}}\label{Sec5}

Module~4 constructs the numerical $\DFJ$ from a snapshot via KDE
and then fits it using a library of analytical functional forms. 
This section presents the technical details, including:
the KDE procedure for deriving the numerical \DFJ\ (\S5.1);
a brief description of the built-in library of \DFJ\ functions, 
%%from which \texttt{GroArnold} selects the best-fitting form for each numerical $\DFJ$ (\S5.2); 
and some considerations regarding the mathematical forms (\S5.2) --- useful 
for users who wish to implement their own custom $\DFJ$ functions;
% and a comparison of the spatial and velocity distributions implied by the fitted DF with those of the original snapshot (\S~\ref{sec_DF_phase_space_validation}).
and an illustration of the \DFJ\ fitting procedure with examples (\S~\ref{sec_M_fit-DFJ}).
\par
% After computing the numerical $\DFnJ$, we can fit the data with various analytical forms and select a best-fitting one. 
%%	The package includes flexible, built-in $\DFnJ$ models for fitting a wide range of user-input galactic properties... 
%%% The key innovations of \texttt{GroArnold} are threefold: ......
%%%   and (iii) it provides a flexible library of analytical $\DFJ$ forms to fit the numerical one.

% Note that some orbits of the particles in a galaxy may be unbounded,
Note that some orbits of the particles in a galaxy may be unbound,
and the TSF method may produce extreme action values for some initial conditions. 
In practice, we therefore screen particles by action bounds during data analysis. 
A particle is retained for computing statistical quantities 
if each component of its action lies in the range $[10^{-2},10^6]\,\mathrm{kpc\,km\,s^{-1}}$. 
For fitting $\DFnJ$, we apply the same action bounds before kernel density estimation.
In all example galaxies we have tested, more than 95\% of the galaxy particles
lie within the adopted screening ranges. 
\par

%
%% After obtaining the numerical distribution function of actions $\DFnJ$, 
%% we fit it with several analytical forms and select the best-fitting one. 
%% As an example, we employ 
%% % a multiple-power-law function (MPLF; see Section~\ref{sec_DFJ-formulae} for details) 
%% to fit the numerical distribution function of actions. 
%

Regarding the \DFJ\ fitting, %
the independent variables $X_i$ are the actions and frequencies estimated by TSF,
i.e., $X_i = (J_\lambda, J_\mu, J_\nu, \Omega_\lambda, \Omega_\mu, \Omega_\nu)$, 
where the first three components are the actions estimated by TSF and the last three are the corresponding frequencies.
Here, $i$ labels the particles. 
The dependent variables $Y_i=\log_{10} f_{N,i}$ are the logarithms (base 10) of the kernel density estimates 
derived from the three-dimensional action distribution of the $N$-body particles. 
The fitting is performed using the \texttt{mpfit} routine implemented in Python.
The above procedure can be repeated over multiple snapshots, allowing the user to estimate actions and fit $\DFnJ$ of multiple snapshots. 
\par

\subsection{Numerical estimation of the distribution function of actions} \label{sec_M_KDE}

When working directly in three-dimensional action space, no additional action state-density factor is required; 
a one-dimensional state-density factor enters only if the actions are compressed to a scalar variable $u=h(\mathbf{J})$. 
See Appendix~\ref{sec_action_state_density} for details. 
\par 

% After estimating the actions $(J_\lambda, J_\mu, J_\nu)$, frequencies $(\Omega_\lambda, \Omega_\mu, \Omega_\nu)$ and angles $(\Theta_\lambda, \Theta_\mu, \Theta_\nu)$ (see Module~3 above),
After estimating the actions $(J_\lambda, J_\mu, J_\nu)$, frequencies $(\Omega_\lambda, \Omega_\mu, \Omega_\nu)$ and angles $(\theta_\lambda, \theta_\mu, \theta_\nu)$ (see Module~3 above),
Module~4 estimates the numerical distribution function of actions by kernel density estimation (KDE) at any target point $\mathbf{J} = (J_\lambda, J_\mu, J_\nu)$ in action space, separately for each particle type (Gadget type). 
The kernel function $K(d;w)$ is the same as the smoothed-particle hydrodynamics (SPH) kernel used for mass-density estimates in \texttt{Gadget2} \citep[their Equation~(A.1)]{Gadget1}: 
%\begin{aligned}
%\end{aligned}
\begin{equation}\label{eq_kernel_gadget}
\begin{split}
&K(d;w)\, =\, \\
&\frac{8}{\pi w^3}
\begin{cases}
1-6(\frac{d}{w})^2+6(\frac{d}{w})^3,\quad & 0\leq\frac{d}{w}\leq \frac{1}{2}, \\
2(1-\frac{d}{w})^3,\quad & \frac{1}{2}\leq\frac{d}{w}\leq 1, \\
0,\quad & \frac{d}{w}>1.
\end{cases}
\end{split}
\end{equation}
%Then the (mass weighted) number density at a target point in action space $\mathbf{J} = (J_\lambda, J_\mu, J_\nu) \equiv J_j\ (j=1,2,3)$ is
Then the kernel estimate of the (mass weighted) distribution function of actions at a target point $\mathbf{J} = (J_1,J_2,J_3) \equiv (J_\lambda, J_\mu, J_\nu)$ is 
\begin{equation}\label{eq_density_kernel}
\begin{aligned}
\begin{split}
f(\mathbf{J}) = \sum\limits_{l=1}^{k} m_l\, K(d_l;w),
\end{split}
\end{aligned}
\end{equation}
% It is governing the smoothing length. Otherwise, by this definition $w$ will be the difference between the kernel center and some particle at $v$ large action 
where $d_l$ is the action-space distance from the target $\mathbf{J}$ to particle $l$. 
We set the kernel support radius $w$ equal to $d_{(k)}$, the distance to the $k$th nearest neighbour in action space: 
\begin{equation}\label{eq_r_h_target}
\begin{aligned}
\begin{split}
d_l &= \sqrt{\sum\limits_{j=1}^{3} (J_{j}-J_{l,j})^2}, \\
w &= d_{(k)} \equiv \max_{l\in\{1,\dots,k\}} d_l.
\end{split}
\end{aligned}
\end{equation}
Here $l=1,2,\dots,k$ indexes the $k$ nearest neighbor particles of the target point in action space, $m_l$ and $\mathbf{J}_l$ are the mass and actions of particle $l$, and $j=1,2,3$ correspond to the three TCE coordinates $(\lambda,\mu,\nu)$. %% - }
\par 

%%%%%%%%%%%%%%%%%%%%%%%%%%%%%%%%%%
%
%
%%%%%%%%%%%%%%%%%%%%%%%%%%%%%%%

\begin{figure*}[tbp]
	\centering
	\includegraphics[width=0.46\textwidth]{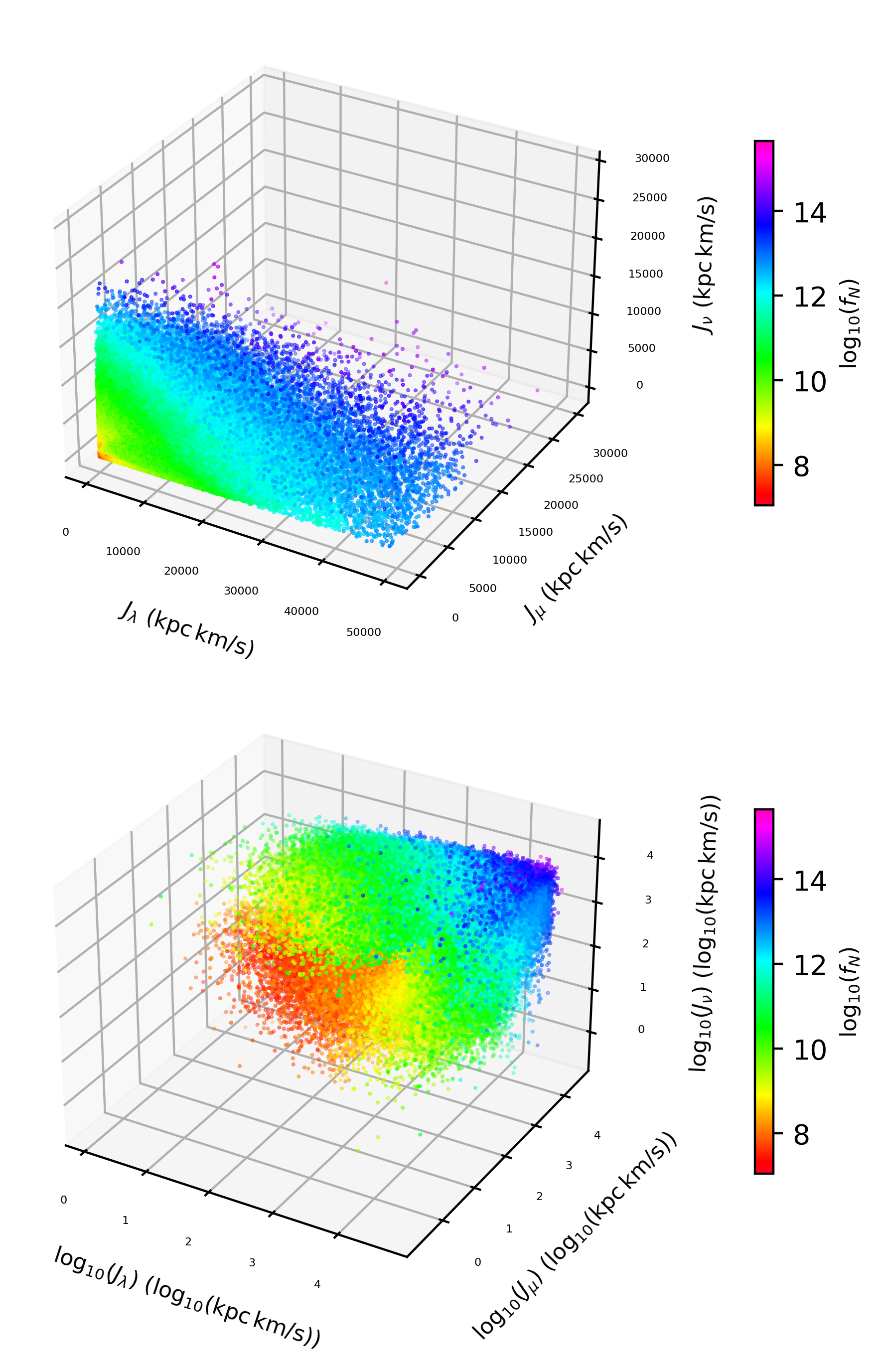} 
	\caption{
		Distribution of particles in the three-dimensional action space of the example galaxy GAL0, color-coded by $\log_{10} f(\mathbf{J})$. 
		The top panel shows the distribution on a linear scale, while the bottom panel shows the same distribution on a logarithmic scale. 
		The colorbar indicates the value of $\log_{10} f(\mathbf{J})$.		
	}
	\label{fig_actions_3d_bounds}
\end{figure*}

%
Fig.~\ref{fig_actions_3d_bounds} shows the particle distribution in three-dimensional action space 
for GAL0 (see the end of \S~\ref{Sec2} for details of this example galaxy), 
over the plotted range $10^{-2}\leq J \leq 5\times 10^4\,\mathrm{kpc\,km\,s^{-1}}$. 
\par 

% After computing the numerical $\DFnJ$, we can fit the data with various analytical forms and select a best-fitting one. 
% For example, a fitting function: a multiple-power-law function (called MPLF, see section \ref{sec_DFJ-formulae} for more details). 
% The input xdata are TSFD actions, $X_i = {J_\lambda, J_\mu, J_\nu, \Omega_\lambda, \Omega_\mu, \Omega_\nu}$; 
% the input ydata are the $\log_{10}$ values of kernel density of the 3D-actions data, $Y_i = \log_{10} f{N,i}(X_i)$.
% %The yerror: 
% % the standard deviation of the action values in 5 near snapshots as the error of actions, 
% % then the  the $\log_{10}$ value of $\DFnJ$ error can be computed by (\ref{eq_error_propagation}), 
% %%  for now: in this paper we do not compute and enable the value of $\DFnJ$ error when fitting. 
%
% The fitting program: mpfit in Python. 
% \par 

%
%\begin{multicols}{2}
%\begin{figure}[H]
%height=14.0cm,
%\end{multicols}
\par

\subsection{Library of \DFJ\ functions}
\label{sec_DFJ-formulae}

%
%% 1. have compiled all the available functions (see Appendix~\ref{sec_append-A_DFJ-formulae}).\\
%% 2. In addition to the \DFJ\ functional forms available in the literature, 
%% we have also devised several functions of our own, such as MPLF and MPLTF ! \\
%% 3. Because the functional forms of are rather sophisticated, 
%% we present some technical considerations involved in constructing f(J), 
%% in the following two subsubsections.
%% {The combinations of actions} and {On the mathematical structure of \DFJ}. 
%%    --- help users who wish to implement their own custom f(J) functions.\\
%

We have compiled a collection of analytical \DFJ\ forms that have been proposed in the literature, 
including those reviewed in Appendix~\ref{sec_append-A_DFJ-formulae}. 
This collection serves as a library from which 
\texttt{GroArnold} iterates over these analytical functions for each numerical $\DFnJ$ to select the best-fitting one. 
\par

In addition to the above functional forms, we have also devised several functions built into \texttt{GroArnold}, 
particularly tailored to the needs of triaxial halo modeling. 
Examples include the multiple power-law function (dubbed MPLF) and the multiple power-law with turning-point function (dubbed MPLTF). 
The latter is used as the fiducial model in the fitting examples presented in this work (see Eq.~\ref{eq_MPLTF}). 
The library is designed to be flexible, and users are encouraged to implement their own custom $\DFJ$ functions.
\par 

Because the functional forms of $f(\mathbf{J})$ can be rather sophisticated, 
we present below two technical considerations that are essential for constructing and using such functions in practice. 
These are: (i) the choice of action combinations and their coefficients (\S~\ref{sec_M_combinations}), 
and (ii) an illustration of the mathematical structure of a representative $f(\mathbf{J})$ form (\S~\ref{sec_M_math}).
\par

%
%
%% \subsection{The combinations of actions} \label{sec_M_combinations}
\subsubsection{The combinations of actions} \label{sec_M_combinations}
%%
%% our consideration and teaching the readers: how to construct a good f(J)
%%

%%% \subsubsection{Approximation of the action-combination variable}
%
Many previously proposed forms of $\DFJ$ are written in terms of combinations of actions, 
sometimes with frequency-dependent coefficients. For example, \citet{Posti15} adopted the form 
\begin{equation}
\label{eq_DFA_P15}
\begin{aligned}
\begin{split}
&f_\mathrm{P15}(\mathbf{J}) = \frac{M_0}{J_0^3} \\
&\quad \frac{\left(1+\frac{J_0}{h_\mathrm{P15}(\mathbf{J},\mathbf{\Omega})}\right)^{(6-\alpha)(4-\alpha)}}{\left(1+\frac{g_\mathrm{P15}(\mathbf{J})}{J_0}\right)^{2\beta-3}} ~, \\
&\mathrm{where} \\
&h_\mathrm{P15}(\mathbf{J},\mathbf{\Omega}) = (J_r\Omega_r+|J_\phi|\Omega_\phi+J_z\Omega_z)/\Omega_r ~, \\
&g_\mathrm{P15}(\mathbf{J}) = J_r+|J_\phi|+J_z ~.
\end{split}
\end{aligned}
\end{equation}
Here $(J_r, J_\phi, J_z)$ are the standard actions in a spherical system. 
%   And in Fig. \ref{fig_actions_2d_ELgh}, we plot the relationship between combinations 
%  ($g$ and $h$ in \ref{eq_gh}) and the energy of each particles in the galaxy GAL0, 
%   it can be shown that $g$ is near $h$. 
% Motivated by this structure, we assume the numerical $\DFnJ$ to be a one-dimensional function of the action-combination variable 
% in the form of Eq.~\ref{eq_gh}, 
% i.e. $f_N(\mathbf{J},\mathbf{\Omega}) \simeq f_N(h(\mathbf{J},\mathbf{\Omega}))$, for the purpose of fitting. 
\par

Motivated by the above function, 
in the current version of \texttt{GroArnold} 
we define several kinds of linear combinations of actions for constructing \DFJ\ functions, 
\begin{equation}\label{eq_gh}
\begin{aligned}
\begin{split}
g(\mathbf{J}) &= J_\lambda + J_\mu + J_\nu ~, \\
h_\mathrm{freq}(\mathbf{J}, \mathbf{\Omega}) &= J_\lambda + J_\mu \frac{\Omega_\mu}{\Omega_\lambda} + J_\nu \frac{\Omega_\nu}{\Omega_\lambda} ~, \\
h_\mathrm{freefixed}(\mathbf{J}) &= J_\lambda + \eta_\mu J_\mu + \eta_\nu J_\nu ~. 
\end{split}
\end{aligned}
\end{equation}
%%% for the authors' and readers' construct/guess f(J) formulae:
%% where $h_\mathrm{freq}$ or $h_\mathrm{freefixed}$ are combination function for actions used in the xxx module.
% 
The function $h_\mathrm{freq}$ combines the actions with coefficients determined by the frequencies, whereas $h_\mathrm{freefixed}$ combines the actions with free coefficients fitted from the data. 
%% Appendix~
Appendix~\ref{appendix_guide} describes the parameter settings for $h = h_\mathrm{freq}$ and $h = h_\mathrm{freefixed}$ adopted in the \texttt{GroArnold} framework. 
A judicious choice of the coefficients in the $h$ functions can reduce 
the scatter of $f_N(\mathbf J,\mathbf\Omega|\mathbf p)$ with respect to $h$. 
\par %%

%% new paragraph:
To provide an intuitive understanding of the role of the $h$ functions in \DFJ\ 
--- particularly for users who wish to customize their own functional forms --- 
we now further explore the three-dimensional action space (as shown in Fig.~6).
%%%
We group the data points of the example galaxy GAL0's particles 
by their $\log_{10} f(\mathbf{J})$ values into bins of width $0.1$ dex; each bin represents an approximate iso-density slice. 
Slices containing fewer than 32 particles are discarded.
\par

\begin{figure*}[tbp]
	\centering
	\includegraphics[width=0.9\textwidth]{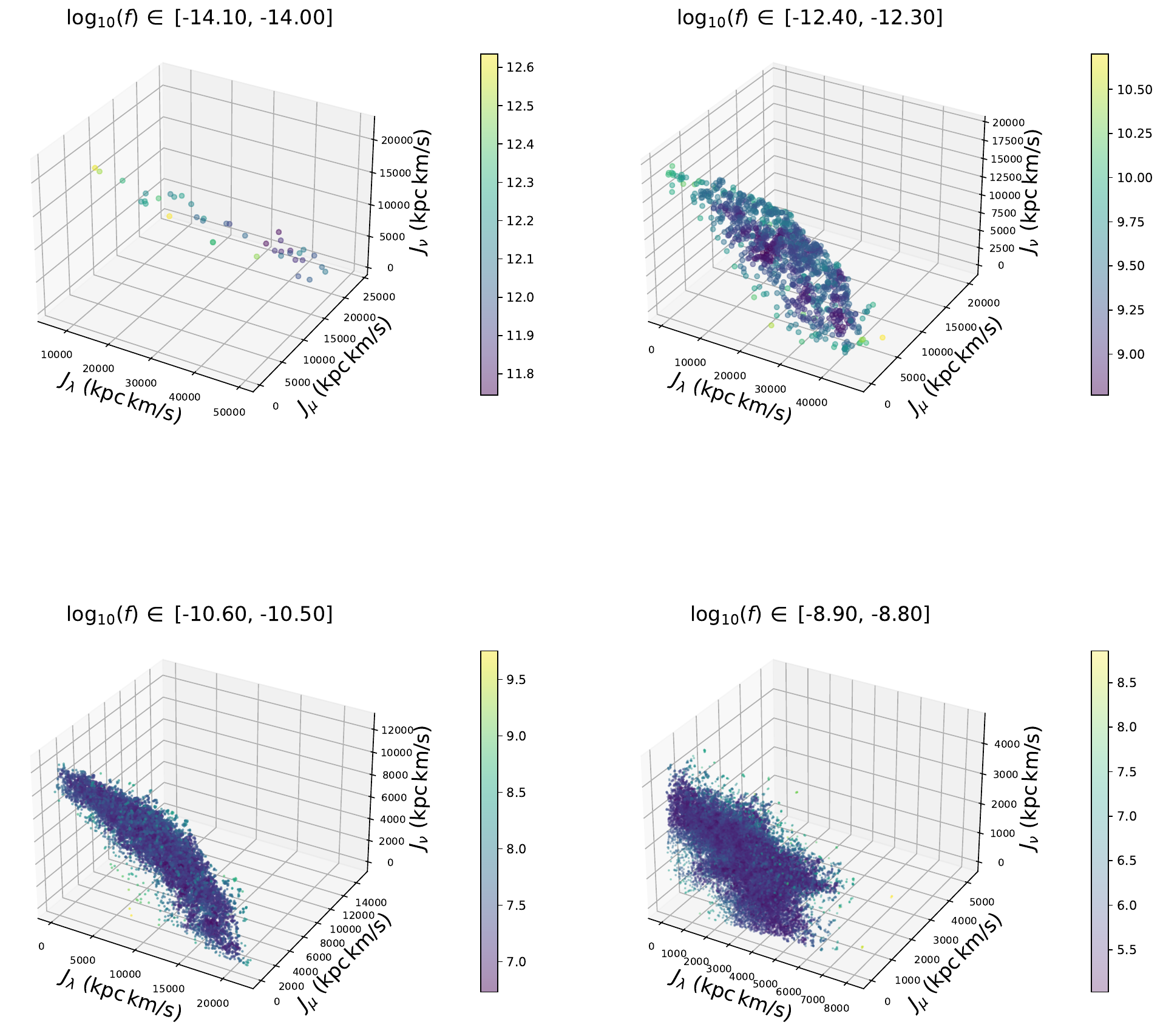}
	\caption{
		Distribution of selected iso-density slices 
		in three-dimensional action space of the example galaxy GAL0's particles.
		Each panel shows the particles within a fixed bin of $\log_{10} f(\mathbf{J})$; 
		the data points in each slice lie approximately on a plane, with some dispersion. 
		The colorbar in each panel indicates the local $\log_{10} f(\mathbf{J})$ value of the particles within that slice.	
	} 
	\label{fig_actions_3d_slices_J} 
\end{figure*}

% %Note that we have ignore the action state density $\mathcal{S}(\mathbf{J})$ 
% %(like the energy state density $g_E$ in statistical physics \citep{BT08}) and we only consider the $f_N(\mathbf{J}, \mathbf{\Omega})$. 
% 
% %Example code X. (actions $\DFnJ$) \par 
% 
% %Example code X. ($\DFnJ$ contour) \par 
% 
% %We partition the data points into multiple groups 
% %in term of the $\log_{10}$ value of $\DFnJ$,
% % with each group corresponding to every 0.1 intervals of $\log_{10} \DFnJ$, 
% % which means these groups near a iso-density slice of $\DFnJ$ (each slice corresponds a range of $\log_{10}$ value of $\DFnJ$). 
% %We discard the slices whose points count is less than 12. 
%
% We group the data points of GAL0's particles 
% by their $\log_{10} f(\mathbf{J})$ values into bins of width $0.1$ dex; each bin represents an approximate iso-density slice. 
% Slices containing fewer than 32 particles are discarded. \par 
% 
Fig.~\ref{fig_actions_3d_slices_J} shows the distribution of several selected iso-density slices in the 3D action space. 
The data points in each slice lie approximately on a two-dimensional surface, which in some cases is close to a plane. 
Assuming that the points in a given slice can be approximated by a plane of the form 
$A_\mathrm{slice} J_\lambda + B_\mathrm{slice} J_\mu + C_\mathrm{slice} J_\nu + D_\mathrm{slice} = 0$,
%we define two (negative) slope for each slice: 
we define two slope parameters for each slice: 
$k_\mu = B_\mathrm{slice}/A_\mathrm{slice}$, describing the variation in the $\mu$ direction relative to the $\lambda$ direction, 
and $k_\nu = C_\mathrm{slice}/A_\mathrm{slice}$, describing the variation in the $\nu$ direction relative to the $\lambda$ direction. 
For simplicity, we can roughly adopt functions that depend only on $h$ to fit $\DFnJ$ (i.e., the above $h_\mathrm{freq}$ or $h_\mathrm{freefixed}$ case; but see \S~\ref{Sec6} for deeper discussions).
\par

%% old: In Fig. \ref{fig_actions_3d_slices2d_J}, we plot the local number-density of data points for several slices in the $(J_\mu, J_\nu)$ 
%% two-dimensional projection of action space. 
%It appears that the points of every slice are roughly on a 2D surface, which is even close to a plane. 
%Thus, if we assume that the points of a slice is near the plane $A J_\lambda + B J_\mu + C J_\nu + D = 0$, 
%we can define two slopes for each slice ``plane'' in the 3D action space --- 
%the slope of $\mu$ to $\lambda$, $k_\mu = B_\mathrm{slice}/A_\mathrm{slice}$, 
%and the slope of $\nu$ to $\lambda$, $k_\nu = C_\mathrm{slice}/A_\mathrm{slice}$. 
%\par

%%% -- So, for simplicity, we adopt a function $\DFnJ$ depends only on $h$ 
%% (either the above $h_\mathrm{freq}$ or $h_\mathrm{frefixed}$ case) to fit $\DFnJ$. \par 
%
%
%In this paper, we display the fit function with $h=h_\mathrm{freq}$ case (fitting 1-$D$ bins). 
%
Specifically, the fitting of the example galaxies that we show in this work 
adopts the fitting functions with $h = h_{\mathrm{freq}}$ (one-dimensional bin fitting), 
just for the purpose of demonstrating the \texttt{GroArnold} workflow. 
The exploration of more accurate and physically motivated functional forms for $f(\mathbf{J})$ 
will be presented in a future study. 
\par

%I suppose one option is to downsample and not display as many points. 
%Or you could use a Gaussian KDE to estimate the density and then color points by the density?}
%
%
%[h]             % 'h' means the image is inserted at the current location
%Scatter plots in 3d action space for selected iso-DF slices. 
%In each panel, points represent particles within a bin of $\log_{10}(\DFnJ)$, 
%forming an approximately planar slice with some dispersion. 
%The color scale indicates the local $\log_{10}(\DFnJ)$ values of the particles within each slice.
%
% figure title
\par

%
%% \subsubsection{Adopted fitting functions}
%% \subsubsection{Illustrating the fitting functions}
\subsubsection{\texorpdfstring{On the mathematical structure of $\DFJ$}{On the mathematical structure of f(J)}}

Honestly speaking,\label{sec_M_math} the functional forms of \DFJ\ are generally rather sophisticated 
(see Equation~\ref{eq_DFA_P15} for an example).
To give users a concrete impression of their mathematical structure, 
we describe and demonstrate a specific form, 
dubbed MPLTF (standing for multiple power-law with turning-point function).
We will use MPLTF as the fiducial model to fit the numerical \DFnJ\ 
of the halo components in the example cases presented in the next subsection.
\par

The mathematical expression of MPLTF is as follows:
\begin{equation}
\label{eq_MPLTF}
\begin{aligned}
\begin{split}
&f_N(\mathbf J,\mathbf\Omega|\mathbf p_\mathrm{all}) = f_N(h|J_0,J_1,J_2,J_3,p_1,p_2,p_3,b) \\
&= \frac{M_\mathrm{sel}}{J_0^3}\left(\frac{h}{J_1}+1\right)^{-p_1} \left(\left(\frac{h}{J_2}\right)^2-b\frac{h}{J_2}+1\right)^{-p_2} \\
&\quad \left(\left(\frac{h}{J_3}\right)^3+1\right)^{-p_3},
\end{split}
\end{aligned}
\end{equation}
where $M_\mathrm{sel}$ is the total mass of particles 
retained after the action-range and the halo particle-type selection, 
$h = h(\mathbf{J},\mathbf{\Omega})$ is the combination function of actions defined in Eq.~\ref{eq_gh}, and $\mathbf p_\mathrm{all}$ denotes all fitting parameters.
For fitting, the mass-weighted KDE values are divided by $M_\mathrm{sel}$; this factor is restored in the reported DF.
This form is similar to the halo $\DFJ$ of Eq.~(14) in \citet{Pascale19}. 
The scale $J_0$ sets the overall normalization through $M_\mathrm{sel}/J_0^3$, while $J_1$, $J_2$, and $J_3$ set the characteristic action scales of the three power-law factors. 
The indices $p_1$, $p_2$, and $p_3$ control the corresponding slopes, and $b$ controls the turning-point behavior of the second factor. 
This additional parameter allows the fit to reproduce the curvature or inflection of the binned $\DFnJ$ data as a function of $h$. 
In the fitting, the parameters are restricted so that the bracketed factors remain positive over the fitted range of $h$. 
\par 

% The fitting form adopted here is a one-dimensional function of $h$, so we first bin the $h$--$f_N$ data and then fit the binned data (see Fig.~\ref{fig_fit_compare7}b).
The fitting form adopted here is a one-dimensional function of $h$, so we first bin the $h$--$f_N$ data and then fit the binned data (see Fig.~\ref{fig_fit_compare7}(a)).
Note that the blank regions in the contours correspond to masked $\log_{10}\,\DFnJ$ values that become NaN, either due to taking the logarithm of negative values in the fitting function or due to values lying outside the valid range as a result of the uncontrolled rapid growth induced by the power-law terms. 
This indicates that these fitting functions with many power-law terms (and exponential functions) might still not be perfect. 
\par

%% %Example code X. (some adviced fit functions avalable)  \par 
%% 
%% Some functions with similar forms are still alternative. 
%% %Here we do not adopt a mixed form of power laws and exponent because the large systematic error of TSF/TSFD actions 
%% and the finite range of actions data do not support to fit the function with power law and exponent well with out parameter degeneracy. 
%% We do not adopt mixed power-law--exponential forms here because the large systematic error of TSFD actions and 
%% the finite range of the action data make such forms difficult to constrain without strong parameter degeneracies. 
%% \par
%% 
%% In summary, we adopt the MPLTF as a practical parametric form for describing the binned halo $\DFnJ$ in our examples, 
%% while noting that its flexibility is limited and that alternative forms may be explored in future applications. 
%% \par 
%

%
%
% \subsection{Fitting analytic forms to the numerical distribution function of actions}
% % \label{sec_fit}
% %% ver 5, 5.4
% %%\subsection{The fitting results: examples}
\subsection{The fitting procedure: examples}
\label{sec_M_fit-DFJ}

In this subsection, we describe the steps to derive the best-fit \DFJ\ formula for a user-specified galaxy (or galactic component), 
and illustrate the fitting outcomes with several concrete examples. 
\par

Galactic triaxial systems span a range of dynamical states and configurations, 
from dispersion-dominated ones to those with considerable rotation.
We therefore construct a suite of galaxy models with varying observational properties.
The procedure is described in \S~\ref{Sec3} and briefly outlined here: 
(i) specify an observation-level initial condition with desired $q_z$ and $\lambda_\mathrm{spin}$; 
(ii) generate and evolve an $N$-body realization to obtain a stable snapshot; 
(iii) apply the alignment procedure; 
(iv) compute the SCF potential coefficients and construct the foci-energy table; 
(v) estimate actions and frequencies for all particles using the TSFD method; 
(vi) compute the numerical $f(\mathbf{J})$ via kernel density estimation; 
(vii) fit both the mass density profile and $f(\mathbf{J})$, and recompute the relevant statistical parameters directly from the snapshot, including those listed in Table~\ref{table_galaxy_params}.
This last step is necessary because, as noted in \S~\ref{Sec3}, 
these parameters may evolve during the simulation 
and need not coincide exactly with the user-input values. 
\par

%%% $1\times 10^6$ halo particles, $5\times 10^5$ disk particles, and $1\times 10^5$ bulge particles. 
Each simulated galaxy contains about $5\times 10^5$ gas particles, $10^6$ halo particles, $5\times 10^5$ disk particles, and $10^5$ bulge particles. 
In the following analysis, we focus on the halo component.
Table~\ref{table_galaxy_params} lists the halo parameters of the evolved snapshots; in particular, the axis ratios describe the shapes after the dynamical evolution discussed in Section~\ref{Sec3}.
The listed parameters are as follows: 
\begin{itemize}
	\item NAME: name of the simulated galaxy model.
	\item PROFILE: dark matter (DM) halo profile type in the initial-condition settings in \texttt{DICE}.
	\item $M_\mathrm{halo}$: mass of the halo, $10^{10}\,M_\odot$.
	\item $r_\mathrm{med}$: median radius of halo particles, $\mathrm{kpc}$.
	\item $\lambda_\mathrm{spin}$: spin parameter of the halo.
	% \item $q_y$: $y$-axis ratio of the halo, computed by Eq.~(\ref{eq_qz}).
	% \item $q_z$: $z$-axis ratio of the halo, computed by Eq.~(\ref{eq_qz}).
	\item $q_y$: $y$-axis ratio of the galaxy, computed by Eq.~(\ref{eq_qz}).
	\item $q_z$: $z$-axis ratio of the galaxy, computed by Eq.~(\ref{eq_qz}).
	\item $q_{J_\mu}$: action-space $J_\mu$ ratio of the halo, computed by $q_{J_\mu} = \sqrt{\frac{\sum\limits_{i} J_{i,\mu}^2}{\sum\limits_{i} J_{i,\lambda}^2} }$, similar to Eq.~(\ref{eq_qz}).
	\item $q_{J_\nu}$: action-space $J_\nu$ ratio of the halo, computed by $q_{J_\nu} = \sqrt{\frac{\sum\limits_{i} J_{i,\nu}^2}{\sum\limits_{i} J_{i,\lambda}^2} }$, similar to (\ref{eq_qz}).
\end{itemize}
\vspace{\baselineskip}
\par
%\bigskip

%The spin parameter. 
The definition of spin parameter of the DM halo component $\lambda_\mathrm{spin}$ (\citet{BT08}, formula (9.89)) is
\begin{equation}\label{eq_spin}
\begin{aligned}
\begin{split}
\lambda_\mathrm{spin} = \frac{L_\mathrm{tot} |E_\mathrm{tot}|^{1/2}}{G M_\mathrm{tot}^{5/2}},
\end{split}
\end{aligned}
\end{equation}
where $M_\mathrm{tot}$ is the total mass, $E_\mathrm{tot}$ is the total energy, and $L_\mathrm{tot}$ is the total angular momentum of the DM halo component. \par 

%The $z$-axis ratio.
We estimate the halo axis ratios $q_y$ and $q_z$ from the particle distribution as follows: 
\begin{equation}
\label{eq_qz}
\begin{aligned}
q_y\!=\!\sqrt{\frac{\sum\limits_i m_i y_i^2}{\sum\limits_i m_i x_i^2}},\,q_z\!=\!\sqrt{\frac{\sum\limits_i m_i z_i^2}{\sum\limits_i m_i x_i^2}},
\end{aligned}
\end{equation}
where $i$ is the index of particles. 
%The $z$-axis ratio $q_z$ (the flattening in the vertical direction) indicates the halo's tendency towards a prolate (large $q_z$) or oblate (small $q_z$) shape. 
The $z$-axis ratio $q_z$ measures the flattening in the vertical direction. 
%In combination with $q_y$, it provides a simple shape diagnostic for distinguishing oblate-like, prolate-like, and more generally triaxial halo configurations. 
Together with $q_y$, it indicates whether the halo is closer to an oblate configuration ($q_y\simeq 1$ and $q_z<1$), a prolate configuration ($q_y\simeq q_z<1$), or a fully triaxial configuration with $1>q_y>q_z$. 
\par

\begin{figure*}[tbp]
	\centering
	\begin{subfigure}[t]{0.46\textwidth}
		\centering
		\includegraphics[width=\linewidth]{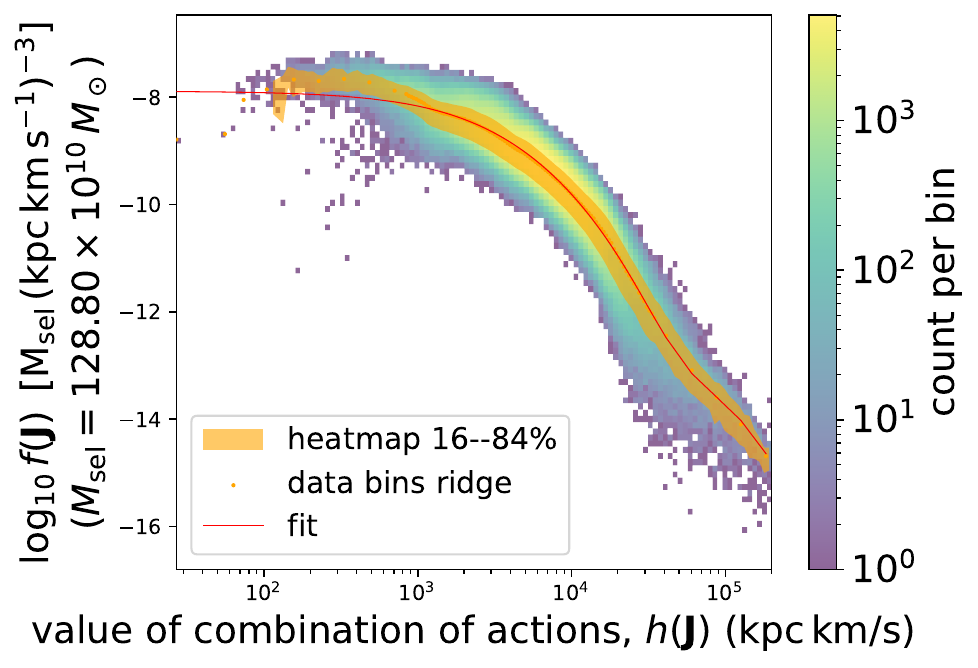}
		\caption{
			Demonstration of the fitting result, expressed in the form of distribution function of actions versus the action combination $h$.
			The colored background shows the particle counts per bin in the $(h, \log_{10} f)$ plane, 
			and the orange band indicates the 16th--84th percentile uncertainty range. 
			The orange points mark the ridge-line data points extracted from the particle distribution, 
			and the red curve is the best-fitting MPLTF model.
		}
		\label{fig_compare5_1}
	\end{subfigure}
	\hspace{0.06\textwidth}
	\begin{subfigure}[t]{0.46\textwidth}
		\centering
		\includegraphics[width=\linewidth]{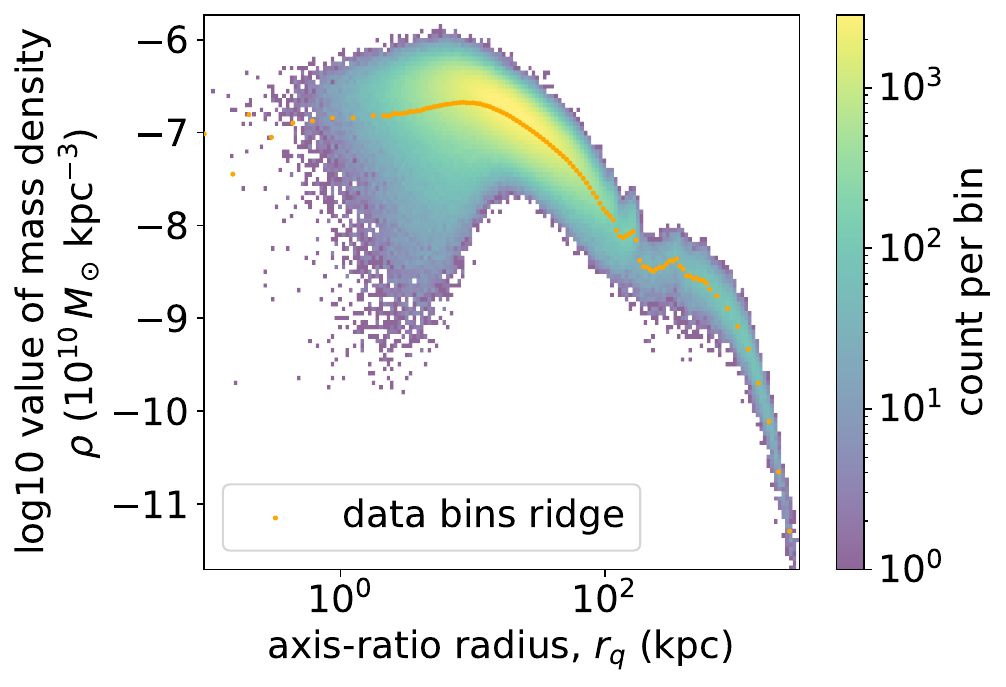}
		\caption{
			The mass density profile $\rho$ as a function of the dimensionless flattened ellipsoidal radius $r_\mathrm{sq}$ (Eq.~\ref{eq_rsq}). 
			The background shows the particle counts per bin in the $(r_\mathrm{sq},\log_{10}\rho)$ plane. 
			The orange points show the ridge-line data points extracted from the particle distribution. 
		}
		\label{fig_compare5_2}
	\end{subfigure}
	\par\medskip
	\begin{subfigure}[t]{0.8\textwidth}
		\centering
		\includegraphics[width=\linewidth,height=0.44\textheight,keepaspectratio]{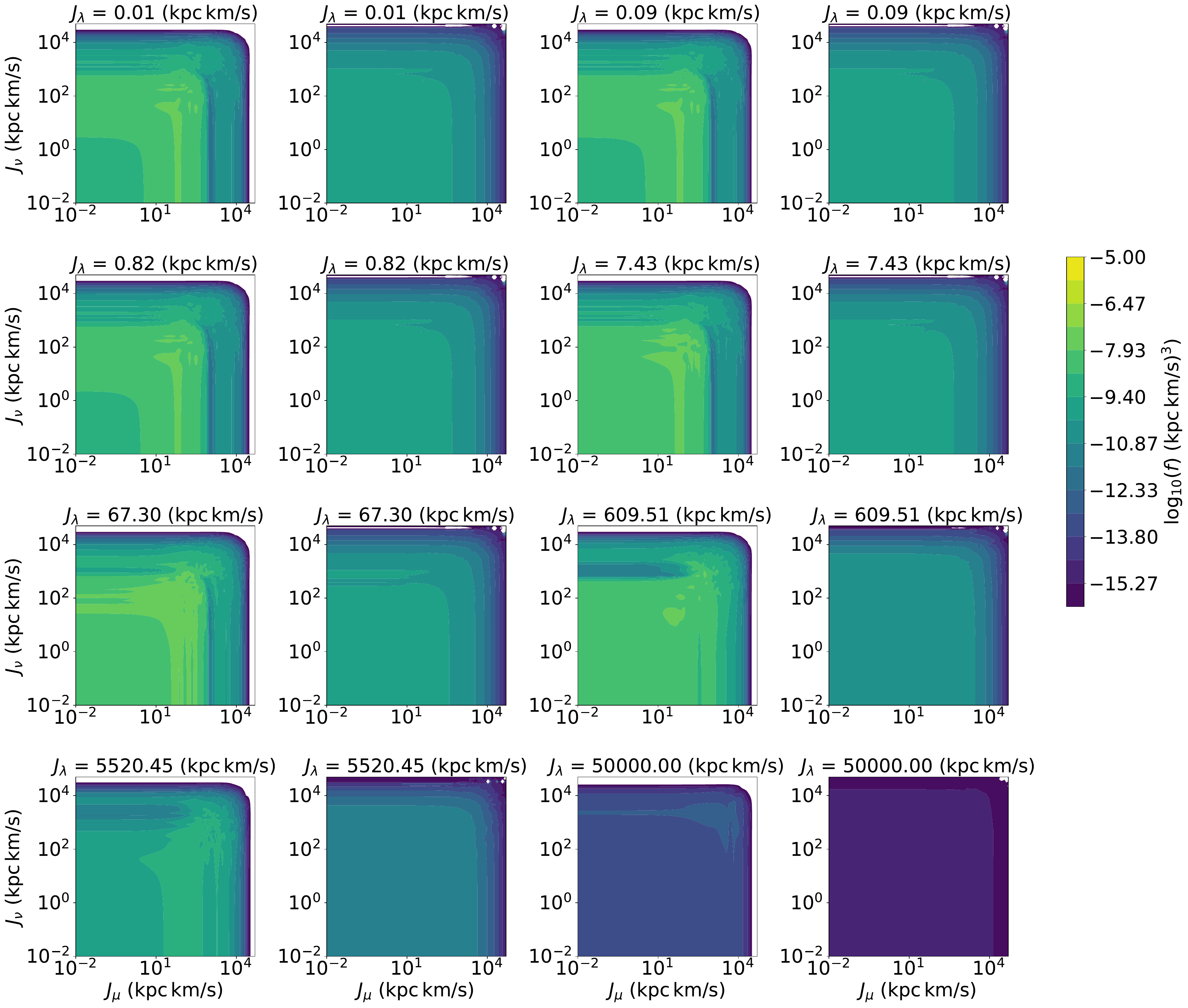}
		\caption{
			Comparison of $f(\mathbf{J})$ contours in the $(J_\mu, J_\nu)$ plane at several fixed $J_\lambda$ slices. 
			In each pair, the left panel shows contours derived from the numerical data, and the right panel shows those from the best-fitting $f(\mathbf{J})$ formula. 
			Both axes are plotted on a logarithmic scale, and the color bar indicates $\log_{10} f(\mathbf{J})$.
		}
		\label{fig_compare5_3}
	\end{subfigure}

	\caption{
		Comparison between the numerical data and the best-fit formula (Eq.~\ref{eq_MPLTF}) 
		for the mass-weighted distribution function of actions $\DFnJ$,
		where $M_\mathrm{sel}$ is the total mass of particles selected from a given component (e.g., a halo) within the fitting range.
		This figure corresponds to the simulated galaxy \texttt{Ein\_spinL\_axisLH0} in Table~\ref{table_galaxy_params}.
	}
	\label{fig_fit_compare5}
\end{figure*}

%%Fig.~\ref{fig_fit_compare5}, as well as \ref{fig_fit_compare8} 
%
Fig.~\ref{fig_fit_compare5} shows the fitting results for the galaxy model \texttt{Ein\_spinL\_axisLH0} (see Table~\ref{table_galaxy_params}), 
% while Figs.~\ref{fig_fit_compare6}--\ref{fig_fit_compare8} in \ref{fig_fit_others} show the corresponding results for the other galaxy models.
while Figs.~\ref{fig_fit_compare6}--\ref{fig_fit_compare8} in Appendix~\ref{fig_fit_others} show the corresponding results for the other galaxy models.
In each figure, panel~(a) compares the numerical $f(\mathbf{J})$ obtained from kernel density estimation with the best-fitting analytical formula, 
plotted as a function of the action combination $h$ (specifically $h_{\mathrm{freq}}$; see Eq.~\ref{eq_gh}). 
%Panel~(b) shows the mass density profile $\rho$ as a function of the dimensionless flattened ellipsoidal radius $r_{\mathrm{sq}}$ (Eq.~\ref{eq_rsq}). 
Panel~(b) shows the mass density profile $\rho$ as a function of the dimensionless flattened ellipsoidal radius $r_{\mathrm{sq}}$ defined in Eq.~\ref{eq_rsq}. 
Panel~(c) provides a more detailed comparison between the numerical and best-fitting $f(\mathbf{J})$ in the $(J_\mu, J_\nu)$ plane for several fixed $J_\lambda$ slices. 
Together, these figures demonstrate that the simple linear combination $h_{\mathrm{freq}}$ 
together with 
the MPLTF fitting function 
(Eq.~\ref{eq_MPLTF}) 
is capable of providing reasonably good fits to $f(\mathbf{J})$ for a variety of triaxial halos. 
\par

As a preliminary exploration, we investigate the connection between the observation-level parameters and $f(\mathbf{J})$. 
%For triaxial halos, there are two key parameters that shape their mass distribution and action-based distribution function $f(\mathbf{J})$: the vertical axis ratio $q_z$ and the spin parameter $\lambda_\mathrm{spin}$. 
For the triaxial halos considered here, we focus on two parameters that affect both the mass distribution and the distribution function of actions: the vertical axis ratio $q_z$ and the spin parameter $\lambda_\mathrm{spin}$. 
%%%s
%Comparing Fig.~\ref{fig_fit_compare5} with Fig.~\ref{fig_fit_compare7}, and Fig.~\ref{fig_fit_compare6} with Fig.~\ref{fig_fit_compare8}, we examine the effect of the spin parameter $\lambda_\mathrm{spin}$ while keeping other parameters nearly identical. 
By comparing Fig.~\ref{fig_fit_compare5} with Fig.~\ref{fig_fit_compare7}, and Fig.~\ref{fig_fit_compare6} with Fig.~\ref{fig_fit_compare8}, we examine the effect of the spin parameter $\lambda_\mathrm{spin}$ while keeping the other parameters nearly fixed. 
We find that the distribution $f(\mathbf{J})$ becomes more dispersed for higher spin. 
%%%
Similarly, by comparing Fig.~\ref{fig_fit_compare5} with Fig.~\ref{fig_fit_compare6}, and Fig.~\ref{fig_fit_compare7} with Fig.~\ref{fig_fit_compare8}, we examine the influence of the vertical axis ratio $q_z$ under otherwise similar conditions. 
The results indicate that the $f(\mathbf{J})$ curve as a function of $h$ exhibits stronger curvature for lower $q_z$.
\par 

%
%%By comparing 
%%Fig.~\ref{fig_fit_compare5} vs. Fig.~\ref{fig_fit_compare7}, 
%%Fig.~\ref{fig_fit_compare6} vs. Fig.~\ref{fig_fit_compare8}, 
%%for different $\lambda_\mathrm{spin}$ (with the other parameters kept nearly identical), 
%%we find that the $\DFnJ$ with higher spin is more dispersed.  
%%\par 
%%
%%By comparing 
%%Fig.~\ref{fig_fit_compare5} vs. Fig.~\ref{fig_fit_compare6}, 
%%Fig.~\ref{fig_fit_compare7} vs. Fig.~\ref{fig_fit_compare8}, 
%%for different $q_z$ (with the other parameters kept nearly identical), 
%%we find that the $\DFnJ$ curve as a function of $h$ is more strongly curved for lower spin.
%%\par

%[h]             
% No blank line means no line breaks
% No blank line means no line breaks %\hfill %Add horizontal blank space appropriately
%% panel C:
% No blank line means no line breaks
\par

%\subsubsection{Compare two kind of spaces}
%Here we compare the relations of actual parameters of DFs in positions and velocitesspace and in actions and frequencies space. \par 
% Example X. (p-p relation) \par 

%\newpage 
%% \clearpage
%% \FloatBarrier
%
%\onecolumn
\setcounter{MaxMatrixCols}{40}
%[htbp]
\begin{table*}[tbh] 
	\caption{
		Parameters of the dark matter halo component in different simulated galaxies. 
	}
	\label{table_galaxy_params}
	\centering
	\small
	\setlength{\tabcolsep}{4pt}
	\begin{tabular}{ccccccccc}
		\toprule
		NAME				&PROFILE	&$M_\mathrm{halo}$	&$r_\mathrm{med}$	&$\lambda_\mathrm{spin}$	&$q_y$	&$q_z$		&$q_{J_\mu}$	&$q_{J_\nu}$ \\
		\midrule
		\texttt{Ein\_spinL\_axisLH0}	&Einasto	&130.15				&16.71				&0.00098					&0.981	&0.948		&0.338			&0.296 \\
		\texttt{Ein\_spinL\_axisLH1}	&Einasto	&130.15				&14.24				&0.00160						&0.991	&0.778		&0.360			&0.260 \\
		\texttt{Ein\_spinH\_axisLH0}	&Einasto	&130.15				&16.60				&0.280						&1.022	&0.834		&0.446			&0.320 \\
		\texttt{Ein\_spinH\_axisLH1}	&Einasto	&130.15				&13.98				&0.331						&0.972	&0.650		&0.493			&0.238 \\
		\bottomrule		
	\end{tabular}
\end{table*}
\par 
%\twocolumn
%

%% Sec6 begin
\section{Discussion}
\label{Sec6}

%The \texttt{GroArnold} framework provides a tool for fitting the number density distribution function of actions to get an analytical $\DFnJ$ form for any user-specified triaxial galaxy or its galactic components. 
The \texttt{GroArnold} framework provides a tool for fitting the mass-weighted distribution function of actions to obtain an analytical $\DFnJ$ form
for a user-specified triaxial galaxy or for individual galactic components. 
%By integrating a series of independent software modules --- ranging from initial condition generation,
%$N$-body simulation, action estimation,
%to the construction and fitting of \DFJ\  --- 
%
The framework enables the action-based modeling of triaxial systems that represent observed galactic structures more realistically than traditional spherical or axisymmetric models, and provides a flexible approach to fitting the $\DFnJ$ in near-equilibrium QSS galaxies (halos), accommodating a wide variety of initial conditions and mass density profiles as the user would like to specify. 
To assess the fidelity of the output DF, below we first compare its predicted spatial and velocity distributions with those of the input $N$-body snapshot.
\par

\subsection{Comparison of input and output spatial and velocity distributions}
\label{sec_DF_phase_space_validation}
% galaxy_general was the generic server run directory, while the GAL0 results were renamed for archival storage and downloaded locally.

%% mock mcmc
{For this comparison, the input snapshot is the GAL0 halo snapshot used for DF fitting, after the preprocessing described in Section~\ref{sec_M_aligenment}.
The output sample consists of particles drawn from its best-fitting DF.
This comparison connects the fit in action space to the density structure and kinematics of the simulated halo. }
\par 

Note that the triaxial St\"ackel Fudge implementation used here does not provide the inverse map from actions and angles to $(\mathbf{x},\mathbf{v})$.
On the other hand, the inverse mappings provided by \texttt{TorusMapper} \citep{BM16_TM} and \texttt{AGAMAb} \citep{Binney2026_AGAMAb} are restricted to axisymmetric potentials.
We therefore use Markov chain Monte Carlo to sample directly in phase space. 
\par 

{At each trial point $(\mathbf{x},\mathbf{v})$, the triaxial St\"ackel Fudge provides the actions $\mathbf{J}_\mathrm{TSF}$ and frequencies $\boldsymbol{\Omega}_\mathrm{TSF}$ needed to evaluate the fitted function $f_\mathrm{fit}$. 
The sampling probability density within the selected phase-space region $\mathcal{S}$ is }
\begin{equation}
\label{eq_DF_phase_space_sampling}
p_\mathrm{samp}(\mathbf{x},\mathbf{v}) = \frac{1}{\mathcal N_{\mathcal S}}f_\mathrm{fit}\!\left[ \mathbf{J}_\mathrm{TSF}(\mathbf{x},\mathbf{v}), \boldsymbol{\Omega}_\mathrm{TSF}(\mathbf{x},\mathbf{v})\right] ~,
\end{equation}
where $\mathcal{N}_{\mathcal{S}}$ normalizes the density over $\mathcal{S}$. 
The probability density is zero outside this region. 
\par 

{For the GAL0 halo example, the sampler evaluates the fitted MPLTF of Equation~(\ref{eq_MPLTF}) through $h_\mathrm{freq}$ as defined in Equation~(\ref{eq_gh}), while keeping the input snapshot's SCF potential fixed.
We compare gravitationally bound halo particles within $200\,\mathrm{kpc}$ in the same coordinate frame, 
and assign the output sample the same total mass as 
that of the halo particles in this region of the input snapshot.
The spatial comparison therefore examines the particle distribution at fixed total mass, and the velocity distributions are compared after normalization.
Appendix~\ref{app_DF_phase_space_comparison} gives the sampling details and further comparisons. }
\par 

\begin{figure*}[tbp]
	\centering
	\includegraphics[width=\textwidth,keepaspectratio]{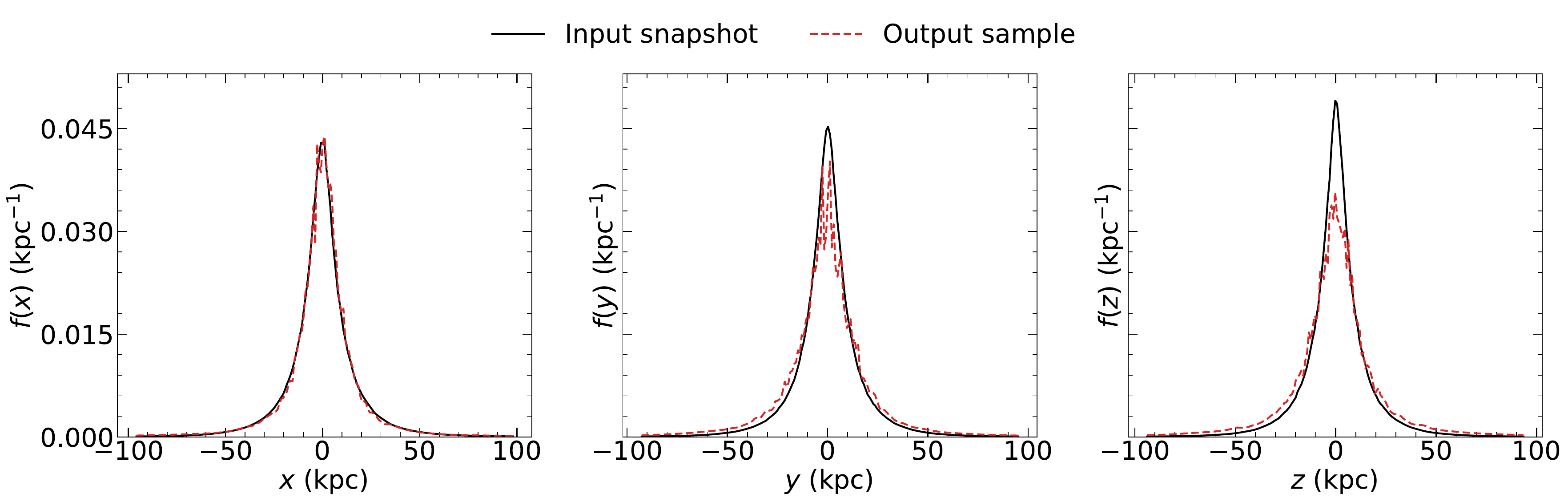}
	\caption{One-dimensional histograms of the GAL0 halo particles showing the spatial distribution along the $x$, $y$, and $z$ axes, 
		obtained by integrating over the other two spatial coordinates. 	
		Black solid curves denote the input snapshot and red dashed curves the output sample.
		Each histogram is normalized over the displayed interval.
		All three panels share the same vertical scale.
	}
	\label{fig_DF_position_marginals}
\end{figure*}

\begin{figure*}[tbp]
	\centering
	\includegraphics[width=\textwidth,keepaspectratio]{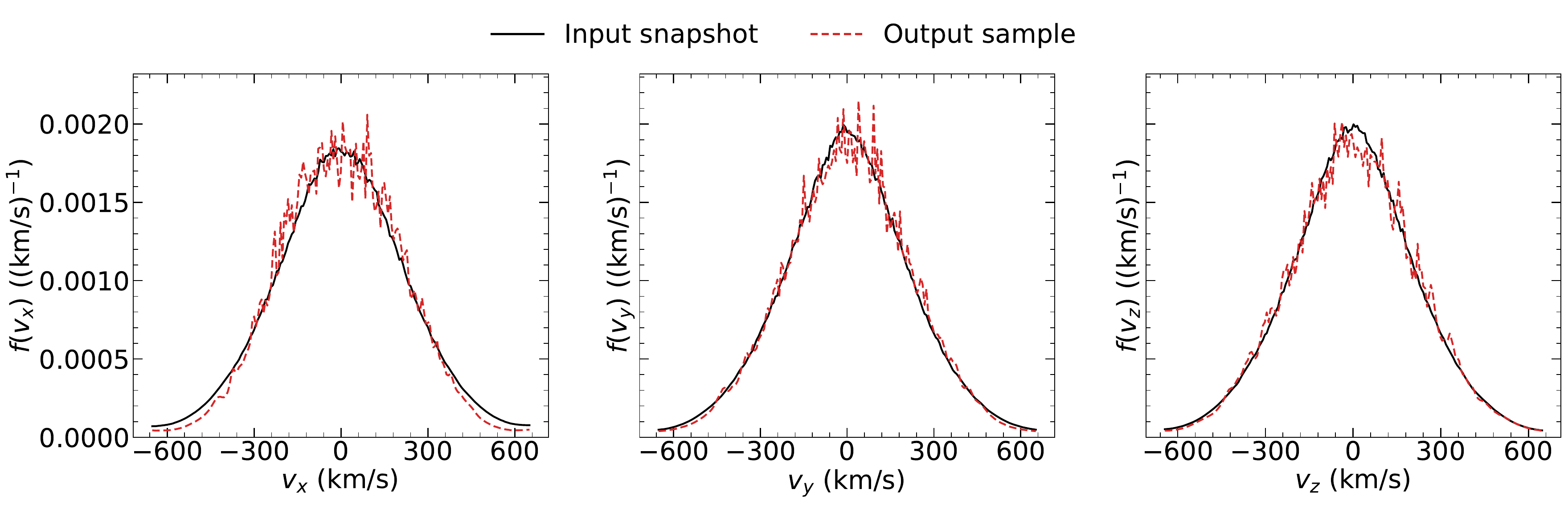}
	\caption{
		One-dimensional histograms of $v_x$, $v_y$, and $v_z$ for the GAL0 halo particles, 
		computed over the full selected spatial region in the same way as for Fig.~\ref{fig_DF_position_marginals}.
		Black solid curves denote the input snapshot and red dashed curves the output sample. 
		Each histogram is normalized over the displayed interval. 
		All three panels share the same vertical scale.
	}
	\label{fig_DF_velocity_marginals}
\end{figure*}

{Figures~\ref{fig_DF_position_marginals} and \ref{fig_DF_velocity_marginals} compare the one-dimensional histograms 
% one-dimensional histograms 
(namely marginal distributions, obtained by integrating over the other two coordinates)
of the three Cartesian position and velocity components for GAL0.
The output sample reproduces the position distributions reasonably well, 
although the central peaks are broader and the half-mass radius is about $21\%$ larger than that of the input snapshot.
These differences are most pronounced within roughly \(\pm10\,\mathrm{kpc}\) along each axis, 
probably reflecting the broadening introduced by errors in 
converting from position--velocity to action--angle coordinates 
in the St\"ackel Fudge action estimation method.
%%converting between position--velocity and action--angle coordinates in the St\"ackel Fudge method.
%%%
For the velocity distributions, the agreement is much better: the output sample matches the input snapshot closely, 
with the largest discrepancy being a reduction of merely $\approx8\%$ in the global velocity dispersion of the $v_x$ component. }
\par

\subsection{Comparison with other similar projects}

\texttt{AGAMA} \citep{V19} provides a broad framework for action-based galaxy modeling, 
including potential construction, orbit integration, action estimation, and self-consistent multicomponent models. 
Its latest extension, \texttt{AGAMAb} \citep{Binney2026_AGAMAb}, 
further improves torus generation and torus-based action finding in axisymmetric potentials, 
including for highly eccentric and nearly circular orbits that are challenging for earlier torus-mapping implementations 
and for the St\"ackel Fudge method of \citet{B12}.
%%%
However, these methods in \texttt{AGAMA} and \texttt{AGAMAb} are formulated for axisymmetric potentials, 
and no general action-estimation method is provided for triaxial potentials. 
By contrast, 
\texttt{GroArnold} transforms position--velocity coordinates into angle--action variables 
not only in spherical and axisymmetric potentials, but also in triaxial potentials, 
using the triaxial St\"ackel Fudge of \citet{SB15} as implemented in \texttt{TACT}.
In addition, \texttt{GroArnold} extends \texttt{TACT} to \texttt{Nbody-TACT}, enabling action estimation directly from $N$-body snapshots.
\par 

A distinctive feature of \texttt{GroArnold} is that 
%we elaborately implement a \emph{stable} $N$-body model for the galaxy that the user specifies, 
we elaborately construct a \emph{stable} $N$-body realization of the user-specified galaxy 
which is not emphasized in other programs such as \texttt{AGAMA}, \texttt{AGAMAb} and \texttt{galpy} \citep{Bovy2015}. 
Since the simulated $N$-body system is stable, 
the best-fit $\DFJ$ likely represents a dynamically stable model. 
This is the sense in which we call the constructed $\DFJ$ ``numerically empirical''. 
\par 

\subsection{Several future developments}

Here we list several \textit{ad hoc} assumptions and limitations of the current version, which will be improved in future developments: 
\par 

(i) Assumptions in Action Estimation: In the action-estimation step, \texttt{GroArnold} employed the TSF method in triaxial confocal ellipsoidal coordinates. 
Although this approach is effective for approximating the actions of particles moving in a triaxial potential, 
the assumption of a non-rotating galaxy (halo) may introduce inaccuracies. 
The fact that the system is not fully synchronized in its rotation leads to a distortion in the effective potential, 
which may affect the accuracy of action estimation, particularly for dynamic systems. 
%If the figure rotation is significant or not well described by a single pattern speed, the effective potential in a rotating frame is not captured by the current implementation, which may reduce the accuracy of the estimated actions. 
\par 

(ii) Handling Resonances and Instabilities: Ignoring possible resonances within the system is a second limitation. 
Although the majority of orbits within the simulated galaxies are stable, 
the lack of resonance considerations might lead to inaccuracies in action estimation, 
particularly for systems exhibiting complex dynamics such as resonant interactions or chaotic behavior. 
We may seek modern methods to handle resonances explicitly to improve the robustness of action estimation. 
\par 

(iii) Fitting Errors and Parameter Uncertainties: The current routine of fitting $\DFnJ$ 
does not account for the full uncertainties associated with the action estimation. 
These uncertainties, particularly the ones arising from time variations in the TSFD actions, 
could propagate into the final $\DFnJ$ fits. 
%Future improvement of the code 
Future versions of the code 
will incorporate more sophisticated error propagation methods %(\ref{eq_error_propagation}) 
during the fitting process. 
\par 

(iv) More accurate and interpretable functional forms for $\DFnJ$ need to be developed. 
\par 

\begin{figure*}[tbp]
	\centering
	\includegraphics[width=0.46\textwidth]{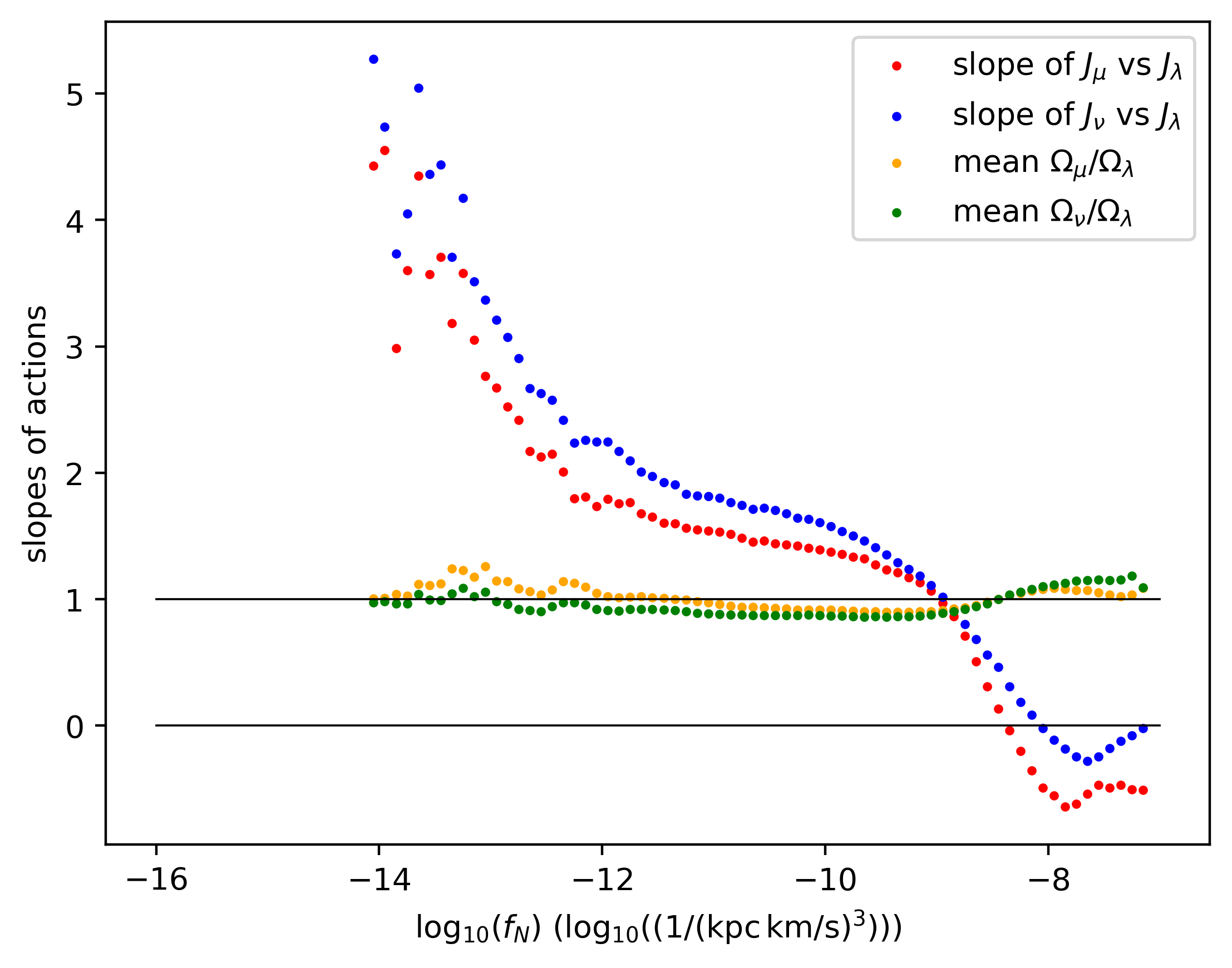}
	\caption{
		Fitted slope parameters of iso-$\DFnJ$ slices as functions of $\log_{10}\DFnJ$, compared with the corresponding mean frequency ratios. 
		The red points show the fitted slope $k_\mu$ of each slice, treated locally as a plane in action space, and the blue points show the fitted slope $k_\nu$.
		The orange points show the mean frequency ratio $\bar{q}_\mu$ for each slice, and the green points show $\bar{q}_\nu$. 
	}
	\label{fig_actions_3d_slopes_J} 
\end{figure*}

Fig.~\ref{fig_actions_3d_slopes_J} shows that the fitted slope parameters $k_\mu$ and $k_\nu$ vary systematically with $\log_{10}\DFnJ$.
We can see that 
%both are not, 
neither is, 
in general, equal to their respective mean frequency ratios within the same slice. 
Nevertheless, the fact that each iso-$\DFnJ$ subset is well described by an approximately planar surface in action space, with well-defined $k_\mu$ and $k_\nu$, indicates that a one-dimensional combination function $h$ that follows the local orientation of these surfaces is meaningful and useful for fitting. 
The mean ratios $\bar{q}_\mu=\langle \Omega_\mu/\Omega_\lambda\rangle_{\rm slice}$ and $\bar{q}_\nu=\langle \Omega_\nu/\Omega_\lambda\rangle_{\rm slice}$ provide a convenient scaling for such coefficients, which motivates the choice $h=h_\mathrm{freq}$ in Eq.~(\ref{eq_gh}). 
\par 

%\begin{figure}[H]
%Fitted slopes of iso-$\DFnJ$ slices in $\log_{10}$ scale and the corresponding mean frequency ratios. 
\par

However, the current $\DFnJ$ functions could be made more flexible. 
In the perfect ellipsoid potential or St\"ackel potential, for a given point in action space $(J_\lambda, J_\mu, J_\nu)$, 
orbit types can be distinguished by analyzing their region on the action surface of constant energy (see action diagram in Figs.~28 and 29 of \citealt{deZ85}). 
The function $h(\mathbf{J})$, represented by a \emph{linear} combination of actions with coefficients related to the corresponding frequencies as shown in Eq.~\ref{eq_gh}, roughly aligns with constant-$\DFnJ$ surfaces in the current implementation. 
However, in a realistic galactic potential, this constant $\DFnJ$ surface may deform, 
indicating that the combination function and its coefficients should be generalized as follows: 
$X(\mathbf{J}) = J_\lambda + \eta_\mu (\mathbf{J}) J_\mu + \eta_\nu (\mathbf{J}) J_\nu$ (instead of the above $h(\mathbf{J})$), 
and the coefficients $\eta_\mu = \eta_\mu(J_\lambda, J_\mu, J_\nu)$ 
and $\eta_\nu = \eta_\nu(J_\lambda, J_\mu, J_\nu)$,
%which lead the combination to be nonlinear, 
which makes the combination nonlinear, 
i.e. a curved surface instead of a plane of constant $\DFnJ$ slice. 
%%%
%% could deviate from their associated frequencies, which is much complicated than the fitting functions in current \texttt{GroArnold} version. 
%This deviation might suggest that a more adaptable $\DFnJ$ of actions such that  
%the coefficients are functions of actions $\eta_j = \eta_j(J_\lambda, J_\mu, J_\nu),\, i=1,2$, 
%which lead the combination to be nonlinear, 
%i.e. a curved surface instead of a plane of constant $\DFnJ$ slice. 
%%%%%
On the other hand, many $\DFJ$ forms have been developed in the literature, 
including the halo $\DFJ$ coefficients of \citet{BV23} and the applications of \citet{SE15} to isophote twisting and radially unstable models. 
It therefore seems meaningful to extend and test these $\DFnJ$ forms in the triaxial case in the future. 
%
%Many $\DFJ$ forms have also been developed in the literature, including ... Extending and testing these forms in triaxial systems is therefore a useful direction for future work. 
\par

(v) Tests with more strongly triaxial models: The modular design allows alternative initial-condition generators or existing $N$-body realizations to be used.
A useful direction for future work is to construct stable models that retain more pronounced triaxial shapes, extending the range of systems on which the framework can be tested.
\par 

%% Sec7 begin
\section{Summary}\label{Sec7}

%% since ``arbitrary'' is risky because not every user-specified model 
%% will necessarily evolve to a usable stable or quasi-stationary snapshot.}
%
With the main goal of facilitating the user's \DFJ-based modeling applications, 
the \texttt{GroArnold} software framework provides a complete end-to-end pipeline 
for extracting and fitting the numerical distribution function of action variables based on $N$-body data, 
and hence obtaining an analytic $\DFnJ$ for %% user-specified 
axisymmetric or triaxial galaxies and dark matter halos that can be specified by the user via an initial configuration file. 
Starting from user-specified properties as observation-level initial conditions, 
\texttt{GroArnold} generates (and evolves) an $N$-body realization to a stable and quasi-stationary state (QSS), 
aligns the QSS snapshot into a well-defined triaxial frame, 
estimates angle--action variables with the $N$-body version of the \texttt{TACT} code 
on a self-consistent field (SCF) potential (namely the \texttt{Nbody-TACT} module), 
and computes the empirical $\DFnJ$ by kernel-density estimation.
Finally, equipped with a suite of built-in analytic forms,%
\footnote{The user can modify or design their own $\DFnJ$ functional forms to fit the simulation-based DF data points.} 
\texttt{GroArnold} then obtains the best-fitting $\DFnJ$ function and its best-fitting parameters for the galaxy or halo 
with the observable properties that the user wants. 
The open-source code of \texttt{GroArnold} (including \texttt{Nbody-TACT}) is available at \url{https://github.com/starlifting1/GroArnold}\/.
\par 

The \texttt{GroArnold} framework has the following highlights:
\begin{enumerate}
	\item We extend the Triaxial St\"ackel Fudge code to discrete gravitational systems, such as the output of an $N$-body simulation, as implemented in \texttt{Nbody-TACT} (Module~3 of \texttt{GroArnold}). 
	This code is based on the open-source package \texttt{TACT} (\url{https://github.com/jls713/tact}\/) developed by \citet{SB16_TACT_code}.
	\texttt{Nbody-TACT} allows action estimation directly from the data of simulation snapshots, instead of analytic gravitational potential models as the original \texttt{TACT} uses. 
	
	\item We integrate stand-alone tools---\texttt{DICE} for $N$-body ICs, \texttt{Gadget2} for dynamics, SCF for potentials, \texttt{Nbody-TACT} for actions, and KDE for densities---into a unified, modular workflow initiated by a single user configuration file. 
	
	%%Action-based distribution functions in explicit functional forms 
	%% can be obtained for any axisymmetric and triaxial gravitational systems 
%
	\item The user can obtain action-based DFs in explicit analytic forms for any axisymmetric or triaxial system (including $N$-body realizations). 
	%The user can obtain action-based DFs in explicit analytic forms for user-specified axisymmetric or triaxial systems, including $N$-body realizations. 
 %The package includes flexible, built-in $\DFnJ$ models to fit a wide range of galactic properties that the user inputs (e.g., density profile including the degrees of coredness or cuspiness, configurational axis ratios, spin parameter, etc.). 
	The package includes flexible, built-in $\DFnJ$ models for fitting a wide range of user-input galactic properties, including density profiles with different degrees of coredness or cuspiness, configurational axis ratios, and spin parameter. 
 Additionally, the user can implement their own fully custom $\DFnJ$ functions.
 	%by comparing its predicted spatial and velocity distributions with those of the input $N$-body snapshot,
	%and confirmed that it reproduces the input with high fidelity. }    
	We also verified the fidelity of the best-fitting $f(\mathbf{J})$ by sampling its predicted phase-space distribution $f(\mathbf{x},\mathbf{v})$ 
	and comparing the output sample with the input $N$-body snapshot in both position and velocity space. 
	We found that the output sample reproduces the input snapshot reasonably well, 
	except for a broadening in the central region of the spatial distribution 
	 $f(\mathbf{x})$ 
	 likely caused by conversion errors from position--velocity to action--angle coordinates in the St\"ackel fudge method.
	% likely caused by conversion errors between position--velocity and action--angle coordinates in the St\"ackel fudge method.
	
\end{enumerate}
\par 

%\subsection{Extending \texttt{GroArnold}}
%\begin{enumerate}
%	\item Alternative IC generators -- supply Gadget-format file; Module 2 will ingest it directly.
%	\item Higher-order potentials -- replace the SCF routine with your favourite basis-function or grid solver and recomplile the Module 3.
%	\item Different fitting laws -- replace fitting method or fitting functions in Module 4.
%\end{enumerate}
%\par 

The \texttt{GroArnold} framework opens up several promising avenues for future research. 
For instance, the framework allows the user to incorporate additional galactic components 
(e.g., stellar disks, bulges and other substructures) as specified in the configuration file, 
enabling the construction of more realistic and structurally rich dynamical models. 
Furthermore, by applying \texttt{GroArnold} to observational phase-space data from large-scale stellar surveys such as Gaia, 
it becomes possible to perform detailed dynamical modeling of 
the Milky Way's internal components---particularly those with triaxial structures---as well as of satellite galaxies in its vicinity. 
In addition, researchers can systematically investigate how the best-fitting forms and parameters of $\DFJ$ 
correlate with observable kinematic quantities, such as the velocity dispersion tensor $\sigma_{ij}$ or the spin parameter $\lambda_\mathrm{spin}$, 
using large suites of $N$-body simulations of galaxies and dark matter halos. 
Collectively, these prospective applications will help deepen our understanding of the assembly and evolutionary pathways of triaxial galactic systems.
\par

%%%%%%%%%%%%%%%%%%%%%%%%%%%%%%%%%%%%%%%%

\begin{acknowledgements}
%\Acknowledgements{
%\section*{Acknowledgements}
%%%%
We thank the anonymous referee for the constructive comments and suggestions that have improved this paper.
We are sincerely grateful to Jason Sanders for his insightful comments and substantial input, which greatly improved this work. 
%
%
%%%%
We thank Shude Mao, Ling Zhu, Zhaoyu Li, Wenting Wang, Jiaxin Han and Yougang Wang for helpful discussions and comments,
and Wenjuan Liu for providing computational resources.
%%%%
We would like to offer an interpretation of the name ``\texttt{GroArnold}'' of our numerical framework: 
in honor of the intellectual heritage of Mikhail~Gromov and the late Vladimir I.~Arnold. 
This work is built upon a certain knowledge of inviscid fluids and symplectic systems, 
% and we wish this dedication serves as a fitting tribute and
and we hope this dedication may be regarded as a fitting tribute and
expresses our profound respect and gratitude for their monumental contributions and their styles, which we deeply admire.
%%%%
This work is supported by the National Natural Science Foundation of China (grant No. 12373013).
%}
\end{acknowledgements}

%\InterestConflict{
%The authors declare that they have no conflict of interest.
%}

%%%%%%%%%%%%%%%%%%%%%%%%%%%%%%%%%%%%%%%%

%\bibliographystyle{aasjournal} %The file to control format of reference and downloaded from the website of journal
%\nocite{*} %Copy exported bibtex content to reference_bibtex.bib; add this line of command to display all reference; delete main.* except for main.tex; compile and run
%\bibliography{reference_bibtex} %The .bib file, copy the bibtex content from the website of reference

%\bibliographystyle{scpma}
%\bibliography{reference_bibtex}

\bibliographystyle{raa}
\bibliography{reference_bibtex}

\clearpage
%%%%%%%%%%%%%%%%%%%%%%%%%%%%%%%%%%%%%%%%

%\begin{appendix}
% Already in two-column mode.

\appendix
%\twocolumn

\section{A brief introduction to (Triaxial) St\"ackel Fudge} \label{sec_M_TSF}

\subsection{A brief introduction to angle--action variables}\label{sec_introduction_actions}

%the intuitive meaning of actions
In an integrable Hamiltonian system, the actions are loop areas in canonically conjugate planes:
\begin{equation}
J_i = \frac{1}{2\pi}\oint p_i\,dq_i,
\end{equation}
so each $J_i$ measures the amplitude of a regular oscillation (or circulation) for every degree of freedom
and is invariant under phase mixing and slow (namely adiabatic) changes of the potential. 
\par 

In an axisymmetric galaxy with an integrable potential in cylindrical coordinates, 
the actions can be chosen as $\{J_R, J_\phi, J_z\}$, 
where $J_R$ measures the amplitude of oscillation in the $R$-direction, 
$J_\phi$ is the $z$-component of angular momentum ($L_z$) and 
measures the amplitude of oscillation in the $\phi$-direction, 
and $J_z$ measures the amplitude of oscillation in the $z$-direction. 
\par 

In a triaxial potential that is (approximately) of St\"ackel form 
in confocal ellipsoidal coordinates $(\lambda,\mu,\nu)$ 
with conjugate momenta $(p_\lambda,p_\mu,p_\nu)$ (see Appendix~\ref{sec_TEC_coordinate}),
the motion separates into three parts, 
and the three actions are given by the one-dimensional oscillation integrals 
between the respective turning points \citep{deZ85,SB15}:
\begin{equation}
\begin{split}
&J_\tau = \frac{1}{\pi}\int_{\tau_-}^{\tau_+} p_\tau(\tau;E,I_2,I_3)\,\mathrm d\tau, \\
&\qquad \tau\in\{\lambda,\mu,\nu\},
\end{split}
\end{equation}
where $\tau_\pm$ are the roots of $p_\tau^2(\tau)=0$ set by energy and the additional isolating integrals. 
Geometrically, $J_\lambda$ measures the radial oscillation across nested ellipsoids (iso-$\lambda$), 
$J_\mu$ measures the oscillation threading one-sheet hyperboloids (iso-$\mu$), 
and $J_\nu$ measures the oscillation threading two-sheet hyperboloids (iso-$\nu$) 
% (see the Section~5 and Fig.~8 of \citet{deZ85} for the orbit families classification in triaxial St\"ackel potential system).
(see Section~5 and Fig.~8 of \citet{deZ85} for the classification of orbit families in triaxial St\"ackel potentials).
\par 

Finally, the fundamental frequencies follow from the Hamiltonian as
\begin{equation}
\mathbf{\Omega}(\mathbf{J}) \;=\; \frac{\partial H}{\partial \mathbf{J}}\;,
\end{equation}
so that angle variables advance linearly, 
$\dot{\mathbf{\theta}}=\mathbf{\Omega}(\mathbf{J})$, 
and resonances occur when $\mathbf{n}\!\cdot\!\mathbf{\Omega}=0$ for some integer vector~$\mathbf{n}$. 
\par 

In the axisymmetric limit of a triaxial St\"ackel potential, two of the three focal parameters coincide 
	(e.g., $\alpha=\beta$ in the standard definition of confocal ellipsoidal coordinates), 
	so the coordinates degenerate from $(\lambda,\mu,\nu)$ to the usual azimuthal angle $\phi$ plus a 2D spheroidal pair $(\lambda,\nu)$ 
	(prolate or oblate spheroidal coordinates). 
	In this limit, the $\mu$--motion becomes a pure circulation around the symmetry axis 
	and its action is just the azimuthal angular momentum \citep{deZ85,SB15}. 
\par 

In the oblate axisymmetric limit (symmetry about the short axis, so $x^2+y^2\equiv R^2$ and $\Phi=\Phi(R,z)$\/), 
	the confocal ellipsoidal coordinates degenerate to the prolate spheroidal system $(\lambda,\nu,\phi)$, 
	in which $\phi$ is cyclic. Consequently, \begin{equation}
J_\phi \equiv \frac{1}{2\pi}\oint p_\phi\,\mathrm{d}\phi = p_\phi = L_z ~,
\end{equation}
and the remaining two St\"ackel actions reduce (up to the conventional loop--orbit normalization) to the usual radial and vertical actions,
\begin{equation}
J_R \equiv J_\lambda \, ,\qquad J_z \equiv J_\nu \, .
\end{equation}
\par 

In a Hamiltonian system, integrals of motion are phase-space functions that remain constant along trajectories and may be exposed by suitable canonical transformations. For an integrable potential, the actions $\mathbf{J}$ provide a particularly useful set of three independent integrals: 
	they label the invariant orbital tori,
	are adiabatic invariants under slow changes of the potential, 
	and admit canonically conjugate angles $\mathbf{\theta}$ that increase linearly in time with frequencies $\mathbf{\Omega}(\mathbf{J})$; 
see below for details. 
\par 

For (quasi-) steady-state systems (usually $\partial f/\partial t = 0$), the Jeans theorem states that the distribution function can be expressed as a function of the integrals of motion, $f = f(\mathbf I)$. 
Furthermore, for systems with regular orbits and non-resonant frequencies, the distribution function can be written in terms of three actions, $f = f(\mathbf J)$, in accordance with the strong Jeans theorem \citep{BT08}. 
\par 

As J.~Binney has emphasized in recent years, the Jeans equation does not fully exploit the information contained in the velocity distribution 
(since it considers only up to second-order moments), and he promotes the use of $\DFJ$ for modeling galaxies.
Here, ``$\DFJ$'' refers to the distribution function expressed in terms of actions. 
In traditional dynamical modeling of the Milky Way, it is usually assumed that the galaxy is in a steady state and can be treated as an integrable system. 
Consequently, each component of the galaxy has a distribution function based on integrals of motion, and a functional form for the $\DFJ$ is fitted to relate the positions and velocities of tracers to the gravitational potential of each component \citep{BR13, Piffl15, Read14, Pablo21}. 
\par 

The advantages of modeling with $\DFJ$ rather than with conventional integrals 
% (e.g., energy $E$, angular momentum $L$, the third integral of axisymmetric galaxy $I_3$, et al.) include the following (see also \citealt{V19}):
(e.g., energy $E$, angular momentum $L$, and the third integral $I_3$ of an axisymmetric galaxy) include the following (see also \citealt{V19}):
\par 

\begin{enumerate}
	\item
	Angle--action variables form a set of canonical coordinates for the (6-dimensional) phase space of a stellar orbit in an integrable mean-field potential.
	%% OLD: Action and angle variables are canonical, i.e. the action has conjugate coordinates or angular variables. 
%
	The motion of a star is on an invariant torus with a constant frequency, 
		% and the transformation from position--velocities to angle--action is volume preserving.
	and the transformation from position--velocity coordinates to angle--action variables is volume preserving.
	The volume of 6D phase space occupied by the stellar motions is just proportional to the volume in 3-dimensional action space.
	%The transformation between the original phase-space coordinate $(\mathbf x,\,\mathbf v)$ and angle--action $(\mathbf \theta,\,\mathbf J)$ has unit determinant, which is convenient for dealing with DFs. 
	
	\item
	Angle--action variables are naturally suited to analyze 
	small-amplitude relaxation as well as 
	perturbations around a (quasi-)equilibrium state \citep{2016MNRAS.457.2569M, 2015A&A...584A.129F, V19}.
	
	\item %3: invariants.
	For gravitational potentials that evolve slowly (on timescales long compared with orbital periods), 
	the action variables $\mathbf{J}$ are adiabatic invariants \citep{BT08}: 
	for regular, non-resonant orbits they remain nearly conserved, with only small corrections, 
	and this approximation breaks down only near resonances or during separatrix crossings. 
	This property makes action-based DFs robust under secular mass growth or gradual redistribution of galactic components, 
	and hence particularly convenient for iterative self-consistent modeling.
	More broadly, we note in passing that collisionless self-gravitating systems, 
	admitting a natural Hamiltonian description through the Vlasov--Poisson equation as noted above, 
	also invite symplectic-geometric and topological perspectives
	that provide a broader mathematical setting for mean-field gravitational dynamics 
	\citep{Gromov1985,Arnold2021}, 
	although such aspects lie beyond the scope of the present work.
	
	%%For slowly varying gravitational potentials, the action variables $\mathbf{J}$ act as adiabatic invariants \citep{BT08}: 
	%%xxxx
	%%This property makes action-based DFs robust under gradual mass growth or redistribution of components, 
	%%and hence particularly convenient for iterative self-consistent modeling. 
	 %Geometrically, actions are defined as integrals over a basis of fundamental cycles on invariant tori and are unique 
	 % only up to integer linear transformations of that cycle basis (a $\mathrm{GL}(n,\mathbb{Z})$ freedom); 
	 % they are therefore best viewed as \emph{canonical adiabatic invariants}, 
	 % not as ``topological invariants'' of the underlying phase-space manifold.
	%% symplectic topology:   
	%%-  Note that Gromov's non-squeezing theorem is a global rigidity result in symplectic topology (expressed via symplectic capacities); 
	%%-   it underscores that symplectic maps are more constrained than mere volume-preserving maps, 
	%%-   NOT action (I.o.M), NOT poincare's integral invariants  \citep{Duistermaat1980, Gromov1985}. 
%
 %%Besides, we note in passing that the Hamiltonian flow of collisionless Boltzmann (Vlasov) systems is symplectic, 
 %%and symplectic systems admit genuine topological invariants beyond the invariants concerning actions discussed above
%
 %%Such invariants offer powerful topological tools for studying mean-field gravitational systems, 
 %%though exploring them lies beyond the scope of this work.  --- too much irrelavant words.

	\item
			%If $E$ appears in the parameter list, the phase space density around the orbit will change in an uncontrolled manner. 
	%One cannot specify the $f(E,...)$ of the star at the beginning of the modeling work in a self-consistent model. 
	%When the distribution function is only a function of the action, the phase space density around each orbit does not change. 
	%So DF of actions has this advantage to model a galaxy with multi-component systems. 
		The orbital energy $E$ is not a quantity that can be specified locally in phase space: it depends on the gravitational potential, which in turn depends on the entire mass distribution of the galaxy \citep{B20}. As a result, a distribution function that depends explicitly on $E$ cannot be specified consistently at the start of a self-consistent modeling procedure, because the value of $E$ changes whenever the potential is updated. This causes the phase-space density associated with an orbit to vary in an uncontrolled way during the iteration. In contrast, the actions $\mathbf{J}$ provide a set of orbital labels that remain invariant under slow changes in the potential. A distribution function of the form $\DFJ$ therefore maintains a fixed phase-space density around each orbit, making it far better suited for constructing multi-component, self-consistent galaxy models. 
 %}
\end{enumerate}
\par 

%% A2:
\subsection{Methods of calculating angle--action variables}\label{sec_action_estimation_methods}

\citet{SB16} reviewed the main angle--action methods in axisymmetric and triaxial potentials and emphasized the difference between convergent and non-convergent methods.
For instance, 
%%before June3.
\citet{B12} introduced a rapid algorithm to estimate the angle--action variables in an axisymmetric potential --- the (axisymmetric) St\"ackel Fudge;
%% several OLD revgreen below. 
\citet{S12} introduced a method for estimating angle--action variables in a realistic Milky Way axisymmetric potential by locally fitting a St\"ackel potential over the region an orbit probes; 
\citet{SB14} introduced a method that recovers the generating function that maps an analytic phase-space torus to the torus to which the orbit is confined by simultaneously solving the constraints provided by each time step (``O2GF'' method in \citealt{SB16}); 
%\citet{SB15} introduced a triaxial St\"ackel Fudge method for action estimation in spherical/axisymmetric/triaxial confocal ellipsoidal coordinates under a spherical/axisymmetric/triaxial potential. 
\citet{SB15} introduced a method for action estimation in spherical, axisymmetric or triaxial confocal ellipsoidal coordinates under a spherical, axisymmetric or triaxial potential --- especially the Triaxial St\"ackel Fudge (TSF) method --- 
in their \texttt{TACT} code (\url{https://github.com/jls713/tact}; \citealt{SB16_TACT_code}), 
which has high computational efficiency at the expense of limited computational accuracy. 
\citet{MB90} introduced ``torus mapping'' to obtain actions for a general axisymmetric potential; 
\citet{BM16_TM} presented the \texttt{TorusMapper} code, which is a non-perturbative technique for creating orbital tori for specified values of the action integrals, 
and which can obtain 
$(\mathbf{x}, \mathbf{v})$ from $(\mathbf{\theta}, \mathbf{J})$. 

\citet{V19} presented the open-source \texttt{AGAMA} library (Action-based Galaxy modeling Architecture), a C++/Python framework that combines flexible gravitational-potential solvers (including expansions built from analytic density profiles or from $N$-body snapshots), orbit integration and orbit analysis, action-based distribution functions and their moments, iterative construction of self-consistent multicomponent galaxy models, and orbit-superposition (Schwarzschild) modeling. 
At present, \texttt{AGAMA} provides built-in transformations between $(\mathbf{x},\mathbf{v})$ and $(\mathbf{\theta},\mathbf{J})$ 
for spherical and axisymmetric potentials, using the axisymmetric St\"ackel Fudge approximation for the latter.
Fully triaxial action estimators (e.g., the triaxial St\"ackel Fudge) are instead offered by other dedicated packages, 
such as \texttt{TACT} \citep{SB15}. 
 \par

	The availability of fast action estimators has made the practical construction of genuinely triaxial action-based models a reality. 
	Notably, 
	the triaxial St\"ackel Fudge of \citet{SB15} generalizes the original St\"ackel Fudge to confocal ellipsoidal coordinates, 
 enabling rapid estimation of actions in arbitrary triaxial potentials. 
	The gain in speed comes at the cost of an approximation whose accuracy varies across different orbit families and regions of phase space, 
	yet it is often sufficient for computing DF moments and for building workable triaxial dynamical models.
		\par

% \subsection{Triaxial confocal ellipsoidal coordinate}\label{sec_TEC_coordinate}
\subsection{Triaxial confocal ellipsoidal coordinates}\label{sec_TEC_coordinate}

A (triaxial) St\"ackel potential is a potential for which the Hamilton--Jacobi equation separates in \emph{confocal ellipsoidal} coordinates; equivalently, the system admits three global isolating integrals (energy plus two more integrals) that are quadratic in the momenta \citep{deZ85, SB16}.%
\footnote{More generally, separability of the Hamilton--Jacobi equation in Euclidean 3-space is possible in several families of orthogonal coordinate systems; 
the confocal ellipsoidal system is the generic choice for genuinely triaxial separable potentials, while spheroidal (axisymmetric) and spherical cases appear as degenerate limits \citep{SB16}.}

The motion of particles in a general St\"ackel potential can be analytically integrated through algebraic operations, which is separable and integrable, such that one can obtain the integrals and actions from the phase space coordinate of a particle \citep{deZ85}. 
%Because the Hamilton--Jacobi equation separates in such potentials, the integrals and actions can be obtained from a particle's phase-space coordinates. 
% \citet{SB15} introduced a rapid algorithm to estimate the angle--action variables in a general triaxial potential -- the St\"ackel Fudge, which is an extension of the ideas in \citep{B12}. 
% Here we make a brief introduction to it. 
\citet{SB15} introduced a rapid algorithm to estimate the angle--action variables in a general triaxial potential -- the St\"ackel Fudge, which is an extension of the ideas in \citet{B12}. 
Here we give a brief introduction to it.
It assumes the target potential is sufficiently close to a St\"ackel potential over the region probed by a given orbit. 
The transformation from the Cartesian coordinates $(x, y, z)$ to $(\lambda, \mu, \nu)$ is
\begin{equation}\label{eq_x2tau}
\frac{x^2}{\tau+\alpha}+\frac{y^2}{\tau+\beta}+\frac{z^2}{\tau+\gamma}=1.
\end{equation}

%The solution $\mathbf\tau = (\lambda, \mu, \nu)$ is the triaxial confocal ellipsoid (TCE) coordinate, 
% The three roots $\boldsymbol{\tau}=(\lambda,\mu,\nu)$ define the triaxial confocal ellipsoid (TCE) coordinates,
The three roots $\boldsymbol{\tau}=(\lambda,\mu,\nu)$ define the triaxial confocal ellipsoidal (TCE) coordinates,
so that, away from the coordinate planes, eight Cartesian points related by sign changes share the same TCE coordinates. 
Set constants $\alpha=-a^2,\, \beta=-b^2,\, \gamma=-c^2$ and suppose $-\gamma\leq \nu\leq-\beta\leq \mu\leq-\alpha\leq\lambda$ without loss of generality. 
\par 

%\begin{figure}[H]
% figure name
\begin{figure*}[tbp]
	\centering
	\includegraphics[height=0.46\textwidth,width=0.46\textwidth]{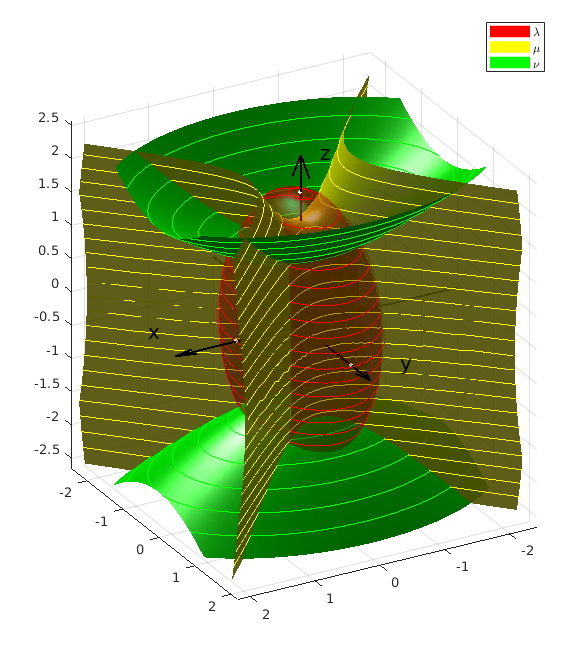} 
	\caption{
		Sketch of the triaxial confocal ellipsoidal coordinate system.
		The red surface is an iso-$\lambda$ surface, the yellow surface is an iso-$\mu$ surface, and the green surface is an iso-$\nu$ surface, shown in three-dimensional Cartesian coordinates.
	}
	\label{fig_ellips_coord_3d} 
\end{figure*}
\par

As $\tau$ varies, the squared semi-axis lengths $a_\tau^2,\, b_\tau^2,\, c_\tau^2$ of the coordinate surfaces change, 
while the focal distances $\Delta_1=\sqrt{a_\tau^2-b_\tau^2}$ and $\Delta_2=\sqrt{b_\tau^2-c_\tau^2}$ remain fixed; 
the resulting family of iso-$\tau$ surfaces therefore forms a set of confocal quadric surfaces in three-dimensional space. 
Fig.~\ref{fig_ellips_coord_3d} is a diagrammatic sketch of the TCE system. 
For $\tau=\lambda$, we have $\lambda+\alpha\geq 0,\, \lambda+\beta\geq 0,\, \lambda+\gamma\geq 0$, so each iso-$\lambda$ surface is an ellipsoid (the red surface in the figure).
% Similarly, since $\mu+\alpha\leq 0,\, \mu+\beta\geq 0,\, \mu+\gamma\geq 0$, the iso-$\mu$ surface represents the hyperboloid of one sheet (the yellow surface in figure); the iso-$\nu$ surface represents the hyperboloid of two sheet (the green surface in the figure).
Similarly, since $\mu+\alpha\leq 0,\, \mu+\beta\geq 0,\, \mu+\gamma\geq 0$, the iso-$\mu$ surface represents a one-sheet hyperboloid (the yellow surface in the figure); the iso-$\nu$ surface represents a two-sheet hyperboloid (the green surface in the figure).
The Cartesian coordinate as well as the spherical coordinate can be seen as when the foci go to the limit of TCE coordinate. 
\par

\subsection{The St\"ackel potential and the triaxial St\"ackel Fudge}\label{sec_SF}

The St\"ackel potential in TCE coordinates can be written as
\begin{equation}
\label{eq_pot_Sta}
\begin{split}
&\Phi_S = -\frac{\zeta(\lambda)}{P_\lambda^2}-\frac{\eta(\mu)}{P_\mu^2}-\frac{\xi(\nu)}{P_\nu^2} \\
&= -\frac{F_\lambda(\lambda)}{(\lambda-\mu)(\lambda-\nu)} \\
&{}-\frac{F_\mu(\mu)}{(\mu-\nu)(\mu-\lambda)}-\frac{F_\nu(\nu)}{(\nu-\mu)(\nu-\lambda)},
\end{split}
\end{equation}
where $F_\lambda(\lambda),\, F_\mu(\mu),\, F_\nu(\nu)$ can be arbitrary functions (see Eq.~(6) of \citealt{deZ85}). 
Here, $P_\tau$ denotes the metric scale factor of the confocal ellipsoidal coordinates for $\tau\in\{\lambda,\mu,\nu\}$. 
%For the convenience of writing, $F(\tau)$ is uniformly used later in this paper to refer to three separation functions $F_\lambda(\lambda),\, F_\mu(\mu),\, F_\nu(\nu)$, which can actually take different specific forms. 
%Note that the three separation functions $F_\lambda(\lambda),\, F_\mu(\mu),\, F_\nu(\nu)$ can actually take different specific forms. 
\par 

Section~4 of \citet{deZ85} and Section~3 of \citet{SB15} 
introduce the details of solving the Hamilton--Jacobi equation (HJE) in a St\"ackel potential 
and the numerical method to estimate frequencies and angles. Here we list the main equations used in the St\"ackel Fudge \citep{SB15}.
\par 

The equations of motion, derived from the HJE, can be written as
% = \frac{\tau^2 E -\tau a + b + F_\tau(\tau)}{2(\tau+\alpha)(\tau+\beta)(\tau+\gamma)} ~ ,
\begin{equation}
\label{eq_pt_ab}
\begin{split}
&p_\tau^2 = \left(\frac{\partial W_\tau}{\partial\tau}\right)^2 \\
&= \frac{\tau^2 E-\tau\mathcal K_1+\mathcal K_2+F_\tau(\tau)}{2(\tau+\alpha)(\tau+\beta)(\tau+\gamma)} ~,
\end{split}
\end{equation}
where $W_\tau$ is the Hamiltonian characteristic function, and $\mathcal{K}_a$ and $\mathcal{K}_b$ are separation constants. 
Then the actions are given by
\begin{equation}\label{eq_Jt}
J_\tau = \frac{1}{\pi} \int_{\tau_{-}}^{\tau_{+}} \mathrm{d} \tau\, p_\tau
\end{equation}
where $(\tau_{-},\, \tau_{+})$ are the roots of $p_\tau=0$, i.e., the roots of the right-hand side of Eq.~\ref{eq_pt_ab}.
\par

For a general gravitational potential $\Phi$, we define an auxiliary function $\chi_\tau$ 
\begin{equation}
\label{eq_chi_tau}
\begin{aligned}
\begin{split}
&\chi_\lambda(\lambda,\mu,\nu) = (\lambda-\mu)(\nu-\lambda)\Phi(\lambda,\mu,\nu) ~, \\
&\chi_\mu(\lambda,\mu,\nu) = (\mu-\nu)(\lambda-\mu)\Phi(\lambda,\mu,\nu) ~, \\
&\chi_\nu(\lambda,\mu,\nu) = (\nu-\lambda)(\mu-\nu)\Phi(\lambda,\mu,\nu) ~.
\end{split}
\end{aligned}
\end{equation}
If $\Phi$ were a St\"ackel potential, these quantities would be given by e.g. 
\begin{equation}\label{eq_chilam}
\begin{split}
&\chi_\lambda(\lambda,\mu,\nu) \\
=\, &F(\lambda)-\lambda\frac{F(\mu)-F(\nu)}{\mu-\nu}
+\frac{\nu F(\mu)-\mu F(\nu)}{\mu-\nu} ~.
\end{split}
\end{equation}
Therefore, for a general potential, one can write
\begin{equation}\label{eq_Ftau_CD}
F_\tau(\tau) \approx \chi_\tau(\lambda,\mu,\nu) + \tau C_\tau + D_\tau ~.
\end{equation}
where $C_\tau$ and $D_\tau$ are constants provided one always evaluates $\chi_\tau$ with two of the ellipsoidal coordinates fixed, i.e., one evaluates $\chi_\lambda$ at fixed $\mu$ and $\nu$.
Substituting these expressions into Eq.~\ref{eq_pt_ab}, we obtain 
\begin{equation}\label{eq_p2_JKtau}
\begin{split}
&2(\tau+\alpha)(\tau+\beta)(\tau+\gamma)p_\tau^2 \\
=\, &\tau^2 E - \tau A_\tau
+ B_\tau + \chi_\tau(\lambda, \mu, \nu) ~,
\end{split}
\end{equation}
where $A_\tau = \mathcal{K}_1-C_\tau$ and $B_\tau = \mathcal{K}_2+D_\tau$ are 
two approximate integrals used by the St\"ackel Fudge 
for each $\tau$ coordinate. 
\par 

\citet{SB15} evaluate $A_\tau,\, B_\tau$ by substituting the initial point $(\mathbf{\tau}_0,\, \mathbf{p}_{\tau0})$, from $(\mathbf x_0,\, \mathbf v_0)$, to Eq.~\ref{eq_p2_JKtau} and the property that in a St\"ackel potential $\partial H / \partial \tau = 0$ when other phase-space variables and the integrals are fixed, then 
\begin{equation}
\label{eq_Btau0}
\begin{split}
&B_\tau = 2(\tau_0+\alpha)(\tau_0+\beta)(\tau_0+\gamma)p_{\tau_0}^2 \\
&\quad {}-\tau_0^2 E+\tau_0 A_\tau-\chi_\tau(\lambda_0,\mu_0,\nu_0) ~,
\end{split}
\end{equation}
and 
\begin{equation}\label{eq_Atau0}
\begin{split}
A_\lambda\, =\, &(\mu_0+\nu_0)E 
+\frac{1}{2}\frac{p_{\mu_0}^2 (\lambda_0-\mu_0)}{P_{\mu_0}^2} \\
&+\frac{1}{2}\frac{p_{\nu_0}^2 (\lambda_0-\nu_0)}{P_{\nu_0}^2} ~, \\[0.3em]
A_\mu \, =\, &(\lambda_0+\nu_0)E 
+\frac{1}{2}\frac{p_{\lambda_0}^2 (\mu_0-\lambda_0)}{P_{\lambda_0}^2} \\
&+\frac{1}{2}\frac{p_{\nu_0}^2 (\mu_0-\nu_0)}{P_{\nu_0}^2} ~, \\[0.3em]
A_\nu \, =\, &(\lambda_0+\mu_0)E 
+\frac{1}{2}\frac{p_{\lambda_0}^2 (\nu_0-\lambda_0)}{P_{\lambda_0}^2} \\
&+\frac{1}{2}\frac{p_{\mu_0}^2 (\nu_0-\mu_0)}{P_{\mu_0}^2} ~.
\end{split}
\end{equation}
In contrast to Eq.~\ref{eq_pt_ab}, the approach for estimating the actions presented in Eq.~\ref{eq_p2_JKtau} 
does not require an explicit form for $F_\tau(\tau)$. 
%% Unlike Eq.~\ref{eq_pt_ab}, the approach to finding the actions presented in Eq.~\ref{eq_pt_ab} 
%% do not require an explicit form for $F_\tau(\tau)$. 
Note that because a Hamiltonian system with a true St\"ackel potential admits only three integrals, 
the six pseudo-integrals $(A_\lambda,\, A_\mu,\, A_\nu,\, B_\lambda,\, B_\mu,\, B_\nu)$ in St\"ackel Fudge 
are not independent.
For more details on the estimation of the actions, angles, and frequencies, see 
Appendix~A of \citet{SB15}.
\par
%%\citet{deZ85} Section 8 and \citet{SB15} Appendix A. \par 

\subsection{Progress in \DFJ\ forms}
\label{sec_append-A_DFJ-formulae}

The analytic functional forms adopted for \DFJ\ so far are mainly motivated by spherical or axisymmetric constructions. 
Consequently, flexible prescriptions for fully triaxial profiles with controllable shape, anisotropy, and truncation 
remain comparatively limited in the literature \citep{B14, Posti15, Cole17, Pascale19, BV23, Binney2026_spheroids}. 
\par 

%Besides from modeling with DF of integrals, 
Many studies have investigated the functional forms of $\DFJ$ for galactic halos. 
For example, \citet{B14} presented an analytic family $\DFJ$ 
%for those with isochrone potential, an extended it to include rotation. 
for systems in an isochrone potential and extended it to include rotation. 
\citet{Posti15} derived a semi-analytical $\DFJ$ model (their Equation~34) 
based on power laws constructed from linear combinations of the actions. %
%%% with power-laws of linear combinations of actions. 
\citet{WE15} proposed a more complicated family of $\DFJ$, which ``enables the velocity anisotropy of the model to be tuned, and so the anisotropy in the inner and outer parts can be specified for the application in hand''. 
%allow the inner and outer velocity anisotropies to be specified independently. 
%
\citet{SE15} proposed that action-based analytic DFs can be constructed --- using an iterative method for self-consistency --- 
to represent systems with triaxial density profiles; 
they presented two examples of such generalizations to triaxiality: 
(i) isophote twisting, based on the \citet{B14} model, 
and (ii) radially unstable models that tend to collapse to triaxial configurations, based on the \citet{WE15} model.
%%% 
% compute from a $\DFJ$ form to potential of its $\DFJ$. 
\citet{Cole17} added an auxiliary function (their Equation~5) to the $\DFJ$ formula of \citet{Posti15} 
to avoid divergence as the action combination approaches 0. 
\citet{Pascale18, Pascale19} presented a $\DFJ$ formula for halos, 
which is similar to that of \citet{Cole17} but includes an exponential tail (see their Equation~5). 
%\citet{BV23} had a new functional model of $\DFJ$ of halo with other coefficients of actions and explains what the parameters control. 
\citet{BV23} adopted a modified double-power-law $f(\mathbf{J})$ for the Galactic dark halo (and other hot components), 
which is related to that of \citet{Posti15} but incorporates two key modifications: 
(i) the phase-space density is regularized to ensure finiteness as $\mathbf{J}\!\to\!0$, 
following the mechanism introduced by \citet{Cole17}; 
and (ii) the linear combination of actions used in earlier double-power-law forms 
is replaced by a more general function of the actions 
to improve the behavior of the azimuthal velocity distribution at small $|V_\phi|$.
	%\citet{BV23} adopted for the dark halo (and other hot components) a modified double power-law $\DFJ$ related to that of \citet{Posti15}, 
	%with two key modifications: (i) the phase-space density is regularized to remain finite as $\mathbf{J}\!\to\!0$ 
	%following the mechanism introduced by \citet{Cole17}, 
	%and (ii) the linear action combination used in earlier double power-law forms is replaced by 
	%a more general function of the actions to improve the behavior of the azimuthal-velocity distribution at small $|V_\phi|$. 
%
In this framework, the parameters control the inner and outer density slopes, the outer truncation, 
as well as the degree of velocity anisotropy and flattening \citep{BV23,Binney2026_spheroids}.
\par 

However, most of these studies proposed $f(\mathbf{J})$ formulae for halos and other galactic components 
by ``deforming'' or generalizing from spherically symmetric gravitational potentials with known analytical $f(\mathbf{J})$ forms --- 
essentially adopting an ansatz, i.e., treating the functional forms thus approximated above 
as the DFs for realistic galaxies, which are often close to axisymmetric or triaxial.
Moreover, such approximation formulae do not necessarily guarantee the dynamical stability of 
the resulting galaxy or dark matter halo models (cf. \citealt{BV23,Binney2026_spheroids}). \\ % 
\par

\section{When a one--dimensional state density is required}\label{sec_action_state_density}
\begingroup\sloppy
The angle--action map $(\mathbf{x}, \mathbf{v})\!\mapsto\!(\mathbf{J}, \mathbf{\theta})$ is canonical, hence it preserves phase--space volume: $d^3x\,d^3v = d^3J\,d^3\theta$ and, after angle--averaging, $dN=(2\pi)^3 f(\mathbf{J})\,d^3J$.
Here we use the number-normalized phase-space convention for $f$, obtained by dividing the action-space mass density in Equation~(\ref{eq_density_kernel}) by $m_{\rm p}(2\pi)^3$, where $m_{\rm p}$ is the common particle mass within a component. This constant rescaling leaves the shape of the DF and the state-density and bin-averaging arguments below unchanged.
Therefore, when working in the full three-dimensional action space, no additional ``state--density'' factor appears. 
Nevertheless, it is often convenient --- for visualization, noise reduction, 
or fitting a low-dimensional representation of the DF --- to compress the actions into a scalar coordinate $u = h(\mathbf{J})$. 
A \emph{state density} then arises because this compression is a projection rather than an invertible change of variables: 
the push-forward of the measure introduces 
\begin{equation}
\label{eq_1d_state_density}
\begin{split}
&S_u(u)\;\equiv\;\int\delta\!\big(u-h(\mathbf J)\big)\,d^3J \;=\;\int_{\Sigma_u}\frac{d\Sigma}{\|\nabla_{\!J}h\|}, \\
&\Sigma_u = \{\mathbf J:h(\mathbf J)=u\},
\end{split}
\end{equation}
\endgroup
so that the number profile in $u$ reads
% M(u) &\;=\; \int f(\mathbf{J})\,\delta\!\big(u-h(\mathbf{J})\big)\,d^3J \\
% &\;=\; S_u(u)\, \overline{f}(u), \\
\begin{equation}
\begin{split}
&N(u)\;=\;(2\pi)^3\!\int f(\mathbf J)\,\delta\!\big(u-h(\mathbf J)\big)\,d^3J \\
&\;=\;(2\pi)^3\,S_u(u)\,\overline f(u), \\
&\overline f(u)\;\equiv\;\frac{1}{S_u(u)}\!\int f(\mathbf J) \delta\!\big(u-h(\mathbf J)\big)\,d^3J.
\end{split}
\end{equation}
Accordingly, $S_u(u)$ is \emph{needed} in two situations:
(i) when one aims to infer a \emph{pointwise} $f(u)$ from binned counts, in which case
% \widehat{f}(u_k) \;\approx\; \frac{M_k}{S_u(u_k)\,\Delta u_k},
\begin{equation}
\widehat{f}(u_k) \;\approx\; \frac{N_k}{(2\pi)^3\,S_u(u_k)\,\Delta u_k},
\end{equation} 
where $N_k$ is the number of particles in bin $k$ and $\Delta u_k$ is its width,
and (ii) when writing a likelihood in terms of the scalar variable $u$, because the sampling PDF is
$p(u)\propto (2\pi)^3 S_u(u)\,f(u)$ (the familiar energy--space analogue is $g(E)$).
By contrast, $S_u(u)$ is \emph{not} required when (a) fitting directly in $3$D action space, or
(b) performing a \emph{forward--modeled} 1--D fit: there one compares the model and data through the
\emph{same} binned functional (bin averages/medians over the identical memberships $B_k$), avoiding any
deconvolution to a pointwise $f(u)$.

\noindent\textit{Special cases.} If $h(\mathbf{J})=\mathbf{a}\!\cdot\!\mathbf{J}+b$ is linear and the admissible
$\mathbf{J}$--domain has a cross--section perpendicular to $\mathbf{a}$ that is independent of $u$, then
$S_u(u)$ is constant (conventionally absorbed into normalization). In the narrow--bin limit, or when
$\log_{10}f(u)$ is nearly linear within each bin, neglecting the variation of $S_u$ has negligible impact.
In general, $S_u(u)$ may be estimated by Monte--Carlo over the selected $\mathbf{J}$--domain:
draw $\mathbf{J}_m$ from a known proposal $\pi(\mathbf{J})$, form $u_m=h(\mathbf{J}_m)$, and use
weighted histograms with weights $w_m\propto 1/\pi(\mathbf{J}_m)$ to recover the shape of $S_u(u)$.

In the one-dimensional fitting branch we compress actions to a scalar 
$u = h(\mathbf{J})$ (here using the legacy frequency--combination),
and we compare \emph{bin--averaged} data to \emph{bin--averaged} model predictions.
Let $B_k$ be the set of particles whose $u_i = h(\mathbf{J}_i)$ fall in bin $k$, 
and let $\mathcal{A}[\cdot]$ denote the same per--bin aggregator used for the data 
(in practice, the mean or median of $\log_{10} f$).
The empirical bin statistic is
\begin{equation}
y_k^{\mathrm{data}} 
= \mathcal{A}\!\left\{\,\log_{10}\widehat{f}(\mathbf{J}_i)\;:\; i\in B_k\,\right\},
\end{equation}
where $\widehat{f}$ is the KDE or binned estimate in action space.
Instead of evaluating an analytic curve at bin centers 
(which implicitly replaces $\mathbb{E}[\,g(U)\,]$ by $g(\mathbb{E}[U])$ 
and can bias fits when the model is curved within a bin),
we \emph{forward--model} the same statistic:
\begin{equation}
y_k^\mathrm{model}(\mathbf p_\mathrm{all}) = \mathcal A\!\left\{\,\log_{10}f\!\big(h(\mathbf J_i);\mathbf p_\mathrm{all}\big) \;:\;i\in B_k\,\right\},
\end{equation}
using the identical bin membership $B_k$ and aggregator $\mathcal{A}$.
Here, $\mathbf p_\mathrm{all}$ denotes the fitting parameters.
We then determine $\mathbf p_\mathrm{all}$ by minimizing the mismatch between $\{y_k^{\mathrm{model}}\}$ and $\{y_k^{\mathrm{data}}\}$ with the same bin weights or uncertainties as the data.
This procedure aligns the estimated targets (a bin--averaged statistic) 
between data and model, avoids within--bin curvature bias, and reduces exactly to the ``fit--to--centers'' shortcut 
in the limit of sufficiently narrow, uniformly populated bins. \\
\par

%%% \FloatBarrier

\section{Fitting results for the remaining galaxy models} \label{fig_fit_others}

\begin{figure*}[tbp]
	
	\centering
	\begin{subfigure}[t]{0.46\textwidth}
		\centering
		\includegraphics[width=\linewidth]{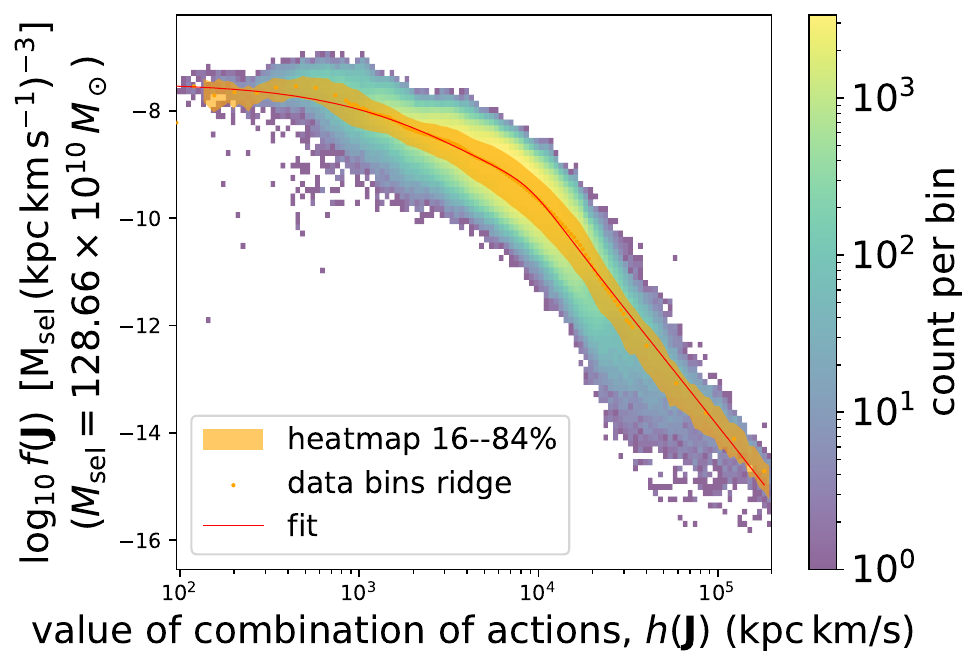}
		\caption{
			Demonstration of the fitting result, expressed in the form of distribution function of actions versus the action combination $h$. 
		}
		\label{fig_compare6_1}
	\end{subfigure}
	\hspace{0.06\textwidth}
	\begin{subfigure}[t]{0.46\textwidth}
		\centering
		
		\includegraphics[width=\linewidth]{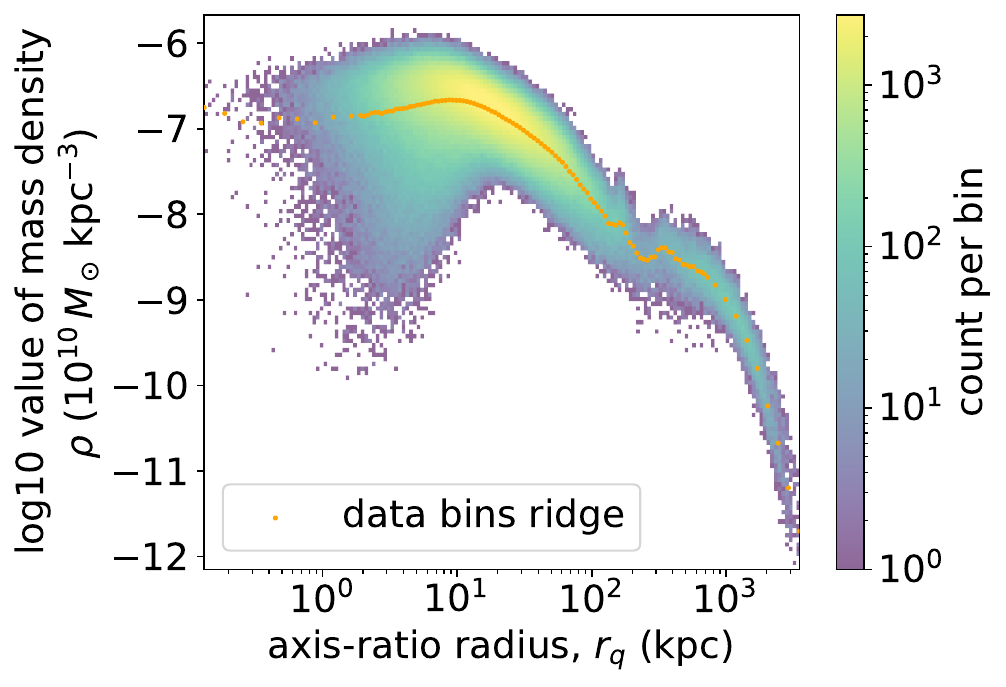}
		\caption{
			The mass density profile $\rho$ as a function of the dimensionless flattened ellipsoidal radius $r_\mathrm{sq}$ (Eq.~\ref{eq_rsq}). 
		}
		\label{fig_compare6_2}
	\end{subfigure}
	\par\medskip
	\begin{subfigure}[t]{0.9\textwidth}
		\centering
		\includegraphics[width=\linewidth,height=0.50\textheight,keepaspectratio]{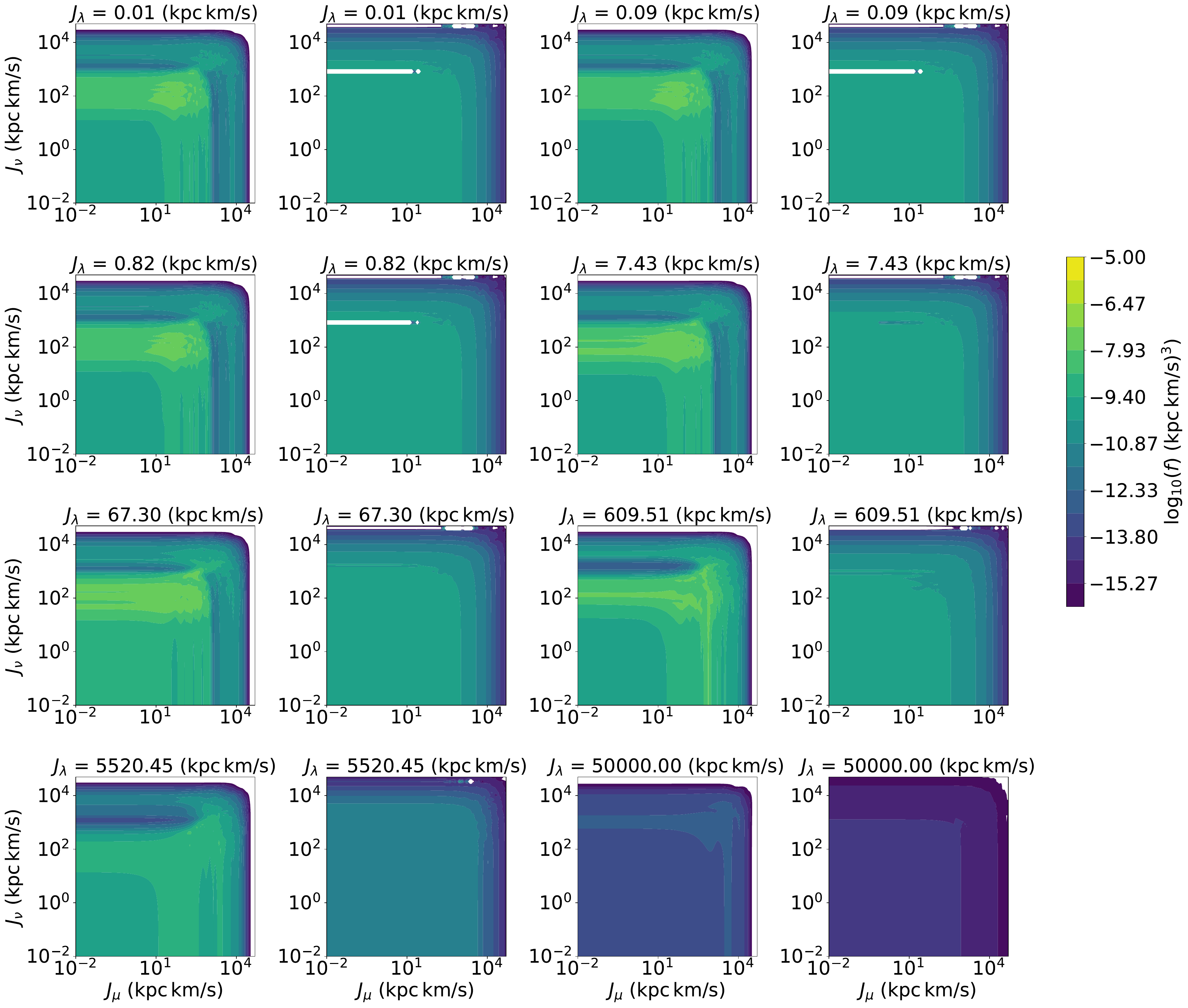}
		\caption{
			Comparison of $\DFnJ$ contours in the $(J_\mu,J_\nu)$ plane at several fixed $J_\lambda$ slices. 
		}
		\label{fig_compare6_3}
	\end{subfigure}

	\caption{
		Comparison between the numerical data and the best-fit formula (Eq.~\ref{eq_MPLTF}) for the mass-weighted distribution function of actions $\DFnJ$.
		As in Fig.~\ref{fig_fit_compare5}, but for the simulated galaxy \texttt{Ein\_spinL\_axisLH1} in Table~\ref{table_galaxy_params}.
	}
	\label{fig_fit_compare6} 
\end{figure*}

\begin{figure*}[tbp]
	
	\centering
	\begin{subfigure}[t]{0.46\textwidth}
		\centering
		\includegraphics[width=\linewidth]{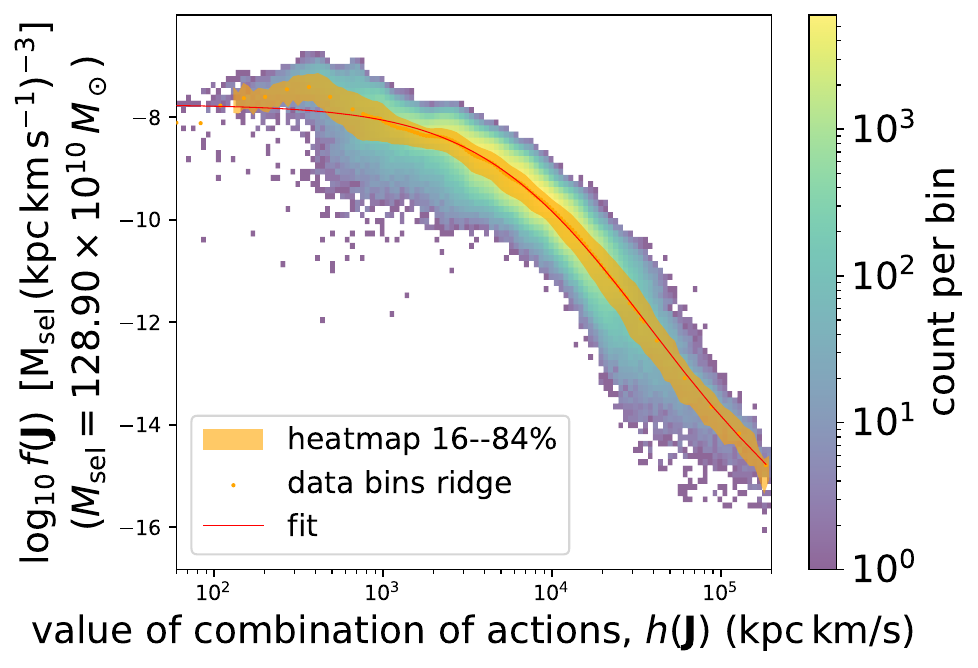}
		\caption{
			Demonstration of the fitting result, expressed in the form of distribution function of actions versus the action combination $h$. 
		}
		\label{fig_compare7_1}
	\end{subfigure}
	\hspace{0.06\textwidth}
	\begin{subfigure}[t]{0.46\textwidth} 
		\centering
		
		\includegraphics[width=\linewidth]{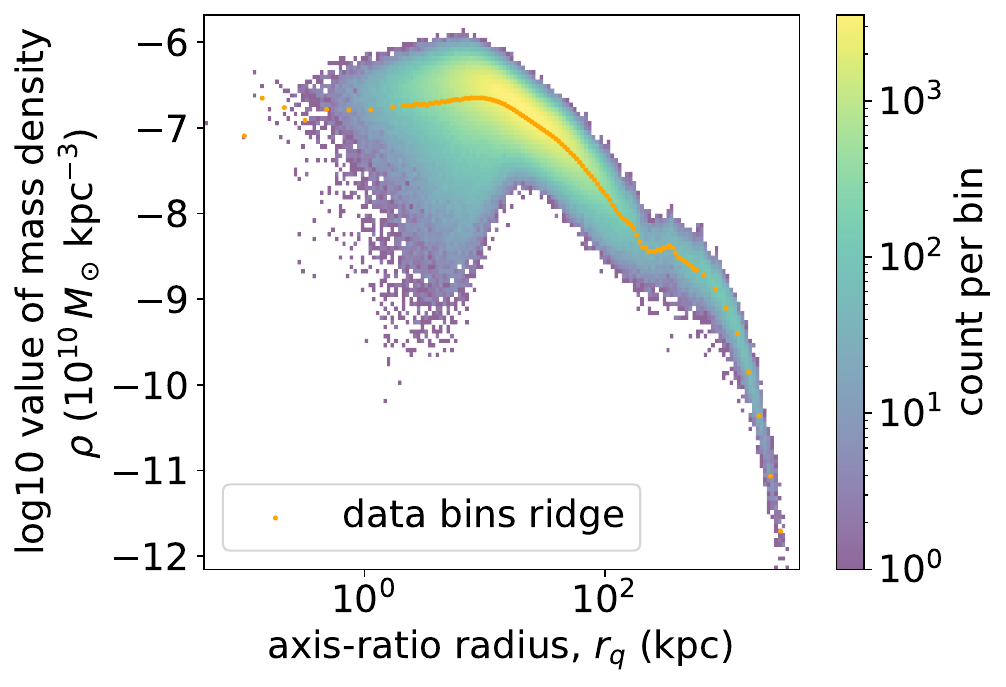}
		\caption{
			The mass density profile $\rho$ as a function of the dimensionless flattened ellipsoidal radius $r_\mathrm{sq}$ (Eq.~\ref{eq_rsq}). 
		}
		\label{fig_compare7_2}
	\end{subfigure}
	\par\medskip
	\begin{subfigure}[t]{0.9\textwidth}
		\centering
		\includegraphics[width=\linewidth,height=0.50\textheight,keepaspectratio]{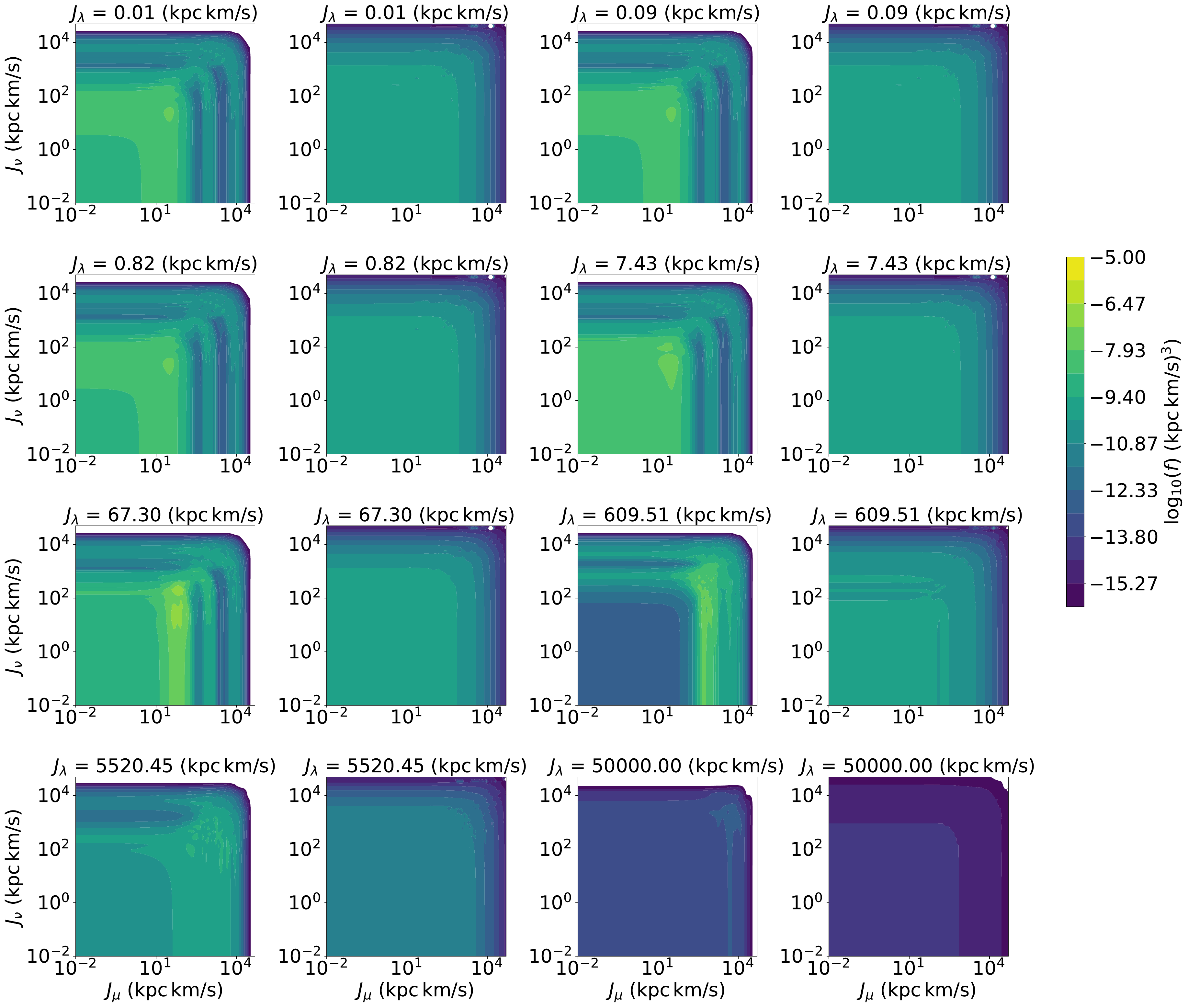}
		\caption{
			Comparison of $\DFnJ$ contours in the $(J_\mu,J_\nu)$ plane at several fixed $J_\lambda$ slices. 
		}
		\label{fig_compare7_3}
	\end{subfigure}

	\caption{
		Comparison between the numerical data and the best-fit formula (Eq.~\ref{eq_MPLTF}) for the mass-weighted distribution function of actions $\DFnJ$.
		As in Fig.~\ref{fig_fit_compare5}, but for the simulated galaxy \texttt{Ein\_spinH\_axisLH0} in Table~\ref{table_galaxy_params}.
	} 
	\label{fig_fit_compare7} 
\end{figure*}

\begin{figure*}[tbp]
	
	\centering
	\begin{subfigure}[t]{0.46\textwidth}
		\centering
		\includegraphics[width=\linewidth]{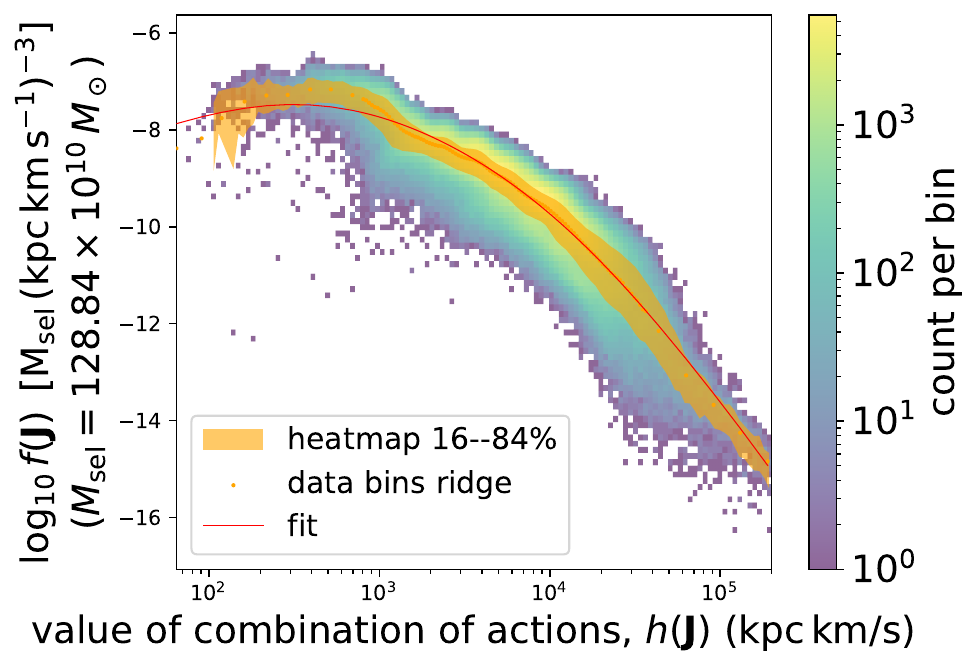}
		\caption{
			Demonstration of the fitting result, expressed in the form of distribution function of actions versus the action combination $h$. 
		}
		\label{fig_compare8_1}
	\end{subfigure}
	\hspace{0.06\textwidth}
	\begin{subfigure}[t]{0.46\textwidth} 
		\centering
		
		\includegraphics[width=\linewidth]{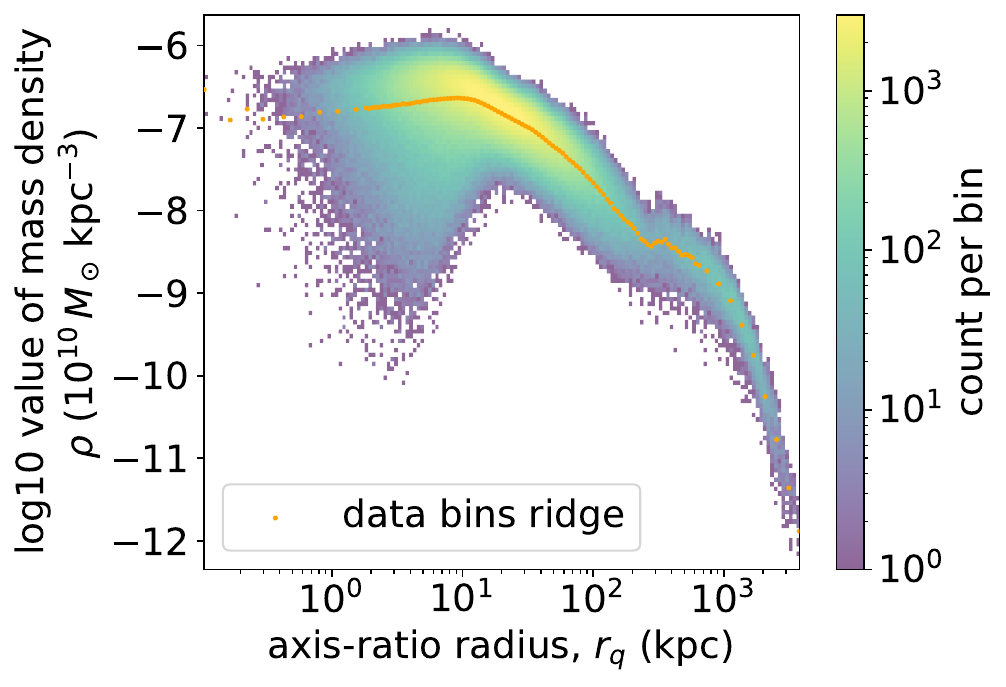}
		\caption{
			The mass density profile $\rho$ as a function of the dimensionless flattened ellipsoidal radius $r_\mathrm{sq}$ (Eq.~\ref{eq_rsq}). 
		}
		\label{fig_compare8_2}
	\end{subfigure}
	\par\medskip
	\begin{subfigure}[t]{0.9\textwidth}
		\centering
		\includegraphics[width=\linewidth,height=0.50\textheight,keepaspectratio]{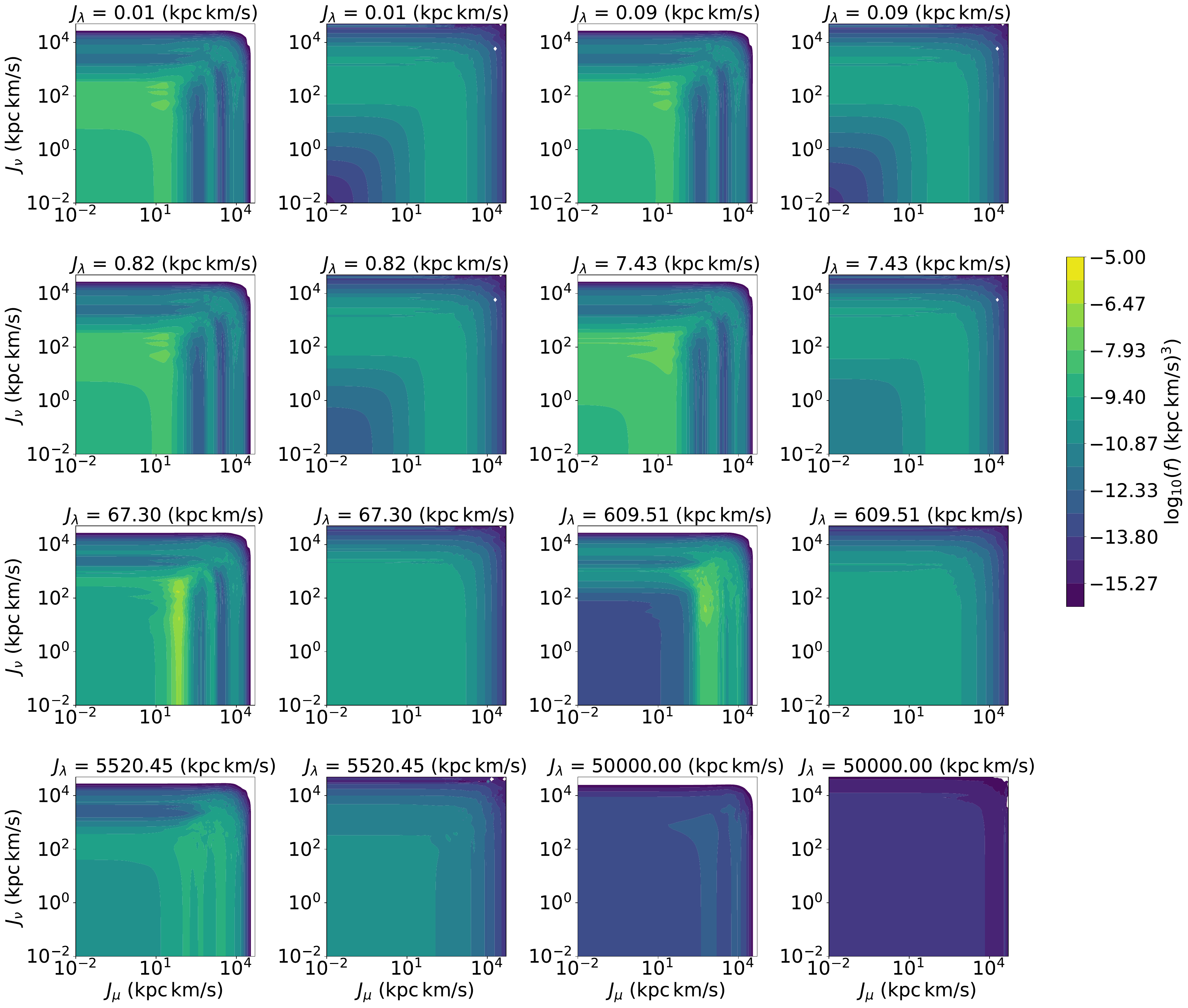}
		\caption{
			Comparison of $\DFnJ$ contours in the $(J_\mu,J_\nu)$ plane at several fixed $J_\lambda$ slices. 
		}
		\label{fig_compare8_3}
	\end{subfigure}

	\caption{
		Comparison between the numerical data and the best-fit formula (Eq.~\ref{eq_MPLTF}) for the mass-weighted distribution function of actions $\DFnJ$.
		As in Fig.~\ref{fig_fit_compare5}, but for the simulated galaxy \texttt{Ein\_spinH\_axisLH1} in Table~\ref{table_galaxy_params}.
	} 
	\label{fig_fit_compare8} 
\end{figure*}

In this Appendix, we present the fitting results 
for the remaining three galaxy models listed in Table~\ref{table_galaxy_params}.
Figures~\ref{fig_fit_compare6}, \ref{fig_fit_compare7}, and \ref{fig_fit_compare8} show the results for \texttt{Ein\_spinL\_axisLH1}, \texttt{Ein\_spinH\_axisLH0}, and \texttt{Ein\_spinH\_axisLH1}, respectively.
\par 

%%% \clearpage
%[h]
%\begin{figure*}[p]
% No blank line means no line breaks
%\hfill %Add horizontal blank space appropriately
%% between panels A and B
%
% \includegraphics[width=\linewidth]{\detokenize{pics__fit_EL1__snapshot_30.fit.txt.massdensity.type_1.png}}
% No blank line means no line breaks %\hfill %Add horizontal blank space appropriately
% No blank line means no line breaks
%Panel (a) shows the one-dimensional fit in $h$, panel (b) shows the corresponding mass density profile, and panel (c) compares the numerical and fitted $\DFnJ$ contours in action space.
\par

%\clearpage
%[h]
%\begin{figure*}[p]
% No blank line means no line breaks
%\hfill %Add horizontal blank space appropriately
%% between panels A and B
%
%panel B
% \includegraphics[width=\linewidth]{\detokenize{pics__fit_EH0__snapshot_30.fit.txt.massdensity.type_1.png}}
% No blank line means no line breaks %\hfill %Add horizontal blank space appropriately
% No blank line means no line breaks
% Comparison between the numerical data and the best-fit formula (Eq.~\ref{eq_MPLTF}) for the number density distribution of actions $\DFnJ$.
%Panel (a) shows the one-dimensional fit in $h$, panel (b) shows the corresponding mass density profile, and panel (c) compares the numerical and fitted $\DFnJ$ contours in action space. 
% figure title
\par

%\clearpage
%[h]
%\begin{figure*}[p]
% No blank line means no line breaks
%\hfill %Add horizontal blank space appropriately
%% between panels A and B
%
%%panel B
% \includegraphics[width=\linewidth]{\detokenize{pics__fit_EH1__snapshot_30.fit.txt.massdensity.type_1.png}}
% No blank line means no line breaks %\hfill %Add horizontal blank space appropriately
% No blank line means no line breaks
% Comparison between the numerical data and the best-fit formula (Eq.~\ref{eq_MPLTF}) for the number density distribution of actions $\DFnJ$.
%Panel (a) shows the one-dimensional fit in $h$, panel (b) shows the corresponding mass density profile, and panel (c) compares the numerical and fitted $\DFnJ$ contours in action space. 
% figure title
\par

%
%%% \clearpage
%\cleardoublepage
%\onecolumngrid
%\onecolumn
\bigskip
\bigskip

\clearpage

\section{The figures of mass contours and potential contours}
\label{contours_of_mass_and_potential}

\begin{figure*}[tbp]

	\centering
	\includegraphics[width=\textwidth]{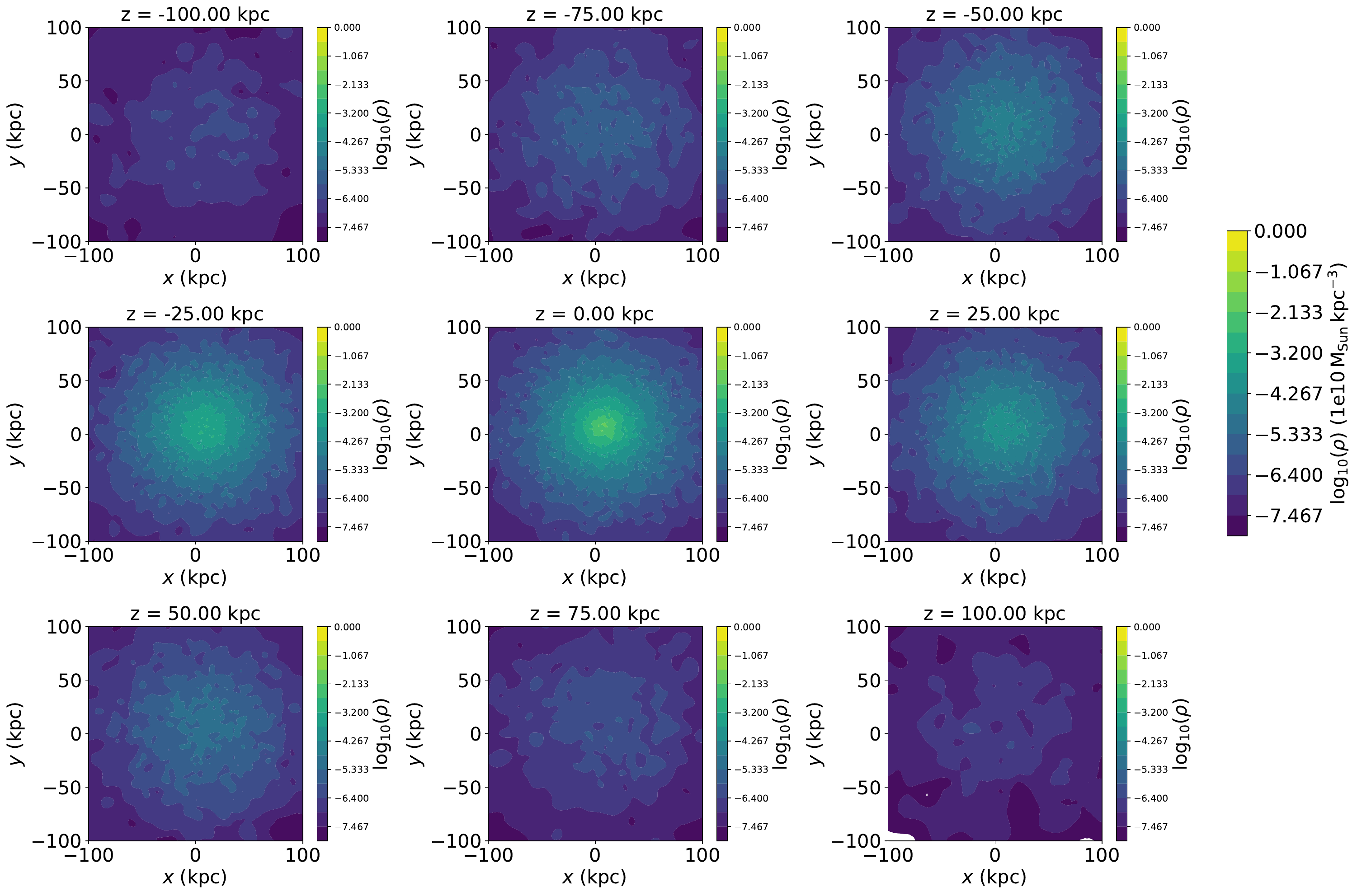}
	\caption{
		Mass density distribution in the $x$--$y$ plane at a series of vertical heights $z$ for the triaxiality-aligned galaxy GAL0. 
		Each panel corresponds to a specific $z$ slice (indicated in the panel), with the plotted $(x,y)$ region covering the range that encloses 98\% of the particles. 
		The colorbar on the right indicates the mass density on a logarithmic scale (in units of $10^{10}$ $\mathrm{M}_\odot\,\mathrm{kpc}^{-3}$).
	}
	\label{fig_mass_contour}
\end{figure*}

\begin{figure*}[tbp]

	\centering
	\includegraphics[width=\textwidth]{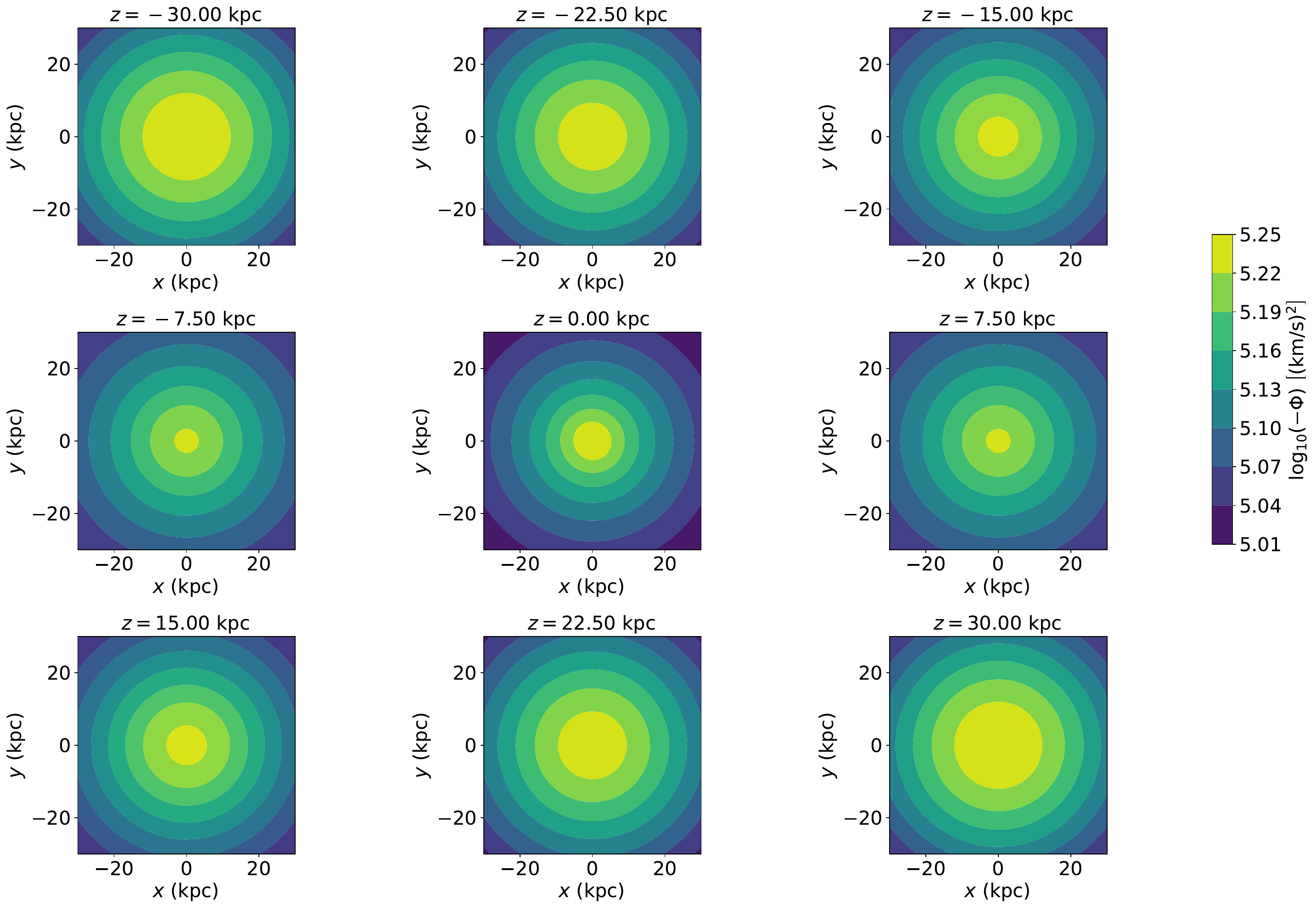}
	\caption{
		Gravitational potential contours derived from the self-consistent field (SCF) expansion for the triaxiality-aligned galaxy GAL0. 
		Each panel corresponds to a specific $z$ slice (indicated in the panel). 
		The colorbar on the right indicates 
		the base-10 logarithm of the absolute value of the potential, $\log_{10}|\Phi|$, 
		where $\Phi$ is in units of $(\mathrm{km\,s^{-1}})^2$.
	}
	\label{fig_potential_contour}
\end{figure*}

Fig.~\ref{fig_mass_contour} shows the direct-summation mass density distributions and Fig.~\ref{fig_potential_contour} shows the SCF potential distributions,
for the triaxiality-aligned galaxy GAL0 (corresponding to \texttt{Ein\_spinL\_axisLH0} in Table~\ref{table_galaxy_params}). 
\par

%% In the main text:
%% Fig.~\ref{fig_mass_contour} of \ref{contours_of_mass_and_potential} demonstrates 
%% the mass density distributions in the $x$--$y$ plane
%% at different $z$ positions for the triaxiality-aligned galaxy GAL0,
%% with the $(x, y)$ range of each panel enclosing $98%$ of the particles.
%[h]
%\begin{figure*}[p]
%% Mass density distributions of the triaxiality-aligned galaxy GAL0,
%% displayed in the in the $x$--$y$ plane at different $z$ positions.
%% %Each panel shows the mass density distribution in a different $z$ plane.
%% The color bar gives $\log_{10}$ value of the mass density.
% The colorbar on the right indicates the mass density on a logarithmic scale (in units of 1e10 $\mathrm{M}_\odot\,\mathrm{kpc}^{-3}$).
\par
\bigskip

%[h]
%\begin{figure*}[p]
%% SCF potential contours of the triaxiality-aligned galaxy GAL0.
%% Each panel shows the SCF potential distribution in a different $z$ plane.
%% The color bar gives $\log_{10}$ log10 of the potential.
\par

\clearpage

\section{Further comparisons of input and output distributions}
\label{app_DF_phase_space_comparison}

\begin{figure*}[tbp]
	\centering
	\includegraphics[width=\textwidth,height=0.66\textheight,keepaspectratio]{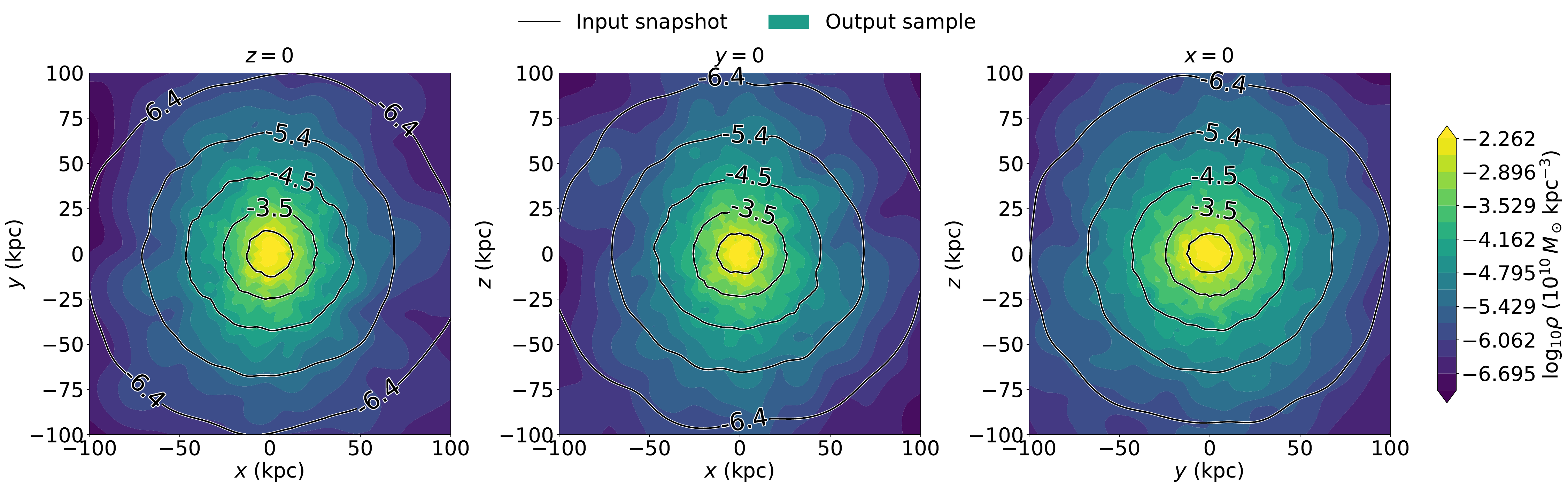}
	\caption{Comparison of mid-plane slices of the spatial density for the GAL0 input snapshot (black contours) and output sample (filled color maps).
		The left, middle, and right panels show the $z=0$, $y=0$, and $x=0$ planes, respectively.
		The samples have the same total assigned mass.
		Both the labels of the contour levels and the color scale denote the base-10 logarithm of 
		the spatial density (in units of $10^{10}\,\mathrm{M}_\odot\,\mathrm{kpc}^{-3}$); 
		the color scale is common to all three panels.	
	}
	\label{fig_DF_spatial_maps}
\end{figure*}

\begin{figure*}[tbp]
\centering
	\includegraphics[width=\textwidth,height=0.66\textheight,keepaspectratio]{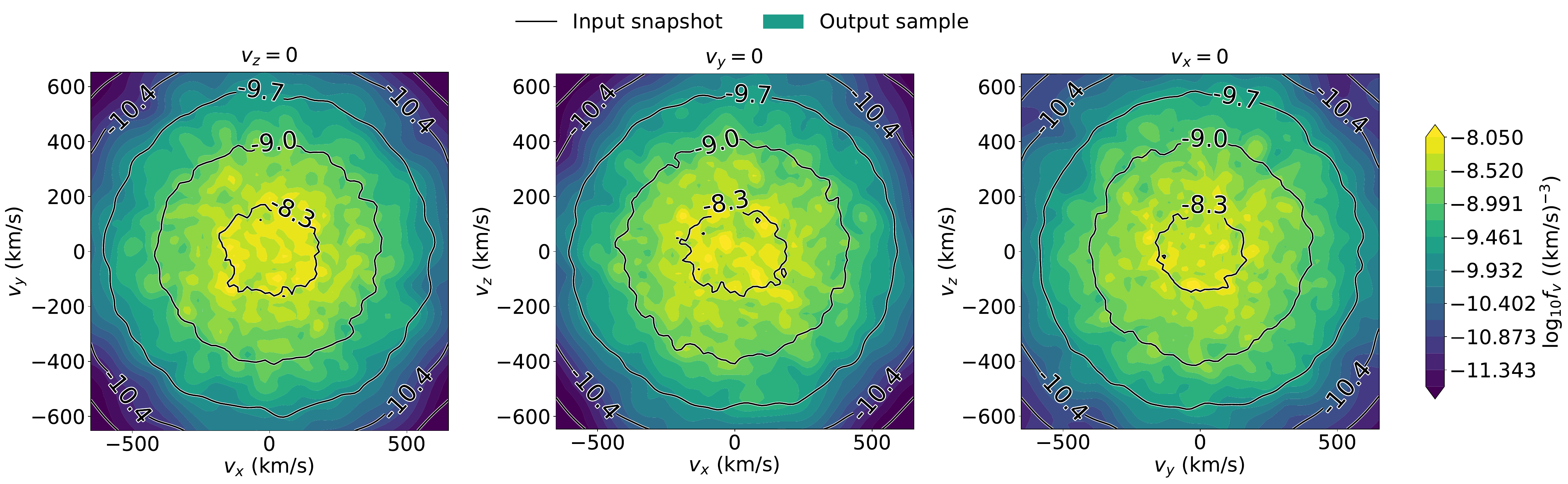}
	\caption{Comparison of mid-plane slices of the velocity probability-density for the GAL0 input snapshot (black contours) and output sample (filled colour maps), integrated over the selected spatial region.
		The left, middle, and right panels show the $v_z=0$, $v_y=0$, and $v_x=0$ planes, respectively.
	 		Both the labels of the contour levels and the color scale denote the base-10 logarithm of 
	 		the velocity probability density (in units of $(\mathrm{km\,s^{-1}})^{-3}$); 
	 		the color scale is common to all three panels.
	}
	\label{fig_DF_velocity_maps}
\end{figure*}

%% mock mcmc
{To supplement the comparison in Section~\ref{sec_DF_phase_space_validation}, Figure~\ref{fig_DF_spatial_maps} 
compares the mid-plane slices of the three-dimensional spatial density for the GAL0 input snapshot and output sample.
Likewise, Figure~\ref{fig_DF_velocity_maps} compares the mid-plane slices of the three-dimensional velocity probability density. }
\par 

{The output sample contains $10^5$ particles generated from the fitted DF with 12 Markov chains after burn-in, subject to the selection in Section~\ref{sec_DF_phase_space_validation} and finite action bounds.%
\footnote{The sampler evaluates $h_\mathrm{freq}$ using the magnitudes of the action and frequency components.
The bounds are $10^{-2}\leq |J_i|\leq5\times10^5$ for each action component and $10^{-2}\leq h_\mathrm{freq}\leq5\times10^5$, in $\mathrm{kpc\,km\,s^{-1}}$.}
For the monitored quantities---radius, energy, $h_\mathrm{freq}$, and the three actions---the maximum split-$\widehat{R}$ is 1.016 and the minimum estimated effective sample size is approximately $1.0\times10^3$. }
% The mock particles retain correlations from the Markov chains. 
\par 

{We apply the same density-estimation method to both the input snapshot and the output sample.
The three-dimensional densities use the compact spline kernel of Equation~(\ref{eq_kernel_gadget}) 
with neighbour counts proportional to the sample sizes.
The spatial-density slices use the common mass normalization specified in Section~\ref{sec_DF_phase_space_validation}. }
\par 

As evidently seen from Figure~\ref{fig_DF_spatial_maps} and Figure~\ref{fig_DF_velocity_maps}, 
the two-dimensional density maps are consistent with the one-dimensional histograms 
shown in Figures~\ref{fig_DF_position_marginals} and \ref{fig_DF_velocity_marginals} of the main text. 
In both spatial and velocity space, the output samples drawn from the best-fitting $f(\mathbf{J})$ closely reproduce the input $N$-body snapshots. 
This agreement confirms that the fitted distribution function provides a reasonably accurate representation of the simulated halo. 
\par

%\clearpage

%
%%% \clearpage
%\twocolumngrid
%\twocolumn

\section{A simple running guide for \texttt{GroArnold} }\label{appendix_guide}

%An example of running the workflow is shown below, supposing your terminal is in \path{GroArnold_framework/install_and_run/}. \par 
\begin{sloppypar}
\texttt{GroArnold} is controlled by the workflow script
\path{GroArnold_framework/install_and_run/workflow_wrapper.py}.
Most user-interface parameters are set in the model directory, e.g. 
\path{install_and_run/initial_conditions/settings_Ein_multicomp_spinL_axisLH_3p3Gyr_1e6count/},
which contains \path{unified_settings.yaml} together with the \texttt{DICE} and \texttt{Gadget2} parameter files.
\end{sloppypar}
\par 

The setting for $h=h_\mathrm{freq}$ in the setting file \path{unified_settings.yaml}
is as follows: 
\begin{Verbatim}[breaklines=true,breakanywhere=true]
is_fit_1d: true
fitting_model: fitting_model_MPLTF_freq
\end{Verbatim}

Users can also choose $h=h_\mathrm{freefixed}$. 
The setting for $h=h_\mathrm{freefixed}$ in the setting file is as follows:
\begin{Verbatim}[breaklines=true,breakanywhere=true]
is_fit_1d: false
fitting_model: fitting_model_MPLTF_freefixed
\end{Verbatim}

\subsection{Running the full pipeline on an example model}

The command below runs the complete workflow for a single setting directory, including
initial-condition generation, $N$-body evolution, snapshot preprocessing, action estimation, and DF fitting and plotting. 
The example assumes your current directory is \path{GroArnold_framework/install_and_run/}.
\par 

For long runs, it can be convenient to 
%detach the process, 
run the process by the following Shell script (in Linux):
\begin{Verbatim}[breaklines=true,breakanywhere=true]
nohup python3 workflow_wrapper.py ./initial_conditions/settings_Ein_multicomp_spinL_axisLH_3p3Gyr_1e6count/ all --no-skip-existing > run_spinL_full.log 2>&1 &
\end{Verbatim}
\par 

\subsection{Outputs}

%By default, results for each model are written under the project data directory \path{GroArnold_framework/GDDFAA/}, 
%  %%The fitted DF products and summary plots are collected in the corresponding \path{fit/} subdirectory of the run, 
%alongside intermediate snapshots, action files, etc., in other subdirectories.
%The fitted DF products and summary plots are stored in the \path{fit/} subdirectory of each run, 
%alongside related outputs (e.g., snapshots and action files) located in neighboring subdirectories. 
%

By default, all outputs for a given model are organized under the working directory \path{GroArnold_framework/GDDFAA/}. 
Intermediate files, such as simulation snapshots and action estimates, are stored in separate subdirectories, 
while the final fitted distribution functions and summary plots are placed in the \path{fit/} subdirectory of each run.
\par

\end{document}